\documentclass[aps,prb,twocolumn,floatfix,superscriptaddress,reprint]{revtex4-1}
\usepackage[utf8] {inputenc}
\usepackage[english] {babel}
\usepackage{graphicx}
\graphicspath{{./}}

\usepackage{dcolumn}
\usepackage{bm}
\usepackage{mathtools}
\usepackage{float}
\usepackage{multirow}
\usepackage{amsmath, amsfonts, amssymb}

\usepackage{lmodern}
\usepackage{textcomp}
\usepackage{t1enc}

\usepackage{graphicx}
\usepackage{setspace}

\usepackage[toc, page]{appendix}
\usepackage{standalone}

\usepackage{verbatim}

\usepackage{natbib}

\begin{document}


\title{\emph{Ab initio} theory of the nitrogen-vacancy center in diamond}


\author{\'Ad\'am Gali}
\affiliation{%
 Wigner Research Centre for Physics, Hungarian Academy of Sciences, PO. Box 49, Budapest H-1525, Hungary}
 \affiliation{%
 Department of Atomic Physics, Budapest University of Technology and Economics, Budafoki út 8., H-1111, Budapest, Hungary}

\date{\today}

\begin{abstract}
Nitrogen-vacancy center in diamond is a solid state defect qubit with favorable coherence time up to room temperature which could be harnessed in several quantum enhanced sensor and quantum communication applications, and has a potential in quantum simulation and computing. The quantum control largely depends on the intricate details about the electronic structure and states of the nitrogen-vacancy center, radiative and non-radiative rates between these states and the coupling of these states to external spins, electrical, magnetic and strain fields and temperature. In this review paper it is shown how first principles calculations contributed to understanding the properties of nitrogen-vacancy center, and will be briefly discussed the issues to be solved towards full \emph{ab initio} description of solid state defect qubits.  
\end{abstract}

\maketitle


\section{Introduction}
\label{sec:intro}
We briefly introduce a prominent solid state point defect, the nitrogen-vacancy center in diamond, which acts as a quantum bit, the elementary unit of quantum information processing. A desiderata is then provided what properties should be determined for understanding solid state defect quantum bits. Finally, the content of this overview is shortly summarized.  

\subsection{Nitrogen-vacancy center - a brief overview}
\label{ssec:nv}
\begin{figure}
\includegraphics[width=0.95\columnwidth]{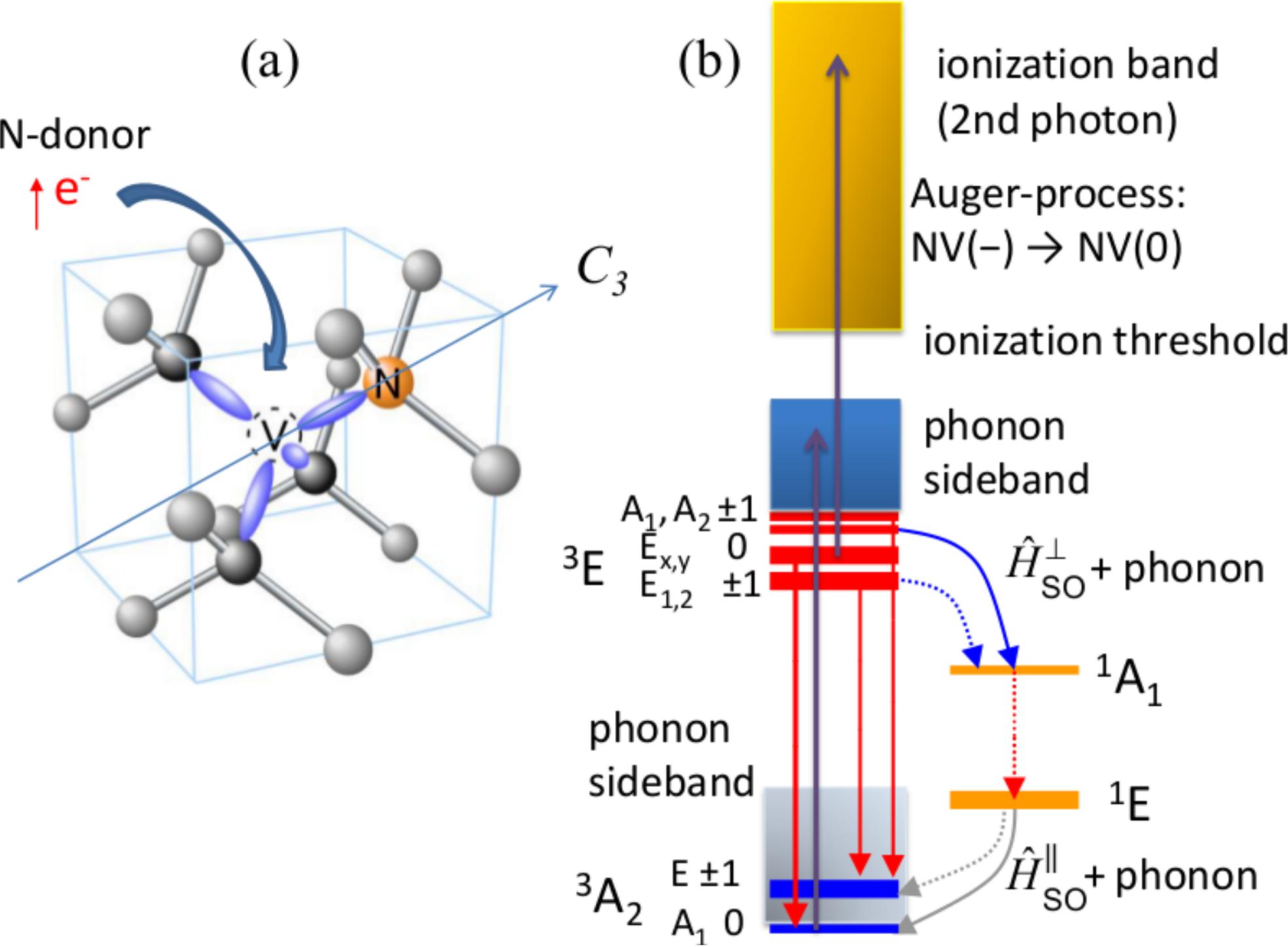}
\caption{\label{fig:struct}(a) Nitrogen-vacancy (NV) defect in diamond. Vacancy is depicted as a circle in the middle of the diamond cage (grey balls are carbon atoms). An electron is provided by the diamond environment, e.g., N-donor, that creates the negatively charged NV defect, i.e., NV($-$), briefly NV center. Three carbon atoms and the nitrogen atom have dangling bonds pointing toward the vacancies. The defect has three-fold rotation axis ($C_3$) about the $\langle111\rangle$ axis of diamond. The eigenstates of NV center can be labeled by $C_{3v}$ symmetry irreducible representations. (b) Electronic structure and decay processes. The double group representations for the triplet states are also depicted with their corresponding spin eigenstates. Red arrows represent emission (weak between singlets). The curved arrows show the intersystem crossing processes that are mediated by phonons and the parallel or perpendicular components of spin-orbit interaction ($\hat{H}_\text{SO}$), where the dominant components are shown by thicker lines. Dark grey arrows show the absorption that may lead to the ionization of NV center. Note that hyperfine interaction occurs between the electron spin and nitrogen nuclear spin (not depicted).}
\end{figure}

Point defects may introduce levels in the fundamental band gap of semiconductors or insulators that radically change optical and magnetic properties of the host material. In particular, these point defects could be paramagnetic, i.e., the electron spin is greater than zero. A primary example of such a point defect is the nitrogen-vacancy (NV) center in diamond~\cite{duPreez1965}. This defect consists of a nitrogen atom substituting a carbon atom near a missing carbon atom in diamond crystal, i.e., vacancy of diamond. It can accept an electron from the environment and can be negatively charged [see Fig.~\ref{fig:struct}(a)]. In this charge state, multiple levels appear in the fundamental band gap of diamond occupied by four electrons. The corresponding electron configurations constitute an $S=1$ ground state and an optically active $S=1$ excited state with below band gap (5.4~eV) excitation energy~\cite{Goss1996, Gali2008, Larsson2008}. As a consequence, the defect has single photon absorption and emission spectra in the visible region in the transparent diamond host. Thus, NV center is also called color center in diamond. The absorption and emission spectra are broad even at cryogenic temperatures caused by the coupling of phonons to the optical transitions. The no-phonon-line or zero-phonon-line (ZPL) optical transition appears at 637~nm (1.945~eV)~\cite{Davies1976}. In particular, the contribution of the ZPL emission to the total emission, i.e., the Debye-Waller factor, is about 0.03, with a relatively large Stokes-shift of about 0.45~eV. This makes possible to excite NV center by green light (typically, 532-nm excitation wavelength) and to detect the emitting photons in the near infrared region of 700-900~nm. The $S=1$ spin in the ground state can be measured by conventional electron spin resonance (ESR) techniques~\cite{Loubser1977}. It is important to note that this defect \emph{always} is a two-spin system as the nitrogen isotopes either have $I=1$ (${}^{14}$N) or $I=1/2$ (${}^{15}$N) nuclear spin. As a consequence, the hyperfine interaction between the electron spin and nitrogen nuclear spin always occurs and nuclear quadrupole coupling appears for the largely abundant $^{14}$N isotope. In addition, hyperfine signature of ${}^{13}$C $I=1/2$ nuclear spins is also observable (natural abundance of 1.1\%) in the corresponding ESR spectrum~\cite{Loubser1978, Gali2008, Felton2009}. The spin Hamiltonian of the NV center in the ground state can be written as
\begin{equation}
\label{eq:Hspin}
\begin{aligned}&
H = \hat{S} \mathbf{g} B + \hat{S} \mathbf{D} \hat{S} + C_Q \hat{I}_\text{N} + \hat{S} \mathbf{A_\text{N}} \hat{I}_\text{N} + \sum_i \hat{S} \mathbf{A}_{\text{C}_i} \hat{I}_{\text{C}_i} 
\\&
- g_\text{N} \hat{I}_\text{N} B - \sum_i g_\text{C} \hat{I}_{\text{C}_i} B \text{,}
\end{aligned}  
\end{equation}     
where the gyromagnetic factors of the electron spin were first observed as isotropic $g_{xx}=g_{yy}=g_{zz}=2.0028\pm0.0003$ (Refs. ~\onlinecite{Loubser1978, He1993}), whereas slight anisotropy, $g_{xx}=g_{yy}=2.0029(2)$ and $g_{zz}=2.0031(2)$, was reported in Ref.~\onlinecite{Felton2009}, which results in the corresponding Zeeman splitting upon external magnetic field $B$. $D$ is the so-called zero-field-splitting tensor, where "zero field" refers to zero magnetic field. In this particular system, the eigenstates of this part of the Hamitonian can be given as $D (S^2_z-2/3)$, where $S_z= \{-1; 0; +1\}$ is the eigenvalue of the electron spin and $D$ zero-field-splitting constant is about 2.87~GHz~\cite{Loubser1978}. This term predominantly arises due to the dipolar electron-spin -- electron-spin interaction, and results in a $D$ energy gap between the $ms=0$ and $ms=\pm1$ levels in the ground state at zero magnetic field when the quadrupole and hyperfine terms are neglected. $C_Q$=$-4.945$~MHz (Ref.~\onlinecite{Pfender2017}) and $A_{zz, \text{N}}=-2.14\pm0.07$~MHz; $A_{xx, \text{N}}=A_{yy, \text{N}}=A_{\perp, \text{N}}=-2.70\pm0.07$~MHz are the quadrupole strength and hyperfine principal values of ${}^{14}$N (Ref. \onlinecite{Felton2009}), respectively, which are always present for each individual NV center. $A$ hyperfine tensor is nearly isotropic for the given nitrogen isotope but can be highly anisotropic for the proximate ${}^{13}$C nuclear spins~\cite{Loubser1978, Gali2008, Felton2009, Smeltzer2011} which generally reduce the symmetry of the spin Hamiltonian. For distant ${}^{13}$C nuclear spins, the simple point dipole -- point dipole approximation for the hyperfine interaction between the electron spin and nuclear spins is valid, however, the Fermi-contact term, i.e., the localization of the electron spin density at the site of nuclear spins, can be significant for the proximate ${}^{13}$C nuclear spins~\cite{Gali2008, Felton2009, Smeltzer2011}. When the magnetic field approaches the zero-field-splitting then the nuclear Zeeman terms of the nitrogen nuclear spin and ${}^{13}$C spins with the $g_\text{N}$ and $g_\text{C}$ nuclear gyromagnetic ratio, respectively, become an important effect~\cite{Ivady2017}.      

A key property of NV center is that the electron spin resonance is correlated with the spin-selective fluorescence intensity~\cite{Gruber1997} and two-photon ionization probability~\cite{Bourgeois2015}, where the latter occurs by subsequent absorption of two photons via the $^3E$ excited state. If the concentration of NV centers is low in diamond and the photo-excitation source is focused into a single NV center then the electron spin resonance of single NV center can be detected either optically~\cite{Gruber1997} or electrically~\cite{Siyushev2019} that are called optically detected magnetic resonance (ODMR) and electrically detected magnetic resonance (EDMR). Since the latter was achieved by measuring the photocurrent it is also called photocurrent based detected magnetic resonance (PDMR). In the ODMR measurements it was found that the fluorescence intensity between the $ms=0$ levels of the excited state and ground state is about 30\% stronger than that between $ms=\pm1$ levels of the excited state and ground state. This is also called ODMR readout contrast. In other words, this readout process converts the electron spin resonance frequency used in electron paramagnetic resonance techniques in the microelectronvolt region into the frequency of optical photons in the electronvolt region. In addition, the number of optical photons emitted in the ODMR process is several orders of magnitude larger per center than the single microwave photon absorbed in the electron spin resonance process which increases the detection sensitivity by large amount. Another consequence of the ODMR process is that optical pumping of the NV center results in almost 100\% population of $ms=0$ level in the ground state~\cite{Loubser1977,  Loubser1978, Childress2006}. It has been recently found that similar PDMR readout contrast is achievable for single NV centers~\cite{Siyushev2019}.  In the PDMR readout, the readout process ends at the neutral NV defect~\cite{Mita1996}. Further optical pumping turns the neutral NV defect into the negatively charged NV defect, i.e., NV center~\cite{Gaebel2006, Aslam2013, Siyushev2013}. 

The readout and electron spinpolarization processes inherently contain flipping of the electron spin during the decay from the excited state to the ground state which is highly selective to the $ms=\pm1$ spin state of the triplet excited state. Selection rules in the decay processes can be efficiently analyzed by group theory. NV center in diamond exhibits $C_{3v}$ symmetry with a symmetry axis lying along the $\langle111\rangle$ directions of diamond~\cite{Davies1976}. By combining the defect-molecule diagram and group theory~\cite{Coulson1957, Lenef1996, Manson2006}, one can predict the character of the triplet ${}^3A_2$ ground and ${}^3E$ excited state as well as the dark singlet states also as a function of external perturbations~\cite{Maze2011, Doherty2011}. Spin-orbit interaction may connect the triplet states and singlet states resulting in non-radiative decay and spin flipping. This type of non-radiative decay is called intersystem crossing (ISC). Understanding ISC is the key for ODMR and PDMR readout.       

The connection between quantum information science and NV centers is intimately bound to the ODMR and PDMR processes and readout. These readout processes make it possible to coherently manipulate single electron spins in solids by microwave fields and optical excitation~\cite{Jelezko2004-1, Childress2006}, where two states of the electron spin of NV center realize a quantum bit, which can be read out and initialized by optical means. These readout mechanisms operate at room temperature~\cite{Gruber1997, Siyushev2019} and even at higher temperatures~\cite{Toyli2012}, and the measurement scheme can be pushed up to 1000~$^\circ$C with pulsed protocols~\cite{Liu2019}. The coherence time of NV center's electron spin in diamond with natural abundant ${}^{13}$C isotopes can reach $\approx$600~$\mu$s~\cite{Childress2006, Maze2008} and up to $2$~ms in ${}^{12}$C enriched diamonds~\cite{Bala2009} as obtained by Hahn-echo measurements, even at room temperature. By control-NOT operation~\cite{Jelezko2004-2, Neumann2010}, the quantum bit information can be written from the electron spin to the nuclear spin. Single-shot readout of the quantum bit was demonstrated~\cite{Robledo2011}, as well as coding of quantum information from the electron spin~\cite{Togan2010} or from the nuclear spin~\cite{Yang2016} to the polarization of the emitted photon or the photon emission itself at a given frequency~\cite{Robledo2011}, which realize spin-to-photon interfaces at cryogenic temperature. This was used to transmit quantum information over 1 km distance~\cite{Hensen2015}. One can conclude that NV center in diamond satisfies DiVincenzo's criteria of quantum information processing~\cite{DiVincenzo2000}. In the followings, we list these criteria that we amend with the NV properties: (1)  Scalable physical system with well characterized quantum bits, i.e., the electron and nuclear spins of single NV center; (2) The ability to initialize the state of the quantum bits to a simple fiducial state, such as initialization of $ms=0$ state by optical means; (3) Long relevant decoherence times, much longer than the gate operation time, i.e., millisecond coherence time of NV center; (4) A “universal” set of quantum gates, e.g., control-NOT operation, as demonstrated for NV center;  (5) A quantum bit-specific measurement capability, i.e., ODMR or PDMR readout; (6) The ability to interconvert stationary and flying qubits, such as spin-to-photon interface of NV center; (7) The ability faithfully to transmit flying qubits between specified locations as demonstrated between remote NV centers. Scalability is still an issue because identical NV centers require Stark-shift tuning of the levels in the excited state~\cite{Tamarat2008, Bassett2011, Pfaff2014}, and despite the technological efforts on creating arrays or clusters of NV centers in diamond~\cite{Meijer2005, Toyli2010, Lesik2013, Yamamoto2013, Haruyama2019}, a robust coherent coupling mechanism between multiple NV centers has not yet been demonstrated for more than three NV centers~\cite{Haruyama2019}.      

NV center in diamond has a favorable electron spin coherence time but this coherence time and the (spin) levels are relatively sensitive to the environment. This is undesirable for building up a quantum computer from NV centers but can be harnessed in the measurement of magnetic~\cite{Maze2008, Balasubramanian2008, Degen2008}, electric~\cite{Dolde2011}, and strain~\cite{Teissier2014, Barfuss2015, MacQuarrie2015prb, Golter2016} fields, and temperature~\cite{Kucsko2013, Toyli2013, Neumann2013} at the nanoscale. The room temperature operation and favorable coherence time of NV center paves the way towards nuclear magnetic resonance of single molecules at ambient conditions~\cite{Staudacher2013, Mamin2013, Devience2015, Haberle2015, Rugar2015, Boss2017, Schmitt2017, Aslam2017, Glenn2018}. Furthermore, the possibility of spinpolarization transfer from the electron spin towards the nuclear spins~\cite{Jacques2009, Gali2009prb13C, Smeltzer2011, Ivady2015} can be employed to hyperpolarize diamond particles~\cite{Chen2015, Alvarez2015, King2015, Scheuer2016, Wunderlich2017, Ajoy2018, Schwartz2018} or external species~\cite{Abrams2014, Fernandez-Acebal2018} attached to the diamond surface. In these quantum sensing and related applications, the NV center should reside close to the surface of diamond.

Theory proposed~\cite{Cai2013, Chou2017} that near-surface NV centers can be also used as a resource to carry out quantum simulations on frustrated quantum magnetism or quantum spin liquid if the surface of diamond is decorated by species with nuclear spins, e.g., $I=1/2$ fluorine~\cite{Cai2013} or $I=1$ ${}^{14}$N (Ref.~\onlinecite{Chou2017}), respectively.   

NV centers can be observed in natural diamonds but controlled preparation is required for the afore-mentioned quantum technology applications. The intentional production of NV defects often starts with a high quality diamond that was grown by chemical vapor deposition with minimal contamination. In the next stage, nitrogen ions are implanted into diamond. After implantation the diamond sample is annealed, in order to remove the damage created by implantation and facilitate the formation of NV center in diamond~\cite{Meijer2005, Rabeau2006, Pezzagna2010, Ofori2012, Lesik2013, Yamamoto2013, vanDam2019, Haruyama2019}. The negative charge state of the NV defect is most likely provided by substitutional nitrogen donor defects that are called P1 center named after its ESR signature~\cite{Smith1959, Loubser1965, Farrer1969}. Shallow NV centers near the surface of diamond are dominantly created by this implantation technique~\cite{Ofori2012}, where the depth of the defects can be controlled by energy of the bombarding ions. It was found that the coherence time and often the photostability of these shallow NV centers are compromised~\cite{Ofori2012}, which might be either related to the quality of the diamond surface or the quality of diamond crystal around the shallow NV centers. Indeed, nitrogen delta-doping technique of diamond growth for creating shallow NV centers improved the coherence time~\cite{Ohno2012}.

In some sensing applications, large ensembles of NV centers are needed. In that case, the starting material could be a heavily nitrogen-doped diamond and then vacancies can be formed by different irradiation techniques (such as electron~\cite{Collins2009, Acosta2009} or neutron irradiation~\cite{Mita1996}, or implantation with inert ions~\cite{Waldermann2007, Huang2013}) in that sample. Finally, annealing is applied to remove the irradiation damages and facilitate the formation of NV defects~\cite{Mainwood1994}. It was found that the optimization of the annealing stages is very important to achieve good coherence properties of the resulting NV centers~\cite{Capelli2019}.   

This brief overview could lead to the impression that the NV center in diamond is probably the most studied point defect in the experiments. Still not all the signatures are well interpreted and understood purely from experimental spectra and simple models. \emph{Ab initio} theory could significantly contribute to understanding the formation, photoexcitation, photoionization processes as well as the ISC processes of NV centers. In particular, results from \emph{ab initio} calculations could guide the idea of improving the coherence properties of NV centers after nitrogen implantation~\cite{Favaro2017}, the notion~\cite{Siyushev2013, Bourgeois2015} and optimization~\cite{Bourgeois2016, Brandt2017, Gulka2017} of PDMR readout of NV center in diamond. We provide a recipe below how to achieve the full \emph{ab initio} description of a solid state defect quantum bit on the exemplary NV center in diamond. 

\subsection{\emph{Ab initio} description of NV center in diamond: a desiderata}
\label{ssec:wishlist}
Two major goals can be identified in the \emph{ab initio} description of NV center in diamond: (i) creation of the NV center and its interaction with the other point defects and diamond surfaces, (ii) determining ionization energies and magneto-optical properties with the corresponding radiative and non-radiative rates, also as a function of external perturbations such as magnetic and electric fields, strain and temperature. 

Issue (i) is a very common target in point defect studies, where the formation energy or enthalpy of the point defects should be determined, and the surface of the host should be studied as a function of the environment, which may influence the surface morphology and termination of the host material. However, issue (ii) requires in-depth investigation of the behavior of a point defect, much deeper than usual in the community of researchers working on point defects in solids, as the thermodynamic properties of point defects are often the main target in the vast majority of such investigations. However, vacancy formation is not necessarily a quasiequilibrium process. ODMR and PDMR signals of the NV center arise from photoexcitation, which is out of thermal equilibrium of the electrons. In addition, the magnetic properties play a central role in solid state defect quantum bits that should be described in details both in the ground state and excited states. 

The list of properties for full \emph{ab initio} description of NV center can be recognized in Fig.~\ref{fig:struct}(b). In the ground state, the zero-field-splitting due to dipolar electron spin - electron spin interaction should be computed. The hyperfine interaction of the electron spin with the nuclear spins is also very important in the entanglement schemes~\cite{Childress2006}, quantum memory~\cite{Fuchs2011, Maurer2012, Yang2016}, quantum error correction~\cite{Waldherr2014}, hyperpolarization~\cite{Jacques2009} schemes as well as in understanding the decoherence of the NV center's electron spin. For instance, by controlling the proximate nuclear spin states around NV center, the coherence time of the electron spin has been pushed beyond one second~\cite{Abobeih2018}. 
Calculation of the nitrogen quadrupole constant can be important in understanding the so-called "dark state" NMR of ${}^{14}$N (Ref.~\onlinecite{Waldherr2011}), which can be used to identify the charge state of the NV defect~\cite{Pfender2017}. 

In the $^3E$ excited state, the spin levels are heavily temperature dependent and show complex features as observed in the photoluminescence excitation (PLE) spectrum~\cite{Batalov2009}. The zero-field-splitting occurs due to the dipolar spin-spin interaction but also spin-orbit interaction takes place. As a consequence, the optically active spin triplet state should be calculated with the corresponding electron dipolar spin-spin interaction and spin-orbit interaction. 

In the ODMR readout process, the calculation of ISC rate requires the spin-orbit matrix element between the triplet states and the corresponding singlet states [see Fig.~\ref{fig:struct}(b)]. This assumes that the optically inactive or dark singlet excited states and levels should be calculated too. 

In the PDMR readout, the probability of photoionization either directly from the ground state or via the real excited state is a key issue. Photoionization could be direct or via Auger-process, thus the photoionization rate of both processes should be calculated. These non-radiative rates compete with the radiative rate from the $^3E$ excited state to the $^3A_2$ ground state, so the latter should be also determined for the sake of a complete description. 

Phonons are involved all of these processes, thus the electron-phonon coupling should be calculated, in order to understand the Debye-Waller factor of the PL spectrum and the ISC processes between the triplet and singlet states~\cite{Goldman2015prl, Thiering2017}.

One should recognize that all the interactions between electron orbitals, electron and nuclear spins, phonons and external fields should be considered and calculated \emph{ab initio} for a complete description of the operation of solid state defect quantum bits in a realistic solid and environment. At least, two main challenges can be identified: (a) calculation of the excited states for sufficiently large models of NV center and (b) treatment of the electron-phonon coupling in the radiative and non-radiative processes. In the followings, recent efforts along these directions are presented together with the other developments and results in the field.  

\subsection{Contents of the review paper}
\label{ssec:toc}
The rest of the paper is structured as follows. The basic \emph{ab initio} methods for studying solid state defect quantum bits are presented in Sec.~\ref{sec:methods}. We then summarize the recent methodology developments for the description of NV center in diamond and the corresponding results in Sec.~\ref{sec:dev_res}. This section starts with the basic modeling and ground state properties in thermal equilibrium, then it continuous with the treatment of excited states and optical properties which includes the discussion of the participation of phonons in the optical transition. Next, the computational methods of the magnetic parameters are discussed, which would complete the description of the magneto-optical properties of NV center in diamond. This is the starting point to determine the radiative and non-radiation rates as described in the next section. Finally, the simulation tools for calculating the various sources of perturbation on the afore-mentioned properties are presented that concludes Sec.~\ref{sec:dev_res}. An outlook is provided for the next steps towards full \emph{ab initio} description of NV center in diamond in Sec.~\ref{sec:outlook}. Finally, the paper is summarized in Sec.~\ref{sec:summary}.

\section{Computational methods}
\label{sec:methods}
The physics of solid state defect quantum bits is the physics of point defects in solids. As a consequence, understanding solid state defect quantum bits means to develop and apply tools to explore the properties of point defects in solids. The most employed \emph{ab initio} technique to study point defects in solids is the plane wave supercell Kohn-Sham density functional theory (DFT) calculation. Two aspects are mentioned in this statement: (i) modeling of point defects in solids, (ii) computational methodology for determining the electronic structure.

\begin{figure}
\includegraphics[width=0.95\columnwidth]{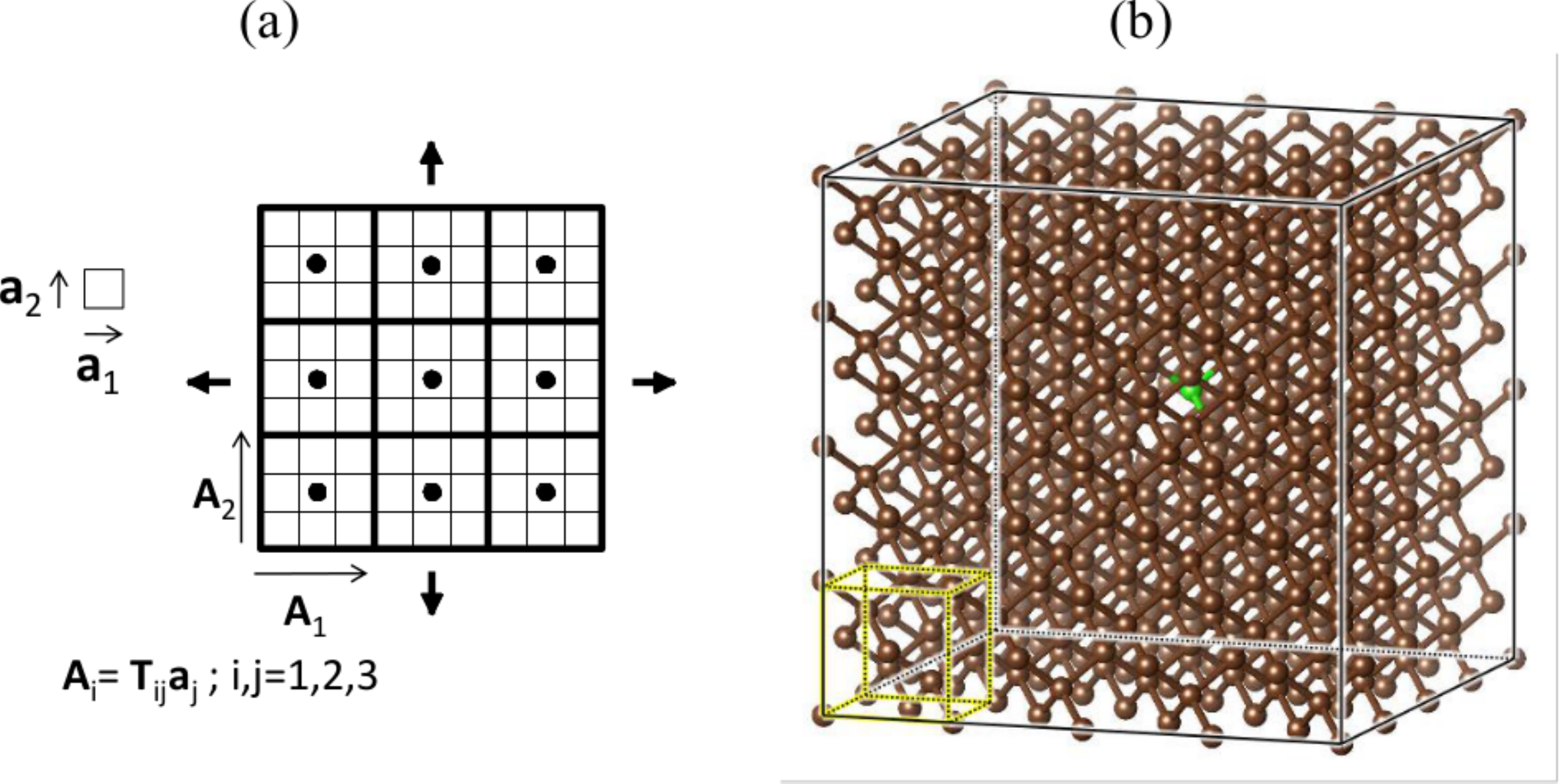}
\caption{\label{fig:supercell}(a) Schematic diagram of the supercell method for modeling of point defects in a solid. The small square is the unit cell. The point defect is represented by a dot which is embedded in $3\times3$ supercell in this partcular example. Generally, the non-singular transformation matrix $\mathbf{T}$ contains only integers which transforms the lattice vectors of the unit cell ($\mathbf{a}$) to those of the supercell ($\mathbf{A}$). The periodic images of the defect may represent a problem, in particular, for charged defects. (b) 512-atom simple cubic supercell of diamond hosting nitrogen-vacancy center (carbon and nitrogen atoms are brown and green balls, respectively). The Bravais unit cell is also shown as small cube.}
\end{figure}

In the supercell model, the point defect is placed in a cluster of the host material which has a periodic boundary in each direction; in other words, the cluster with the defect is a unit cell (see Fig.~\ref{fig:supercell}). If the size of the cluster is sufficiently large then the defect can be considered as isolated. In practice, the concentration of the defect in the modeling is much higher than that in the experiments because of the limits of the computational capacity. That may lead to dispersion of the defect levels in the fundamental band gap which is a clear sign of the interaction between the periodic images of the defect. We note that the k-points in the Brillouin-zone (BZ) of the supercell are folding into the BZ of the primitive cell~\cite{Evarestov1975}, thus integration of reduced number of k-points in the BZ of the supercell may result in converged wave functions and electron charge density. This reduced number of k-points can be generated by Monkhorst-Pack (MP) scheme~\cite{MP1976}. For sufficiently large supercells, single $\Gamma$-point BZ sampling suffices. We note here that only $\Gamma$-point calculation guarantees that all the symmetry operations appear for the respective wave functions and charge density, which is an important issue in the investigation of degenerate orbitals and levels. In addition, a practical advantage of $\Gamma$-point calculations is that the wave functions are real which reduces the computational capacity and time. In recent years, the NV center is often embedded into a simple cubic supercell of diamond lattice that originally contains 512 atoms as shown in Fig.~\ref{fig:supercell}(b). In this case, $\Gamma$-point sampling is near the absolute convergent k-point set but larger supercell might be needed for highly accurate calculations (e.g., spin-orbit coupling in Ref.~\onlinecite{Thiering2017}). We note that the need of relatively large diamond cluster for accurate calculation of NV center presently excludes to use quantum chemistry methods that are based on the extension of Hartree-Fock method because of the intractable computational capacity and time, or those can be applied with compromising the accuracy by the small size of the diamond cluster~\cite{Zyubin.JCompChem.09, Delaney2010}. Therefore, another approach has to be applied.  

Kohn-Sham (KS) DFT~\cite{Kohn1965} has been very powerful method to determine the ground state of solids. The literature is very rich about KS-DFT (e.g., Ref.~\onlinecite{Book_Parr}), which is not repeated here. Briefly, the total energy of the system, $E$ is a functional of the electron charge density $n(r)$ of the interacting electron system, where $n(r)$ can be expressed by non-interacting KS single particle wave functions, $\phi_i$ as $n(r) = \sum_i c_i \left|\phi_i\right|^2$ and $c_i$ is the occupation number of $\phi_i$ KS wave function.  The key expression in KS DFT is the exchange-correlation potential and functional, which is universal for a given number of electrons of the system. In theory, there exists such an exact exchange-correlation functional. However, this functional is unknown and is approximated in practice. The most simple but unexpectedly successful approximation is the local density functional theory (LDA), where the exchange-correlation functional is calculated from that of the homogeneous electron gas at each point~\cite{C-A80}. A powerful extension of LDA was achieved by taking the generalized gradient approximation (GGA) of the electron charge density. In particular, the Perdew-Burke-Ernzerhof (PBE) functional is often applied in the family of GGA functionals~\cite{PBE}. These functionals can be extended to spinpolarized electron systems, where the functionals will also depend on the spin state. In that case the spin density, $n_s(r) = n_\uparrow (r) - n_\downarrow (r)$ with $n_\uparrow (r)$ and $n_\downarrow (r)$ spin-up and spin-down densities, respectively, whereas $n(r)= n_\uparrow (r) + n_\downarrow (r)$, in which the corresponding spin-up and spin-down KS wave functions and spin densities are varied independently in the sense that the wave functions are not bound to form the spin eigenstate of the system. If the final solution is not a spin eigenstate of the system then the solution suffers from "spin contamination". In practice, this "spin contamination" is tiny in LDA or PBE calculations. A key problem of these functionals is that they suffer from the so-called self-interaction error which results in too low band gap of semiconductors or insulators. In particular, the calculated band gaps of diamond is about 4.2~eV in LDA or PBE DFT calculations. On the other hand, the electron charge and spin density of the ground state can be well calculated for many systems including point defects in solids such as NV center in diamond~\cite{Gali2008, Gali2009prb13C, Smeltzer2011}. Nevertheless, improvement in the applied functional or correction to the self-interaction error was needed, in order to calculate the ionization energies and excitation energies of the NV center in diamond and other solid state quantum bits. This will be discussed in the next section.

It is important to note that the external potential of the system in KS DFT equations is the potential of ions, i.e., the Coulomb potential of carbon and nitrogen atoms of the NV center in diamond. This method inherently treats ions in a semiclassical fashion in the sense that the ions are classical particles but the vibration of these classical particles in the self-consistent adiabatic potential energy surface (APES) can be calculated at quantum mechanical level. In other words, we apply Born-Oppenheimer approximation which separates the electronic and ionic degrees of freedom. This approximation works generally well for many systems but may fail for degenerate orbitals and levels, where vibrations may effectively couple those states. We will show below that the description of the double degenerate ${}^3E$ excited state of the NV center requires to go beyond Born-Oppenheimer approximation (BOA). In the APES, the global minimum energy can be found by minimizing the quantum mechanical forces acting on the ions that can be calculated analytically in KS DFT by applying the Hellmann-Feynman theorem. This is called geometry optimization procedure for a given electronic configuration. The vibrations can be calculated in the quasiharmonic approximations by moving the ions out of equilibrium, and fitting a parabola to the resulting energy differences around the global energy minimum. This procedure sets up a Hessian matrix which can be diagonalized to obtain the vibration (phonon) modes and eigenvectors.

In practice, $\phi_i$ KS wave functions should be expanded by known basis functions or calculated numerically on a grid. In the former case, a natural choice for three-dimensional system with periodic boundary conditions is the plane wave basis set, i.e. solution of Bloch-states. An advantage of the plane wave basis set is that the numerical convergence can be straightforwardly checked by adding more-and-more plane waves to the basis set. The disadvantage of plane wave basis set is that very short wavelength (high kinetic energy) plane waves are needed to produce the strongly varying wave functions of the core orbitals (such as C 1s orbital) in a small radius near the ions, which would mean intractably large basis set. In order to reduce the computational cost, the ionic Coulomb potentials are replaced so that the effect of the ionic Coulomb potential and the shielding of the core electrons are combined into soft potentials that act on the valence electrons, and the core electrons are not explicitly calculated in the KS DFT procedure. Bl\"ochl worked out the so-called projector augmentation wave (PAW) method~\cite{Blochl1994}, which produces a soft potential for the valence electrons but can fully reconstruct the all-electron, meaning the core electron plus accurate valence electron, solution in the core region of ions. This is particularly important for calculating accurate hyperfine constants of point defects in solids~\cite{Blochl2000}. Thus, plane wave supercell calculation with PAW method is a very powerful method for highly numerically convergent KS DFT calculations.  

We shortly mention here another modeling approach that was applied to NV center in diamond (e.g., Ref.~\onlinecite{Goss1996}). In that approach, the diamond cluster is terminated by hydrogen atoms, i.e., molecular cluster model. As the translation symmetry disappears in this model, the inversion symmetry cannot be maintained in a spherical molecular cluster of diamond. NV center has no inversion symmetry, thus this property is not necessarily a disadvantage. On the other hand, the surface may introduce polarization of bonds and extra surface states that are artifacts in modeling isolated NV centers in a perfect diamond host. By applying sufficiently large clusters the polarization of bonds may disappear at the central part of the cluster where the defect is placed. On the other hand, diamond is peculiar in the sense that special surface states appear in hydrogenated diamond (see also Sec.~\ref{ssec:surface}). Shockley in his fundamental work~\cite{Shockley1939} already predicted that (hydrogenated) diamond should have negative electron affinity. As a consequence, surface mirror image states~\cite{Cole1969} or Rydberg states~\cite{Mulliken1964, Voros2009, Kaviani2014} appear in hydrogenated diamond clusters that produce deep empty levels at about 1.7~eV below the conduction band of diamond. Those empty states may mix with the NV center's empty states which makes the excitation calculation of NV center in diamond problematic. The maximum probability of the Rydberg wave functions can be found outside of the diamond cluster~\cite{Voros2009, Kaviani2014}, thus those states can only be described by wave functions that are extended and not very much localized on the atoms. In molecular cluster models, the orbitals are often expanded by linear combination of atomic orbitals (LCAO) that are localized around the ions. Non-orthogonal gaussian type orbitals~\cite{Boys1950} (GTO) are applied in the LCAO calculations as the corresponding integrals in the KS DFT equations can be computed very efficiently compared to those of Slater-type orbitals. These GTO orbitals can be chosen to be well localized that can describe the relatively localized wave functions of the NV center but are not able to describe the Rydberg states properly, thus the empty levels from Rydberg states do not appear in the band gap of diamond. By using this trick, non-converged basis set for the surface Rydberg states, NV center in hydrogenated diamond clusters can be modeled as isolated defect in diamond. On the other hand, the valence band and conduction band edges converge very slowly as a size of the molecular cluster towards those of the perfect diamond crystal, i.e., quantum confinement effect (see Ref.~\onlinecite{Kaviani2014} specialized to NV center in diamond), thus the molecular cluster model does not generally enable the calculation of ionization energies and thresholds. However, it can be an acceptable model for calculating the electron and spin density of the NV center and related properties.  We note that all-electron basis can be applied in these GTO calculations without significant increase of the computational time with respect to that of valence-electron calculations, thus the hyperfine tensors can be directly calculated in this approach~\cite{Nizovtsev2018}. 

\section{Method developments and results}
\label{sec:dev_res}

In this section, we collect the recent developments on \emph{ab initio} calculation of NV center in diamond. We start with the basic ground state properties in bulk diamond and diamond surface, and then we continue with the excited state and related magneto-optical properties, which are used to calculate the corresponding decay rates and coupling parameters to external perturbations.

\subsection{Formation energies and charge transition levels}
\label{ssec:form}

\begin{figure}
\includegraphics[width=0.85\columnwidth, keepaspectratio]{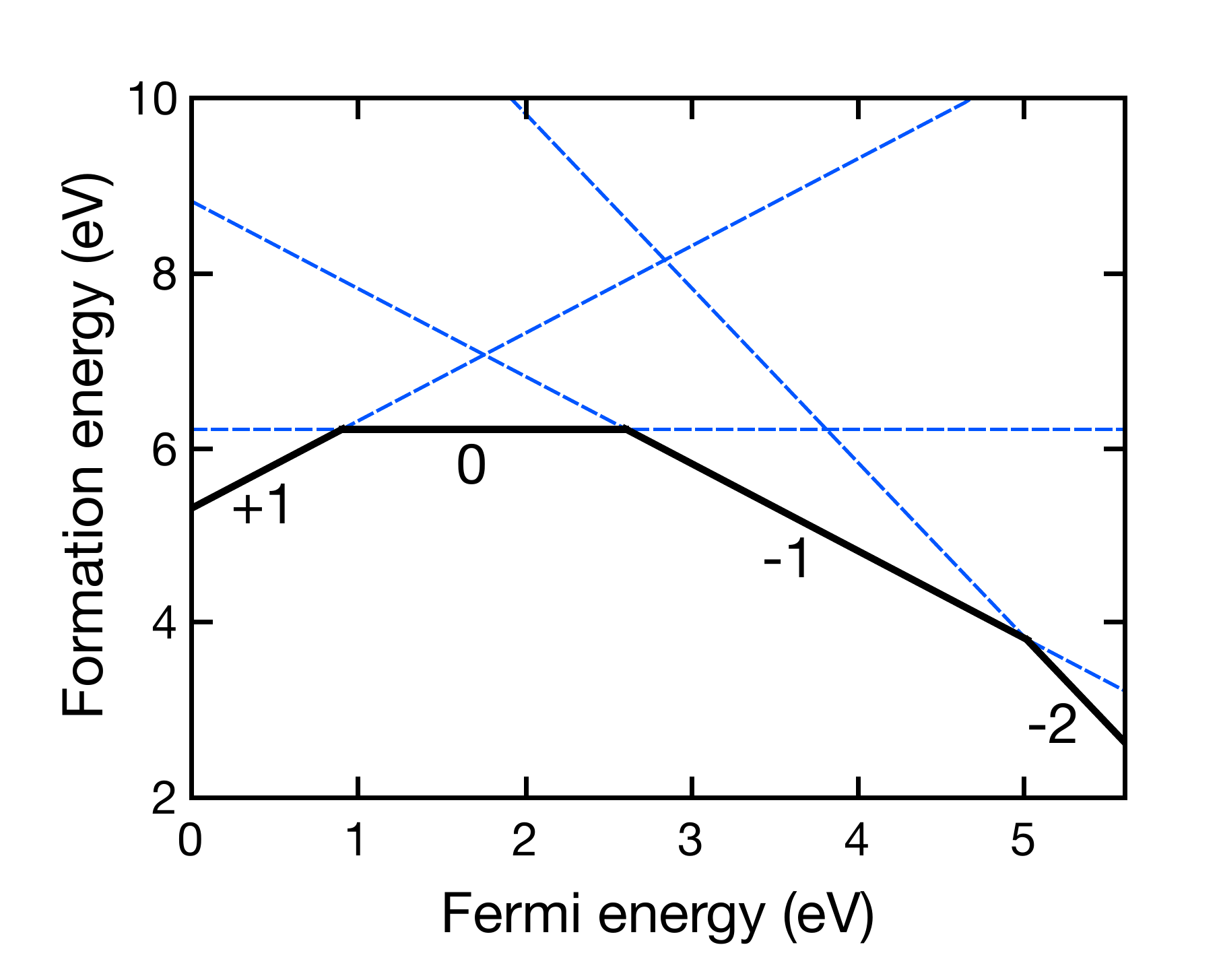}
\caption{\label{fig:NVocc}Formation energy of nitrogen-vacancy in diamond (Ref.~\onlinecite{Deak2014}). The chemical potential of nitrogen was set to the energy of nitrogen atom in the nitrogen molecule at $T$=0 K. The crossing lines correspond to the adiabatic ionization energies. The Fermi-level is aligned to the top of the valence band. We note that these data differ, in particular, for the donor level from Ref.~\onlinecite{Weber2010} that was obtained in a small 64-atom supercell.}
\end{figure}

The formation energy of the defect ($E_\text{form}^q$) in the charge state $q$ can be calculated from the total energy of the defect ($E_\text{tot}^q$) and the corresponding chemical potential of the atoms ($\mu$) constituting the defect and the electron~\cite{Northrup1996}, which can be written for 512-atom diamond supercell with NV center as
\begin{equation}
\begin{aligned}&
E_\text{form}^q (E_\text{F}) = E_\text{tot}^q(\text{C512:NV}) - 510/512\times E_\text{tot}(\text{C512}) -\mu_\text{N} \\&
+ q (E_\text{F} + E_\text{V}) + E_\text{corr}^q \text{,}
\end{aligned}
\end{equation}
where the chemical potential of the carbon atom can be expressed from the total energy of the perfect diamond lattice [$E_\text{tot}(\text{C512})$], whereas the chemical potential of the nitrogen atom ($\mu_\text{N}$) depends on the growth conditions. It can be set to the half of the total energy of the nitrogen molecule as a reference~\cite{Deak2014}. $E_\text{F}$ is the Fermi-energy referenced to the valence band top $E_\text{V}$. $E_\text{corr}^q$ is the correction of the total energy for charged supercells, because the charged supercells are neutralized by a jellium background in the plane wave supercell calculations that can all interact with their periodic images~\cite{Makov1995} with resulting a shift in the total energy. For medium localization of defect states, such as NV center in diamond, Lany-Zunger correction~\cite{Lany2008} and Freysoldt correction~\cite{Freysoldt2009} yields equivalent and relatively accurate results in sufficiently large supercells~\cite{Komsa2012}. The formation energy of the defect provides at least two very important quantities: (i) the concentration of the defect ($N_i$) in thermal equilibrium can be calculated as 
\begin{equation}
N_i = N_i^0 \exp \left(-\frac{E_\text{form}^q}{k_\text{B}T}\right) \text{,}
\end{equation}
where $N_i^0$ is the density of $i$ sites in the perfect lattice and $k_\text{B}$ the Boltzmann constant, and $T$ is the temperature in kelvin; (ii) the adiabatic ionization energies or occupation level of the defect between $q$ and $q+1$ charge states can be calculated as 
\begin{equation}
E(q|q+1) \equiv E_\text{form}^q (E_\text{F}) = E_\text{form}^{q+1} (E_\text{F}) \text{,}
\end{equation}
which gives the position of the Fermi-level with respect to $E_\text{V}$, where the concentrations of the defects in charge states $q$ and $q+1$ are equal, $E(q|q+1) = E_\text{tot}^{q+1} +E_\text{corr}^{q+1}  -  E_\text{tot}^q - E_\text{corr}^q $. Although, the controlled formation of NV center is not a thermal equilibrium process, still formation energies and the corresponding concentration of defects can provide information about their abundance, in particular, when occurrence of various defects in diamond is studied and compared with each other. Regarding the adiabatic ionization energies, it is highly important to apply such a functional which is able to reproduce the experimental band gap, otherwise the calculated ionization energies with respect to the band edges are not comparable with the experimental data. As was explained above LDA or PBE DFT predicts about 1.2 eV lower band gap than the experimental one. On the other hand, it has been found by applying intensive tests on defects Group-IV semiconductors and diamond~\cite{Deak2010} that the so-called HSE06 screened and range separated hybrid functional~\cite{HSE06}, i.e., mixture of screened Fock-exchange and PBE exchange functionals, is able to reproduce the experimental band gaps and the ionization energies are also reproduced within 0.1~eV accuracy. The test provided good results also for NV center in diamond~\cite{Deak2010, Deak2014}. This is a huge improvement over the accuracy of LDA and PBE functionals. The calculated HSE06 formation energy of NV center in diamond is illustrated in Fig.~\ref{fig:NVocc}. 
These calculations were carried out for substitutional nitrogen (N$_\text{s}$), carbon vacancy, nitrogen di-interstitial (N$_2$), N$_2$V (two nitrogen atoms substitute carbon atom near an adjacent vacancy), divacancy (V$_2$), and NVH defect, where a hydrogen atom saturates one of the carbon dangling bonds in the NV defect. By calculating the formation energy of these defects and applying the charge neutrality condition, the concentrations of defects were determined under quasi thermal equilibrium conditions at a given temperature that might occur in chemical vapor deposition process~\cite{Deak2014}. It was found that the concentration of NV center will be very low at any nitrogen concentrations because of the favorable formation of either N$_2$ or NVH. NV centers can be rather created in nitrogen-doped diamond after irradiation which creates carbon vacancies (V). The formation energy of the defects can be used to calculate the energy balance of defect reactions, i.e., to study defect chemistry in diamond. As an example, at relatively low Fermi-level positions (such as at $E_\text{V}$+2.0~eV, where V is neutral and mobile), combination of two vacancies results in V$_2$ with energy gain of 4.2~eV, whereas the energy gain of combining   
N$_\text{s}$ and V to form NV is only 3.3~eV. This means that rather divacancy forms than NV center when single vacancies diffuse in the diamond crystal upon annealing. However, it is much likely (by about 5~eV) to remove a carbon atom near N$_\text{s}$ than that from perfect lattice. These results implied that the formation of NV occurs, when the vacancy is formed very close to N$_\text{s}$. Divacancies are paramagnetic, electrically and optically active defects (see Ref.~\onlinecite{Deak2014} and references therein) which can be detrimental for the charge state stability and photostability of the NV center~\cite{Deak2014}. It was proposed that high temperature annealing at around 1200-1300~$^\circ$C may reduce the concentration of vacancy clusters that can improve the properties of NV centers~\cite{Deak2014}. 

This result could motivate a recent experimental effort to hinder the formation of divacancy or vacancy clusters by engineering the charge state of the single vacancies so that they are not mobile any more~\cite{Favaro2017}. In particular, in boron-doped diamond the single vacancies become positively charged or even double positively charged~\cite{Deak2014}, and the formation of vacancy clusters should be significantly reduced. To this end, F\'avaro de Oliveira and co-workers produced a thin boron-doped layer near the region where nitrogen ions were implanted~\cite{Favaro2017}. Indeed, this method resulted in an improved charge stability and coherence properties of NV centers in diamond~\cite{Favaro2017}.
We note here that atomistic molecular dynamics simulations with using empirical potentials showed in this study~\cite{Favaro2017} that indeed divacancies and larger vacancy aggregates are formed near NV center after N ion implantation when the vacancies are neutral, and the proximate vacancy aggregates' spin causes the decoherence of the NV center's electron spin~\cite{Favaro2017}. 

\subsection{Diamond surface}
\label{ssec:surface}

\begin{figure*}
\includegraphics[width=0.8\textwidth]{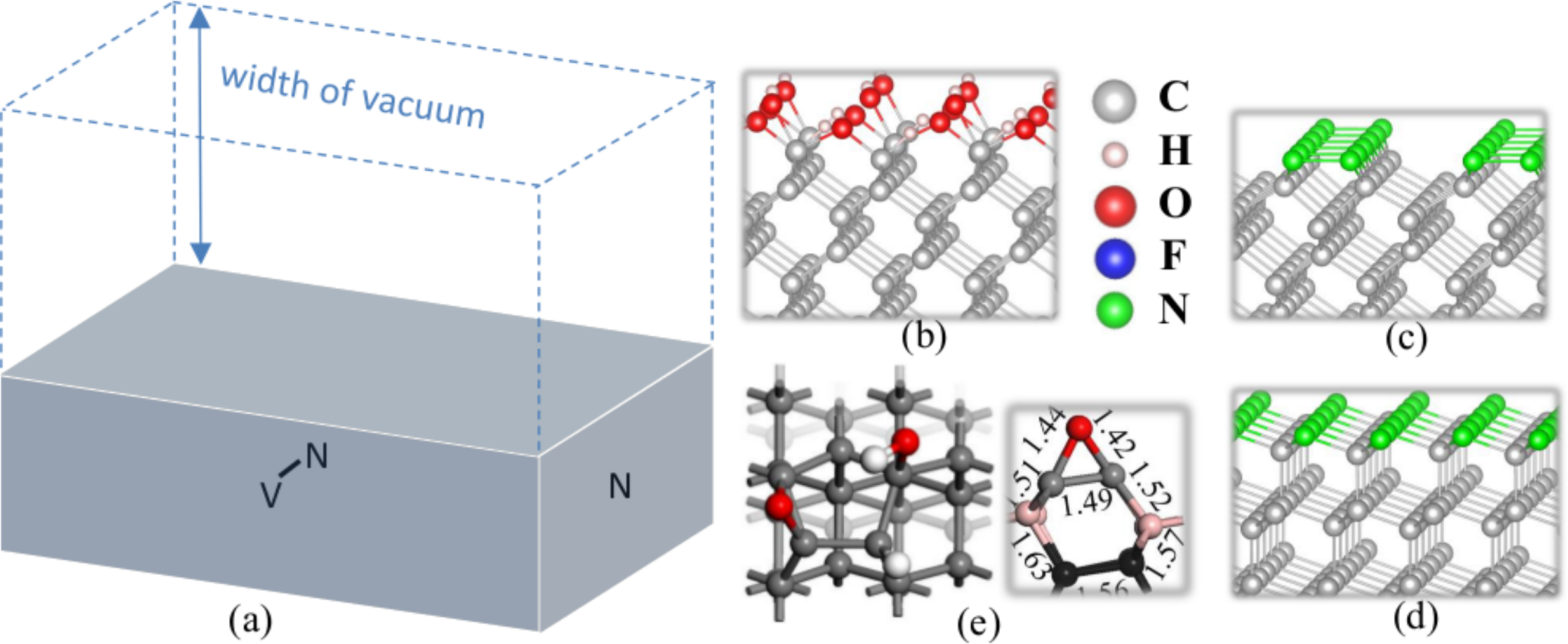}
\caption{\label{fig:surface}(a) Schematic diagram of the slab modeling of point defects at the surface: NV is an acceptor, whereas N is a donor. The entire slab is neutral. (b) Proposed smooth oxygenated (100) diamond surface for NV sensor applications. (c) Nitrogen terminated (100) diamond surface. (d) Nitrogen terminated (111) diamond surface.  (e) Mixed oxygen terminated (113) diamond surface with an epoxy C-O-C bonding configurations shown also from side view with typical bondlengths in \AA ngstr\"om unit. The data are taken from Refs.~\onlinecite{Kaviani2014, Chou2017, Li2019}.}
\end{figure*}

The shallow implanted NV centers below diamond surface may suffer from the vacancy cluster formation as explained above (e.g., Ref.~\onlinecite{Dhomkar2018}) but the presence of diamond surface can also pose a problem. Typically, NV centers were implanted into (100) diamond surface. It was found that shallow NV centers in hydrogen terminated diamond are not stable. This can be explained by the fact that hydrogenated (100) diamond has an electron affinity of about $-1.3$~eV~\cite{Maier2001} that had been predicted by Shockley~\cite{Shockley1939}. When water absorbs to this diamond surface then it creates a band bending of diamond~\cite{Grotz2012} which shifts down the Fermi-level and converts NV center to neutral NV defect~\cite{Fu2010, Rondin2010, Hauf2011}. As a consequence, stability of NV centers requires diamond surface with positive electron affinity. An additional criterion is that no surface related level should appear in the fundamental band gap of diamond. \emph{Ab initio} DFT supercell calculations can predict the electron affinity of surfaces of solids by comparing the calculated conduction band minimum ($E_\text{C}$) with respect to the vacuum level. In the supercell modeling of surfaces, a slab model is the only option, e.g., diamond slab with (100) bottom and top surfaces~\cite{Chou2017mrs}. It is often desirable to choose the same type of termination at the bottom and the top of the slab, in order to avoid an artificial polarization across the slab, and the size of the vacuum region should be sufficiently large for convergent electron affinity calculations [see Fig.~\ref{fig:surface}(a)]. Again, the choice of the functional is crucial. In early calculations, LDA or PBE DFT calculations were applied with too low band gap~\cite{Tiwari2011}. In order to obtain realistic results, the calculated $E_\text{V}$ was fixed and $E_\text{C}$ was shifted to reproduce the experimental band gap, i.e., scissor correction was applied to $E_\text{C}$~\cite{Tiwari2011}. However, HSE06 functional calculations with accurate band structure calculation showed~\cite{Kaviani2014} that $E_\text{V}$ should shift down and $E_\text{C}$ should shift up by about the same amount in the correction of PBE band edges. Thus, the trends could be well produced the LDA calculations in Ref.~\onlinecite{Tiwari2011} but accurate results are expected from HSE06 electronic structure calculations~\cite{Kaviani2014}. 

In experiments, oxygenation is applied in order to stabilize the charge state of shallow NV centers in (100) diamond. It was found that alcohol groups (-OH) will not turn the negative electron affinity of hydrogenated diamond surface to positive electron affinity. Rather, ether bridges (C-O-C bonds) are responsible for this issue~\cite{Kaviani2014}. However, ether bridges will stiffen the diamond lattice with introducing defect levels in the gap of diamond, and these groups most likely roughen the diamond surface with creating optically and electrically active defects. Kaviani and co-workers suggested to apply smooth and well controlled oxygenation techniques that would result in a mixture of C-H, C-OH and C-O-C bonds at the diamond surface with no stress at the diamond surface but with positive electron affinity and clean band gap [see Fig.~\ref{fig:surface}(b)]. It was also found~\cite{Kaviani2014}, in agreement with Ref.~\onlinecite{Tiwari2011}, that smooth fluorine termination provides a large positive electron affinity (100) diamond surface, although, an empty band appears just below $E_\text{C}$ according to HSE06 DFT~\cite{Kaviani2014}. Motivated by experimental efforts~\cite{Stacey2015}, nitrogen terminated diamond surface was also studied from first principles, and positive electron affinity was predicted for larger than 0.5 monolayer of N-N groups rather than C-H groups at the surface [see Fig.~\ref{fig:surface}(c)]. Indeed, a recent study has found improved NV properties on nitrogen terminated diamond (100) surface over oxygenated one, although, the improvement was not striking~\cite{Kawai2019}. Most likely, the nitrogen termination of (100) diamond surface was not perfect in that experiment. HSE06 DFT calculations predicted~\cite{Chou2017} that nitrogen-terminated (111) diamond surface is more favorable than that on (100) diamond surface, as nitrogen can naturally replace the top C-H layer on (111) diamond without introducing any strain~\cite{Chou2017}. Thus, it is likely that (111) nitrogen-terminated diamond surface may host NV centers with excellent properties for sensing applications [Fig.~\ref{fig:surface}(d)]. This is appealing as preferential alignment of NV centers along the (111) axis can be realized~\cite{Edmonds2012, Michl2014, Lesik2014, Tahara2015}, up to 99\%~\cite{Ozawa2017, Osterkamp2019}, in the growth process of diamond which is desirable for magnetometry and related applications. However, the growth of high quality (111) diamond at sufficiently high rate is still a big challenge. Alternatively, (113) diamond can be grown with 79\% preferential alignment of NV centers at a considerable growth rate~\cite{Lesik2015, Chouaieb2019}. HSE06 DFT calculations have recently obtained a surprising result that nitrogen-termination is not preferred for this surface because of introducing surface-related bands into the band gap of diamond but rather oxygen termination may result in an excellent environment for hosting NV centers for quantum sensor applications~\cite{Li2019}. On (113) diamond surface, oxygen can form so-called epoxy bonds with the surface carbon atoms that are stable according to the \emph{ab initio} simulations [Fig.~\ref{fig:surface}(e)]. This bonding situation does not frustrate the top carbon layers, thus it produces clean band gap and positive electron affinity. 

The calculated electron affinities in seeking positive electron affinity are presented in Table~\ref{tab:elaff}. We note that we do not provide here a comprehensive list of surface terminators creating negative electron affinities that are not relevant for NV sensor applications.
\begin{table}
\caption{\label{tab:elaff}Calculated electron affinities of various diamond surfaces in eV unit. Mixed means smooth diamond surface with combination of C-H, C-OH and C-O-C as shown in Fig.~\ref{fig:surface}(b) for (100) surface. On (113) surface, C-O-C creates an epoxy structure that differs from ether-like bonds on (100) surface. Depending on the type of surface termination the (100) surface is $(2\times1)$ reconstructed whereas $(113)$ surface is always $(2\times1)$ reconstructed. These data and data for other mixed types of surface termination can be read in Refs.~\onlinecite{Kaviani2014, Chou2017, Li2019}.}
\begin{ruledtabular}
\begin{tabular}{r|cccccc}
surface  & C-H & C-OH & C-O-C & mixed & C-N & C-F \\\hline
(100)    & $-1.7$ & $-0.6$ & $+2.4$ & $+0.5$ & $+3.5$ & $+3.0$ \\
(111)    & $-1.6$ & $+0.2$ & - &  - & $+3.2$ & $+3.6$ \\
(113)    & $-1.8$ & $-0.1$ & $+2.2$ &  $+0.5$ & $+3.6$ & $+3.3$ \\
\end{tabular}
\end{ruledtabular}
\end{table}

Previously, we mentioned above that preferential alignment of NV centers have been reported during in-growth process of NV centers in the chemical vapor deposition of (111) and (113) diamond. We note here that a single DFT study predicted~\cite{Karin2014} that preferential alignment of NV centers might be achieved in bulk diamond by inducing 2\% biaxial strain during the diamond growth at about 970~$^\circ$C temperature, however, this has been not verified in the experiments. The preferential alignment of NV centers in (110) and (111) diamond surfaces was studied by \emph{ab initio} calculations~\cite{Atumi2013, Miyazaki2014}. It was found that nitrogen prefers to reside on carbon layer beneath the top carbon layer of the surface~\cite{Atumi2013}, and the NV centers will be preferentially aligned at the kink of the diamond terrace during the growth of (111) diamond under hydrogen-rich environments~\cite{Miyazaki2014}.

Realistic diamond surfaces may contain steps, voids, and other defects that were not considered in previous modeling. An exception is the carboxyl group (double carbon-oxygen bonds) at the oxygenated (100) diamond surface which creates levels in the band gap diamond with localized states that can absorb light in the visible and perturbs NV measurements~\cite{Kaviani2014}. Thus, carboxyl groups should be eliminated from oxygenated diamond surface. Recently, very surprising results have been reported on (100) diamond surface that could have a direct relation to surface related charge or spin noise~\cite{Stacey2019}. \emph{Ab initio} modeling tentatively identified stable pairs of sp$^2$ C-C bonds in the void defect, i.e., single surface carbon-vacancy at the top of diamond surface which introduces an empty level about 1.4~eV above $E_\text{V}$ as found by surface sensitive analysis techniques~\cite{Stacey2019}. This empty state may be filled by electrons during illumination of diamond surface that provides a surface charge and spin. Since this defect is very general and relatively abundant on (100) diamond surface that might be the source of surface noises felt by near-surface NV centers~\cite{Stacey2019}. The interaction of NV center with a nearby acceptor defect has been recently modeled in a bulk supercell by DFT calculations~\cite{Chou2018}. The bulk model was chosen to avoid modeling problems of the slab calculations. The acceptor defect was chosen to be the neutral NV defect for simplicity~\cite{Chou2018}. A common sense among experimentalists is that NV center is an atomic like defect with very localized orbitals. Chou and co-workers showed in the study of interaction of NV center with neutral NV acceptor~\cite{Chou2018} that the NV orbitals spread to 4~nm from the center of the NV defect in the plane of the three carbon atoms nearest to the vacancy. If the acceptor defect is closer than about 7~nm distance from NV center with similar wave function extension then they can directly interact quantum mechanically without any illumination of NV center, which can lead to a decrease in the spin coherence time of the NV center~\cite{Chou2018}. The calculated quantum mechanical tunneling rate between the defects could well explain the experimental data in diamond sample with high density of NV defects~\cite{Choi2017}, which highlights the predictive power of DFT methods. Recently, it has been found in experiments~\cite{Bluvstein2019} that shallow NV center can ionize in dark on experimentally relevant timescales which can be understood as tunneling to a single local electron trap as the mechanism behind this process. 

NV centers were also considered close to diamond surfaces by using slab models. Modeling of NV center in slab models faces several problems: i) charge correction of the negatively charged defect is not straightforward as the potential of a point charge in a surface is only conditionally convergent, thus handling of charged slab supercell is painstaking, ii) artificial polarization might appear due to the bottom of the slab (double surface), and iii) the finite width of the slab may introduce quantum confinement effect. Basically, the total energy of NV center in the slab model may be converged by increasing the lateral size of the slab model~\cite{Chou2017mrs} and by applying total energy correction techniques that were recently proposed in the literature~\cite{Lozovoi2001, Lozovoi2003, Komsa2012, Scivetti2013, Wang2015, Vinichenko2017, Bal2018, Freysoldt2018, Tahini2018, Smart2018}. The sufficiently large lateral size is necessary to minimize the artificial interaction of the defect with its periodic images too~\cite{Pinto2012} as usual in three-dimensional bulk modeling. The other two problems may be minimized by adding the same surface termination at the bottom and the top of the slab and by the use of sufficiently large width of the slab~\cite{Kaviani2014}. According to intensive tests, about 2.2-nm width produces minute quantum confinement effect~\cite{Kaviani2014}.

Kaviani and co-workers invented to do a workaround in the problem of charged slab supercell by replacing it to another but readily solvable problem~\cite{Kaviani2014}: a neutral slab model is used for the negatively charged NV defect (NV center) at the expense that another defect enters the slab. Namely, the N$_\text{s}$ donor will donate an electron to the neutral NV acceptor defect by creating a pair of NV center and positively charged N$_\text{s}$. If these defects are placed into the same layer of the slab and the slab has a cubic-like shape then the dipole-dipole interaction between the periodic images of the defect pairs can be minimized. Rather, the presence of ionized N$_\text{s}$ near the NV center may shift the levels or split the degenerate levels of NV center. Extensive tests showed that if the defects are placed at least 7.5~\AA\ far from each other then the degeneracy of the corresponding orbitals is maintained and the constant shift in the KS levels can be well monitored and corrected~\cite{Kaviani2014}. In a recent study, the pair of NV center and ionized N$_\text{s}$ was analyzed in detail with arriving the same conclusion that 7.5~\AA\ distance between the two defects suffices to avoid the splitting of the degeneracy of the corresponding orbitals, and NV center can be well approximated as isolated~\cite{Lofgren2018}. These achievements makes possible to directly simulate NV center with correct total energies that is necessary for modeling direct interaction of NV center with surface species~\cite{Kaviani2014}.

We mention in this context that direct tunneling between the nitrogen donor and neutral NV defect has been recently discussed in details in type 1b diamond~\cite{Manson2018}, which study largely extends the original idea from Collins~\cite{Collins2002} about the electron transfer between these nearby defects. We note that one conclusion from the \emph{ab initio} study of the interaction of NV center and neutral NV acceptor defect that the rate of tunneling depends on the actual orientations of the nearby defects because of the direction of the spatial extension of the orbitals~\cite{Chou2018}. This should hold for the nitrogen donor -- NV defect pair too as the neutral nitrogen donor orbitals shows special spatial extension due to the giant Jahn-Teller distortion~(e.g., Refs.~\onlinecite{Deak2014, Londero2018}), thus accurate determination of the tunneling rate as a function of distance and relative orientation between the two defects requires \emph{ab initio} calculations.

\subsection{Excited states}
\label{ssec:exc}

\subsubsection{Electronic solution}
\begin{figure}
\includegraphics[width=0.9\columnwidth, keepaspectratio]{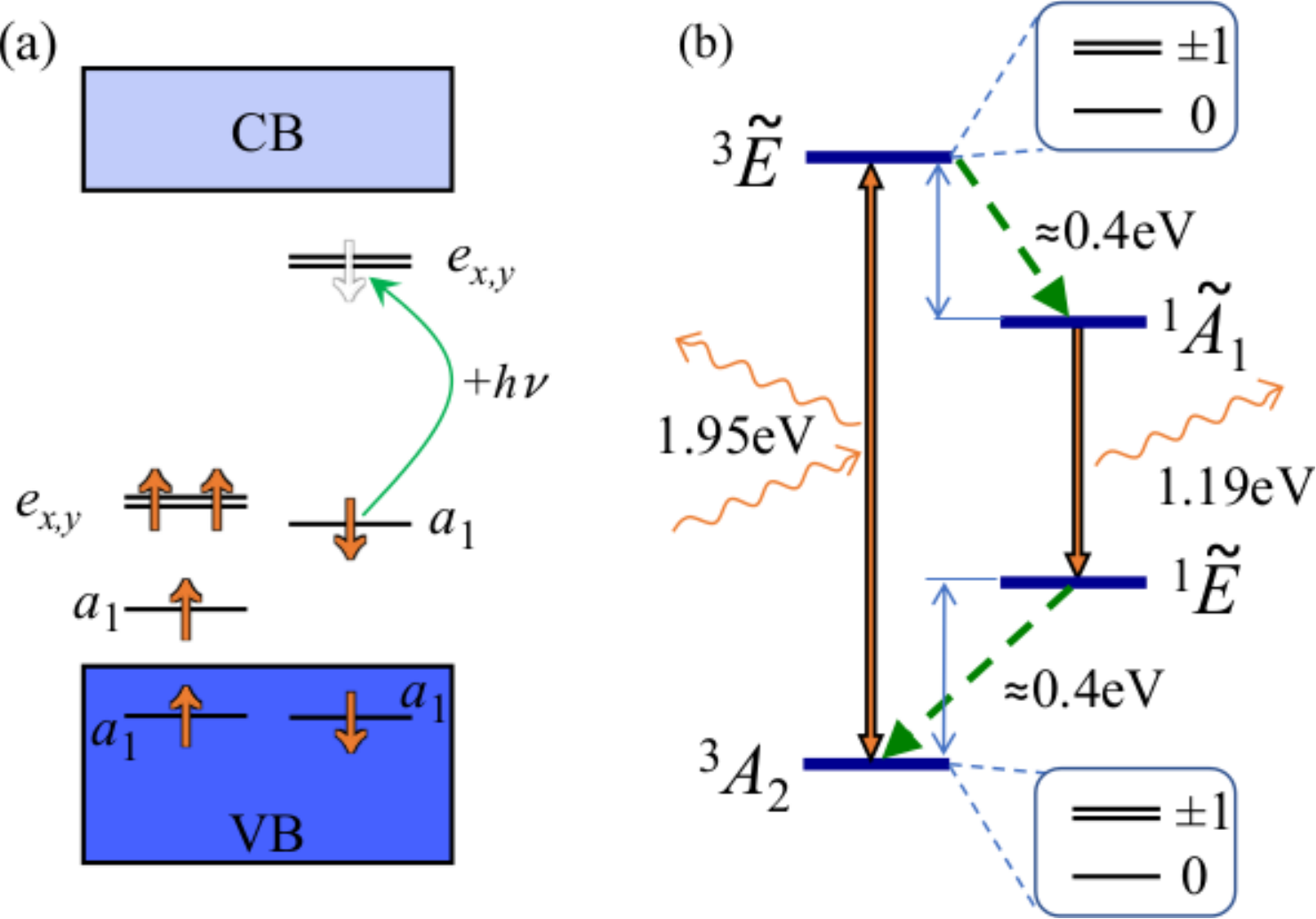}
\caption{\label{fig:exclev} (a) Single-particle scheme of the electronic structure of NV center. The spinpolarization between spin-up and spin-down electrons in the Kohn-Sham density functional theory results in different levels for spin-up and spin-down orbitals. VB and CB labels valence band and conduction band, respectively. The fundamental band gap of diamond is 5.4~eV. The green arrow represents the optical transition within single-particle scheme. (b) Level structure of NV center with taking the geometry relaxation in the corresponding electronic states into account. The character of the wave functions are depicted. Tilde labels the vibronic or polaronic nature of the state which is particularly strong for ${}^3E$ state and ${}^1E$ state.}
\end{figure}

The quantum bit operation and readout works via optical excitation, thus understanding the absorption and decay from the excited state is highly important.  
To this end, the first task is to calculate the excited states and levels. This is far from trivial. The extension of wave function of NV center in diamond requires sufficiently large models for accurate calculations~\cite{Thiering2017, Chou2018}. In addition, if diamond band edges are also involved in the photoexcitation process, then the diamond host should be modeled at equal footing with the NV center itself. As a consequence, the highly accurate but extremely costly wave function based quantum chemistry methods have only limited accuracy as these methods can be applied on small molecular cluster models~\cite{Zyubin.JCompChem.09, Delaney2010}. KS DFT, including HSE06 DFT, which can be used in large supercells in practice, has built-in limitations because of the approximations in the KS DFT functionals, in particular, the highly correlated singlet excited states, e.g., ${}^1A_1$ state, cannot be directly calculated by KS DFT correctly. Highly correlated states may be recognized in the combination of defect-molecule picture and group theory as multi-determinant states~\cite{Goss1996, Manson2006, Maze2011, Doherty2011}. Defect states in solids may be viewed as highly correlated electron states in contact with a bath of extended states. In a recent work, the wave function method is embedded into DFT framework, in order to synthesize the advantages of both methods and follow the character of the system~\cite{Bockstedte2018}. For the chosen orbitals (typically defect orbitals) the electron-electron interaction is exactly calculated as Coulomb interaction between the electrons, i.e., configurational interaction (CI), whereas the interaction of the other electrons are treated with HSE06 DFT. The interaction between the chosen orbitals and the extended states is treated within random phase approximation in a way that the corresponding dielectric function calculated with the constraint of extended states  (cRPA)~\cite{Bockstedte2018}. This CI-cRPA method does not contain any fitting parameters, and it is transferable in the sense that the number of orbitals in the CI active space can be systematically increased. This method can be used to analyze the character of the wave functions and the features of the other methods. In particular, the position of the ${}^1A_1$ level is very sensitive to the screening of the electrons from the diamond bands~\cite{Bockstedte2018}. As a consequence, pure quantum chemistry CI method in tiny diamond cluster models yield too high ${}^1A_1$ level~\cite{Delaney2010}. In a diamond cluster Hubbard model calculation it was shown that the ${}^1A_1$ state inherits double excitation from the lower $a_1$ level to the upper $e$ level in the gap~\cite{Ranjbar2011}, which provides an important insight about the nature of the ${}^1A_1$ state, although, the limitation of the diamond cluster model resulted in a false position of the ${}^1A_1$ level in this calculation~\cite{Bockstedte2018}. For this reason, the very popular many-body perturbation method on top of DFT calculations, called GW+BSE (see Ref.~\onlinecite{Onida2002} and references therein) fails to properly describe ${}^1A_1$ level~\cite{Ma2010, Choi2012} because BSE can typically describe excited states with combination of single excitation configurations. The extended Hubbard model calculation fit to the GW calculation of the in-gap defect levels provides accurate singlet and triplet levels of NV center~\cite{Choi2012, Bockstedte2018}. We note here that NV center is exceptional in the sense that the accurate description of the electronic structure of other solid state quantum bits requires to consider resonance orbitals from the valence band in the CI active space, for which the fitting procedure is ill defined~\cite{Bockstedte2018}. The final conclusion is that the ${}^1A_1$ level resides about 0.4~eV below the excited state ${}^3E$ level. We note that the electronic ${}^1E$ state is basically stable against distortion [it can be described as a single Slater-determinant in an appropriate basis of $e_{x,y}$ orbitals in Fig.~\ref{fig:exclev}(a)~\cite{Larsson2008, Thiering2018}], and it becomes slightly unstable against distortion because of the appearance of ${}^1E^\prime$ character in the $^1E$ state~\cite{Thiering2018, Bockstedte2018}. The predominant simple character of the ${}^1E$ state makes the ${}^1E$ level almost insensitive to the choice of the computational method that lies about 0.4~eV above the ground state ${}^3A_2$ level. The final level diagram is shown in Fig.~\ref{fig:exclev} in which the global energy minimum of the corresponding APES is taken into account (see below).
The final picture well accounts for the measured ZPL energy between the singlet states of about 1.19~eV~\cite{Rogers2008}, and derivation from the measured non-radiative rates from the ${}^3E$ level to the ${}^1A_1$ combined with a phenomenological model on the phonon participation and density of states~\cite{Goldman2015prl} that concluded about 0.4~eV gap between these two levels. 

The CI-cRPA method also justified the co-called constraint occupation DFT or $\Delta$SCF method for calculating the  ${}^3E$ excited state~\cite{Goss1996, Gali2009} of which level is relatively insensitive to the size of the diamond cluster and different computational methods. In the constraint occupation DFT the electronic configuration is set to build up the Slater-determinant of the ${}^3E$ excited state as shown by the green arrow in Fig.~\ref{fig:exclev}(a). A big advantage of $\Delta$SCF method is that the forces acting on the ions can be straightforwardly calculated, thus the APES of the ${}^3E$ state can be mapped by KS DFT method. This is of high importance in understanding the nature of the interaction of electrons and phonons in the excited state and its role in the optical spectrum and non-radiative decay.

\subsubsection{Strong electron-phonon coupling: vibronic wave functions}

\begin{figure}
\includegraphics[width=\columnwidth, keepaspectratio]{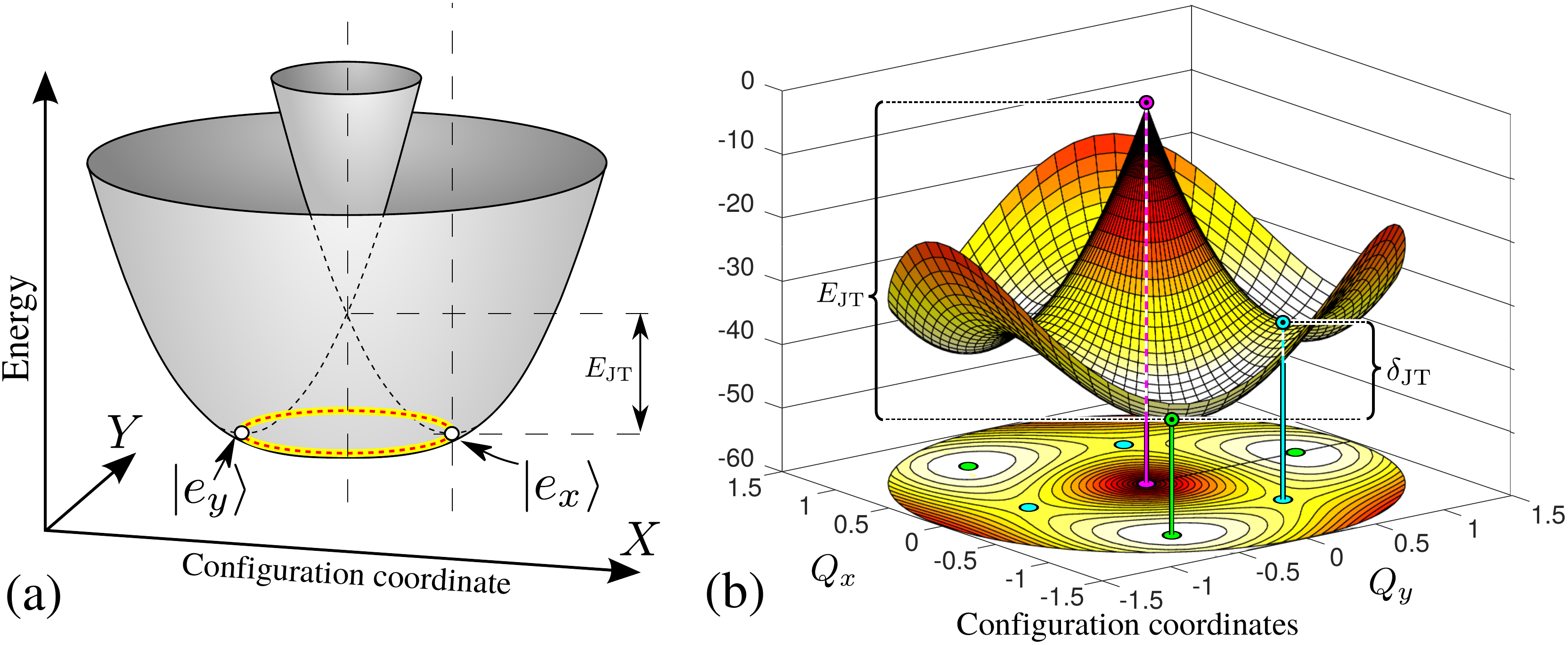}
\caption{\label{fig:DJT_APES}(a) Schematic diagram of adiabatic potential energy surface (APES) of typical $E\otimes e$ linear dynamic Jahn-Teller (DJT) system along the appropriate configuration coordinate. $E_\text{JT}$ is the Jahn-Teller energy which separates the total energy at the high symmetry geometry (conical intersection point) and the lowest energy in APES. (b) Calculated HSE06 DFT APES of NV center in diamond in the electronic ${}^3E$ excited state. The barrier energy ($\delta_\text{JT}$) among the three minima with $C_{1h}$ symmetry breaks the axial symmetry of DJT that can be handled by the so-called quadratic Jahn-Teller Hamiltonian (see Ref.~\onlinecite{Thiering2017}). $Q_x$ and $Q_y$ are the configuration coordinates associated with the effective double degenerate $e_x$ and $e_y$ vibration mode of $\hbar \omega_e$=77.6~meV, respectively.}
\end{figure}

Experiments indicated~\cite{Fu2009, Ulbricht2016} that dynamic Jahn-Teller (DJT) effect occurs in the ${}^3E$ excited state which is double degenerate and have two components, $\left|E_{\pm}\right\rangle= \frac{1}{\sqrt{2}}(\left|E_x\right\rangle \pm i \left|E_y\right\rangle)$. Symmetry breaking $e_{x,y}$-type phonons distort the system and couples strongly to the ${}^3E$ electronic state~\cite{Zhang2011, Thiering2017}. This is a so-called $E\otimes e$ DJT system for which the APES shows a sombrero shape (two shifted parabolas with minima at distorted $C_{1h}$ geometries which crosses at a conical intersection point at the high $C_{3v}$ symmetry as shown in Fig.~\ref{fig:DJT_APES}) rather than a simple parabola with the minimum at the high symmetry point~\cite{Abtew2011, Thiering2017}. By calculating the APES of ${}^3E$ state and applying the $E\otimes e$ DJT theory from Bersuker~\cite{Bersuker2006}, one can setup an electron-phonon Hamiltonian with a single effective phonon mode for which the electron-phonon parameters can be derived from the calculated APES~\cite{Abtew2011, Thiering2017}. The limitation of the single effective phonon mode is briefly discussed in the Outlook section. 

DFT HSE06 calculations yielded about 42~meV Jahn-Teller energy and about 9~meV barrier energy between the global minima of APES~\cite{Thiering2017} that might be superior over the DFT PBE results yielding smaller values~\cite{Abtew2011}. Because of the finite barrier energy, the DJT should be solved in quadratic Jahn-Teller approximation.
The exact solution can be expanded into series as
\begin{equation}
\label{eq:tilde_Psi}
\left|\widetilde{\Psi}_{\pm}\right\rangle =\sum_{nm}\left[c_{nm}\left|E_{\pm}\right\rangle \otimes\left|n,m\right\rangle +d_{nm}\left|E_{\mp}\right\rangle \otimes\left|n,m\right\rangle \right]\text{,}
\end{equation}
where the expansion is convergent with maximum of four oscillator quanta ($n+m\leq4$), $\left|n,m\right\rangle$ is the occupation representation of $e_x$ and $e_y$ vibrations, respectively, and $c_{nm}, d_{nm}$ coefficients are obtained from the solution of the $E\otimes e$ DJT electron-phonon Hamiltonian~\cite{Thiering2017}. We note that other systems may require a larger number of oscillator quanta (e.g., the product dynamic Jahn-Teller system of neutral silicon-vacancy qubit in Ref.~\onlinecite{Thiering2019}). The tilde over $\Psi$ labels the vibronic or polaronic nature of the wave function: the electron-phonon wave function cannot be factorized into electronic and phonon wave functions as assumed in BOA but will be linear combination of such wave functions. Thus, one has to go beyond BOA in DJT systems, i.e., in strongly coupled electron-phonon systems. The vibronic spectrum starts with an $\widetilde{E}$ ground state, then it is followed by $\widetilde{A}_1$, $\widetilde{A}_2$, and $\widetilde{E}$ vibronic levels by 39~meV, 57~meV, and 90~meV, respectively. We note that the $\widetilde{A}_1$ and $\widetilde{A}_2$ levels split due to the quadratic Jahn-Teller interaction. The vibronic solution will have a serious consequence in the effective spin-orbit splitting between the ${}^3E$ spin levels and ISC processes towards ${}^1A_1$ state, and it also manifests in the PL spectrum. We note that the singlet counterpart of $\left|^3E\right\rangle$, $\left|^1E^\prime\right\rangle$, experiences the same type of DJT effect as $\left|^3E\right\rangle$.  

\begin{figure}
\includegraphics[width=0.9\columnwidth]{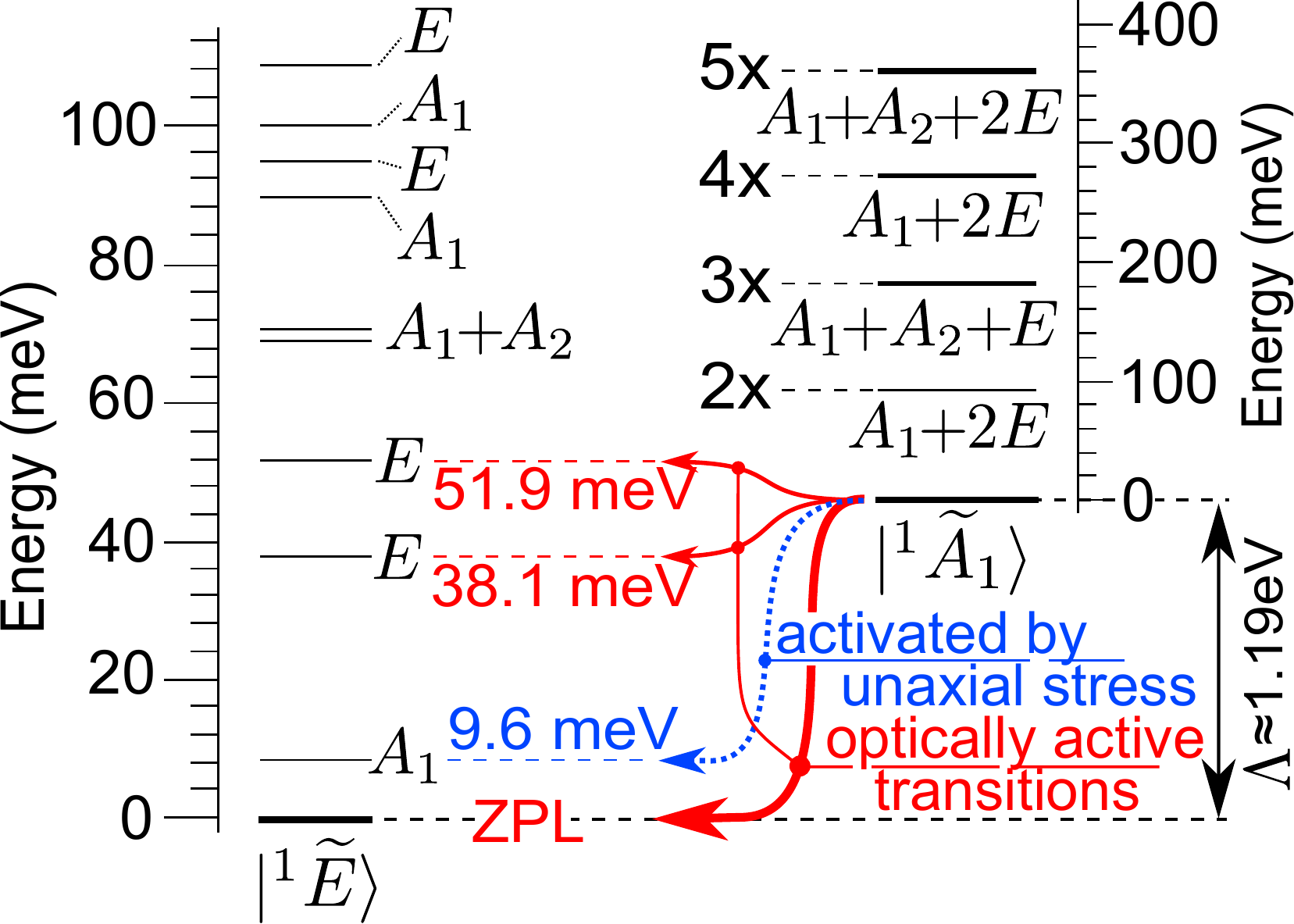}
\caption{\label{fig:el_ph_spectra}The vibronic levels of $\left|^{1}\widetilde{E}\right\rangle$ and $\left|^{1}\widetilde{A}_{1}\right\rangle$. The selection rules for the photoluminescence spectrum is indicated. Here the ZPL energy of 1.19~eV between $\left|^{1}\widetilde{E}\right\rangle$ and $\left|^{1}\widetilde{A}_{1}\right\rangle$ is not scaled for the sake of clarity. The calculated effective phonon frequency of the PJT+DJT Hamiltonian is $\hbar \omega_e$=66~meV. The vibronic levels of $\left|^{1}\widetilde{E}\right\rangle$ do not follow the ladder structure at all, whereas the vibronic levels of $\left|^{1}\widetilde{A}_{1}\right\rangle$ do show the ladder structure but significantly larger (92~meV) than $\hbar \omega_e$. These results are taken from Ref.~\onlinecite{Thiering2018}. The optical transition activated by uniaxial stress is observed in Ref.~\onlinecite{Rogers2015}.}	
\end{figure}

Phonons may couple non-degenerate states too which is often called pseudo Jahn-Teller effect (PJT). This occurs for the electronic $\left|^1A_1\right\rangle$ and $\left|^1E\right\rangle$, despite from the fact, that the energy spacing between the two is more than 1~eV~\cite{Thiering2018}. The dynamic coupling between these two states may be rationalized by invoking the symmetry breaking $e_{x,y}$ phonon modes which distort $\left|^1E\right\rangle$ to $\left|^1A^\prime\right\rangle$ and $\left|^1A^{\prime\prime}\right\rangle$, and $\left|^1A_1\right\rangle$ becomes $\left|^1A^\prime\right\rangle$, so the $\left|^1A^\prime\right\rangle$  component of the distorted $\left|^1E\right\rangle$ can couple to $\left|^1A_1\right\rangle$ by Coulomb interaction. According to DFT+CI-cRPA method, the contribution of $\left|^1A_1\right\rangle$ into $\left|^1E\right\rangle$ is about 2\% in the distorted geometry~\cite{Bockstedte2018}. At high $C_{3v}$ symmetry the contribution of $\left|^1E^\prime\right\rangle$ to $\left|^1E\right\rangle$ is about 10\% which is an electron-electron correlation effect. As a consequence of this fact, $\left|^1E^\prime\right\rangle$ brings DJT effect into $\left|^1E\right\rangle$ which is damped by the contribution factor. The full electron-phonon Hamiltonian contains the PJT and damped DJT effect that looks like a superlinear DJT Hamiltonian; even though the PJT and DJT Hamiltonian is written in linear approximation for the sake of simplicity, the final equation is not linear because PJT and DJT distorts the system in a different manner~\cite{Thiering2018}. The solution of PJT+DJT Hamiltonian is a set of polaronic wave functions. The combined $\left|{}^{1}\widetilde{A}_{1}\right\rangle \oplus \left|{}^{1}\widetilde{E}_{\pm}\right\rangle$ states may transform as $E$, $A_1$ and $A_2$. The $\widetilde{A}_2$ vibronic states do not play a significant role, thus we only show the expressions for the $^1\widetilde{E}_\pm$ and $^1\widetilde{A}_1$ vibronic states as follows
\begin{widetext}
\begin{subequations}
	\begin{align}&
	\left|^{1}\widetilde{E}_{\pm}\right\rangle =\sum_{i=1}^{\infty}\left[c_{i}\left|^{1}\bar{E}_{\pm}\right\rangle \otimes\left|\chi_{i}\left(A_{1}\right)\right\rangle +d_{i}\left|^{1}A_{1}\right\rangle \otimes\left|\chi_{i}\left(E_{\pm}\right)\right\rangle +f_{i}\left|^{1}\bar{E}_{\mp}\right\rangle \otimes\left|\chi_{i}\left(E_{\mp}\right)\right\rangle +g_{i}\left|^{1}\bar{E}_{\pm}\right\rangle \otimes\left|\chi_{i}\left(A_{2}\right)\right\rangle \right]
	\label{eq:Epolaron}
	\\&
	\left|^{1}\widetilde{A}_{1}\right\rangle =\sum_{i=1}^{\infty}\left[c_{i}^{\prime}\left|^{1}A_{1}\right\rangle \otimes\left|\chi_{i}\left(A_{1}\right)\right\rangle +\frac{d_{i}^{\prime}}{\sqrt{2}}\left(\left|^{1}\bar{E}_{+}\right\rangle \otimes\left|\chi_{i}\left(E_{-}\right)\right\rangle +\left|^{1}\bar{E}_{-}\right\rangle \otimes\left|\chi_{i}\left(E_{+}\right)\right\rangle \right)\right]
	\label{eq:Epolaron2}
	\end{align}
	\label{eq:spin-orbit}
\end{subequations} 
\end{widetext}
that govern the shape of the phonon sideband in the optical spectra of singlets. We label the symmetry adapted vibrational wavefunctions, e.g., $\left|\chi_{1}\left(A_{1}\right)\right\rangle =\left|00\right\rangle$, or in general, by $\left|\chi_{i}\left(\dots\right)\right\rangle$. The $g_i$ coefficients are generally tiny and can be ignored. On the other hand, the non-zero $d_i$ and $c_i^\prime$ ($f_i$ and $d_i^\prime$) coefficients drive the ISC process, and they are also responsible for the shape of PL spectrum of the singlets. The resultant electron-phonon spectra are depicted in Fig.~\ref{fig:el_ph_spectra} and the values of the coefficients are listed in Ref.~\onlinecite{Thiering2018}. The lowest energy dark $A_1$ excited vibronic level appears above the ground state $\left|^{1}\widetilde{E}\right\rangle$ because of PJT effect and plays a crucial role in the temperature dependent lifetime of $\left|^{1}\widetilde{E}\right\rangle$.

\subsection{Optical properties}
\label{ssec:optprop}

\begin{figure}
\includegraphics[width=0.95\columnwidth, keepaspectratio]{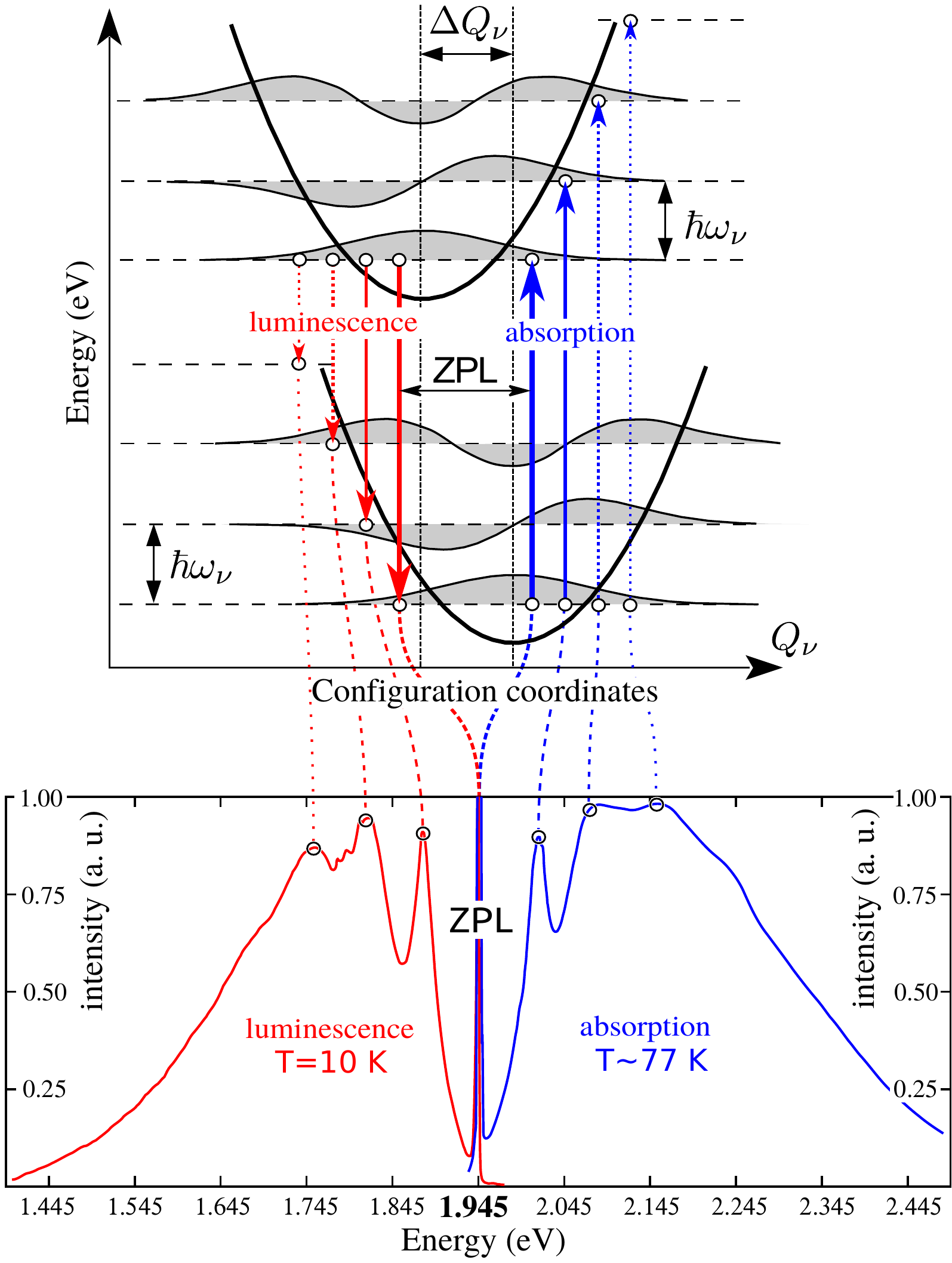}
\caption{\label{fig:opticalexc} Upper panel: Total energy against configuration coordinate sketches the adiabatic potential energy surface (APES). Configuration coordinate is related to the position of ions. $h\omega_\nu$ is the energy quantum of the (effective) $\nu$ phonon mode in quasiharmonic approximation, i.e., parabolic APES. In the Huang-Rhys approximation the APES of excited state (upper parabola) and that of the ground state (lower parabola) are the same but the minima shifted by $\Delta Q_\nu$. The wave functions of the phonons are shown, and the overlap of the phonon wave functions of the two APES will appear as phonon sideband in the optical spectra. Lower panel: Absorption and emission spectra of nitrogen-vacancy in diamond at low temperatures. The phonon modes coupled strongly to the optical transition can be identified as bumps in the phonon sideband of the corresponding spectrum. ZPL is the zero-phonon-line optical transition. Note that the two spectra are not identical in the phonon sideband as features at around $-0.15$~eV from ZPL are visible in the luminescence spectrum that are missing at $+0.15$~eV from ZPL in the absorption spectrum. These features are explained by the vibronic nature of the ${}^3E$ state (see Refs.~\onlinecite{Gali2011njp, Thiering2017}).}
\end{figure}

Similarly to molecules, vibrations or phonons can contribute to optical transitions of point defects in solids. The optical transitions of molecules are often modeled by Franck-Condon theory which applies BOA. One can sketch the APES in the electronic ground and excited state together with the phonon states and levels as shown in Fig.~\ref{fig:opticalexc}. The most simple case is that the APES can be described as a parabola both in the electronic ground and excited state, so Franck-Condon theory perfectly works. Principally, the two APES parabolas may have different tangents or effective vibration modes but they may be similar in the electronic ground and excited state. Huang-Rhys theory assumes that two APES parabolas are identical except a shift in the minimum of the parabolas. In this case, the derivation of the phonon lineshape in the optical excitation spectrum is simplified. In the experiments, the Debye-Waller factor can be measured ($DW$) which has a direct relation to the Huang-Rhys factor $S = - \ln DW$ which measures that how many effective phonons participate in the optical transition on average. Generally, the larger the distance between the two APES parabolas, i.e., the larger movement of ions upon optical excitation, the more likely to incorporate phonons in the optical transition. NV center has a broad absorption and emission spectrum (see Fig.~\ref{fig:opticalexc}) even at low temperatures, and $DW\approx0.03$ at cryogenic temperature which means that about 3.5 effective phonons participate in the radiative decay. 

Construction of the absorption and emission spectra in the Huang-Rhys approximation requires the calculation of APES both in the electronic ground state and excited state, as well as the vibration modes of the point defect in the electronic ground state. DFT calculations showed that quasilocal vibration modes appear at around 65~meV in which the carbon and nitrogen atoms vibrate the most around the vacancy~\cite{Gali2011njp, Zhang2011}, and these modes appear in the absorption and PL spectra as bump features (see Fig.~\ref{fig:opticalexc}). DFT $\Delta$SCF method can be used to calculate the APES in the electronic excited state. By forcing $C_{3v}$ symmetry in the electronic excited state, Alkauskas and co-workers calculated the PL spectrum of NV center in diamond~\cite{Alkauskas2014}. They developed an embedding technique, in order to calculate the contribution of long wavelength phonons~\cite{Alkauskas2014}, and they obtained a good agreement with the experimental data, except a small feature in the phonon sideband of the PL spectrum (see Fig.~\ref{fig:opticalexc}) as pointed out in Ref.~\onlinecite{Thiering2017}. It was shown above that DJT occurs in ${}^3E$ state which dynamically distorts the symmetry. By taking one of the parabolas in the sombrero APES, i.e., the distorted $C_{1h}$ geometry, the PL spectrum can be calculated within Huang-Rhys theory, and all the features in the phonon sideband of the PL spectrum were well reproduced~\cite{Thiering2017}. \emph{Ab initio} theory revealed that about 10\% of the phonon sideband emission is associated with the $e_{x,y}$ phonon modes that are responsible for the dynamic distortion in the ${}^3E$ state~\cite{Thiering2017}. This proves that the DJT feature indeed appears in the PL spectrum of NV center. Furthermore, the DJT nature of ${}^3E$ state was considered to be responsible for the anomalous temperature dependence of the ZPL width~\cite{Abtew2011}. 

The afore-mentioned asymmetry in the phonon sideband in the PL and absorption spectra is attributed to the DJT nature of the excited state. We note that a characteristic double peak in the absorption spectrum was associated with a Jahn-Teller feature~\cite{Davies1976} but \emph{ab initio} theory~(see Ref.~\onlinecite{Alkauskas2014} and private communication with Audrius Alkauskas) produces this double peak by the contribution of $A_1$ phonons. The reason behind this asymmetry may be identified as the consequence of the dynamics of ions in DJT systems. By taking the low temperature limit for the sake of simplicity, the electron occupies the lowest energy $\widetilde{E}$ vibronic level in the ${}^3E$ excited state upon illumination. This state may be viewed as the electron continuously tunnel from one of the global $C_{1h}$ minima to the other. As the spontaneous emission of the photon from this excited state is an instantaneous process, the electron just stays in one of these global $C_{1h}$ minima at any time when the radiative decay starts. The ${}^3A_2$ ground state has a high symmetry, thus the phonon sideband of the PL process should be calculated as if the excited system was frozen in the distorted C$_{1h}$ symmetry and it arrives to the high C$_{3v}$ symmetry in the electronic ground state. As there are continuum of $E$ phonons that distort the symmetry from C$_{3v}$ to C$_{1h}$ in the ${}^3E$ state, therefore the Huang-Rhys theory can be used to calculate the shape of the PL spectrum. On the other hand, the absorption process is different. The rate of phonon absorption may go with the radiative lifetime of the excited state which is associated with the optical transition dipole moment. This rate is $\Gamma_\text{rad} \approx 13$~MHz for NV center in diamond (e.g., Ref.~\onlinecite{Goldman2015prl} and references therein). However, the tunneling rate of the electron in the ${}^3E$ excited state is much faster. Bersuker~\cite{Bersuker2006} analyzed this for the lowest energy solutions of $E\oplus e$ DJT system, and the $\Gamma_\text{tunnel}$ goes as $\Gamma_\text{tunnel} \propto \Delta E / h$, where $\Delta E$ is the energy gap between the first excited state and ground state vibronic levels and $h$ is the Planck-constant. This formula yields $\Gamma_\text{tunnel}\approx 100$~GHz. The exact solution is 112.6~GHz for the ${}^3E$ state in NV center, where the corresponding equations are Eqs.~(7a-c) in Ref.~\onlinecite{Thiering2016} (see also references therein). As a consequence, $\Gamma_\text{tunnel} \gg \Gamma_\text{rad}$, therefore, the phonon sideband of the absorption spectrum should be calculated as the combination of the $A_1$ phonons in the Huang-Rhys approximation (predominant contribution) plus direct calculation from the ground state vibration function towards the high symmetry polaronic solution caused by the $E$ phonons. This has not yet been published, to our knowledge, for the optical transition between the triplets but similar spectrum was published for the optical transition between the singlets~\cite{Thiering2018} that will be discussed below. 

The excited states and excitation spectrum of the triplets can be basically calculated by time-dependent DFT (TD-DFT) too~\cite{Runge1984, Casida1995}. The usual linear response approximation was applied in the TD-DFT calculations of NV center~\cite{Casida1995, Gali2011}. We note that the accurate calculation of the spectrum requires the proper choice of the DFT functional. In 1.4 nm diamond molecular cluster model, the PBE0~\cite{PBE0} hybrid density functional provides quantitatively good results within TD-DFT~\cite{Gali2011}. The quantum mechanical forces acting on the ions in the electronic excited state can be calculated within TD-DFT formalism. The experimental Stokes-shift of NV center
of about 0.45~eV could be well reproduced by PBE0 TD-DFT calculations~(see Supplementary Materials in Ref.~\onlinecite{Vlasov2014}) despite the limitations of a molecular cluster model as explained above. This shows the predictive power of TD-DFT method. TD-DFT calculations can be combined with molecular dynamics simulation to monitor the evolution of the ionic motions in the electronic excited state. It was found that phonons move the system from the phonon excited state to the zero point energy (lowest energy phonon level) after photoexcitation from the electronic ground state within 50~fs which agrees well with the observed decay time from pump-probe PL measurements~\cite{Ulbricht2018}. The reason behind the peculiar ultrafast motion is tentatively associated with the DJT nature of the ${}^3E$ excited state. The observed femtosecond electronic depolarization dynamics of NV center~\cite{Ulbricht2016} was associated with the nonadiabatic transitions and phonon-induced electronic dephasing between the two components of the ${}^3E$ state based on \emph{ab initio} molecular dynamics simulations. 

\begin{figure}
\includegraphics[width=0.8\columnwidth]{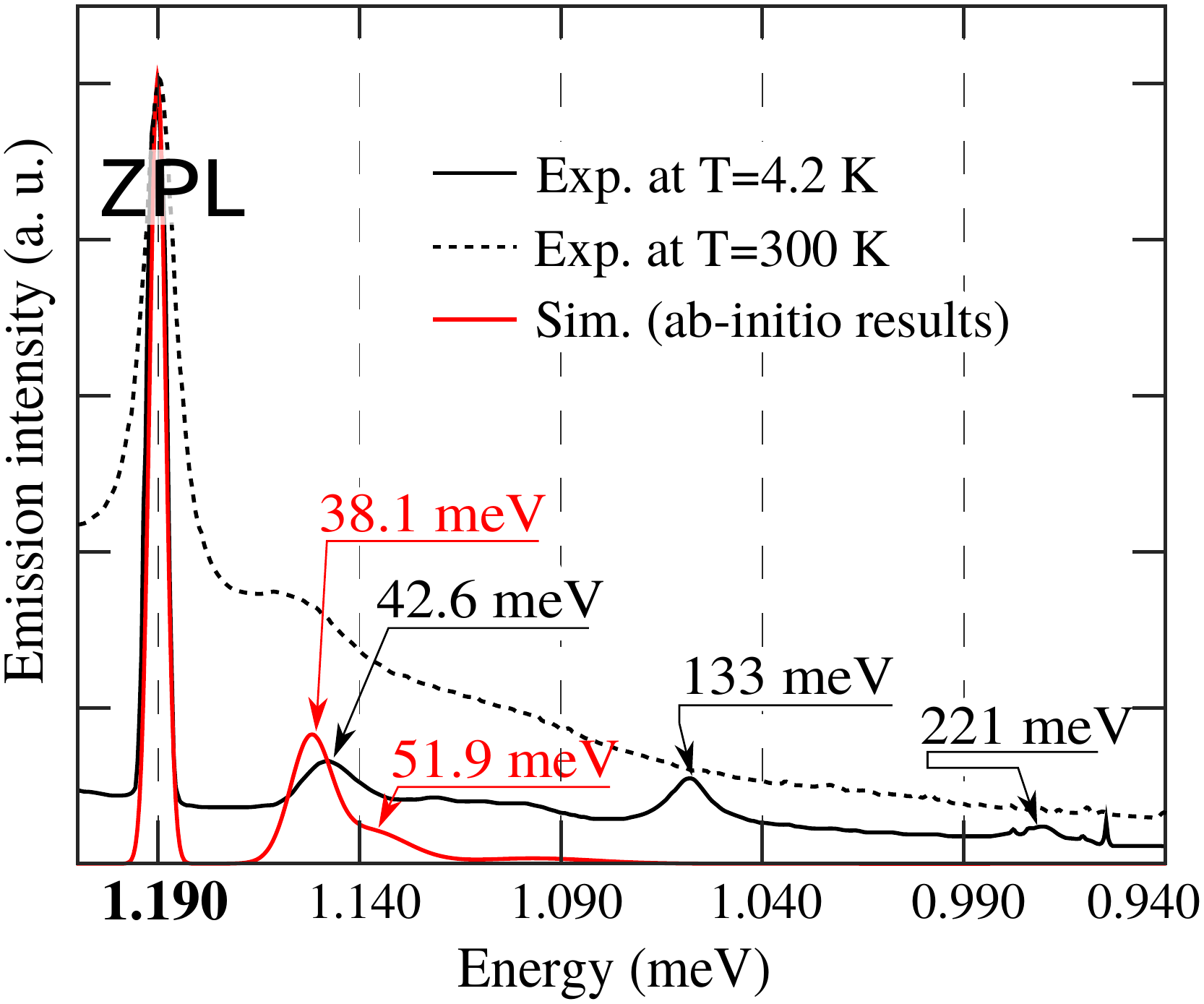}
\caption{\label{fig:PL_singlet}Experimental photoluminescence spectrum of the singlets at low (black solid line) and room (dotted black line) temperatures compared to the simulated spectrum from \emph{ab initio} solution (red curve from Ref.~\onlinecite{Thiering2018}). We note that the experimental spectra show a substantial and minor background at room and cryogenic temperature, respectively. The simulation curve does not include background signal. The ZPL energy is now set to zero, in order to easily read out the position of vibration features in the spectrum. 2~meV, 5~meV, and 10~meV gaussian smearing was used for the linewidth of the ZPL, first and second vibronic emissions, respectively, where the width of the ZPL and vibration bands were read out from the experimental spectrum recorded at cryogenic temperature. This theory does not account for the features at 133~meV and 221~meV. These features seem to disappear at room temperature PL spectrum, and they may not belong to NV center.}	
\end{figure}

We now discuss the PL spectrum of the singlets, in particular, the phonon sideband. The starting point is that the final state, $\left|^{1}\widetilde{E}\right\rangle$, is very far from the quasiharmonic vibration spectrum, thus the usual Franck-Condon approximation does not hold. In addition, the $\Gamma_\text{tunnel}\approx 31$~GHz in the vibronic ground state of $\left|^{1}\widetilde{E}\right\rangle$, which is several orders of magnitude faster than the inverse radiative lifetime of the ${}^{1}\widetilde{A}_{1}$~\cite{Bockstedte2018}. Therefore, the optical transition dipole moment should be calculated directly between the polaronic states which can be written as~\cite{Thiering2018}
\begin{equation}
I\left(^{1}\widetilde{A}_{1}\rightarrow{}^{1}\widetilde{E}^{(n)}\right)=\left|\left\langle ^{1}\widetilde{A}_{1}\right|e\hat{r}_{x}\left|^{1}\widetilde{E}^{(n)}\right\rangle \right|^{2} \text{.}
\label{eq:Opt}
\end{equation}
It was found from direct calculation of the intensities in Eq.~\eqref{eq:Opt} that the optical transition to the first vibronic $A_1$ state of ${}^{1}\widetilde{E}$ state is not allowed. However, there is a significant optical transition dipole toward the split $E$ vibronic states around 45~meV. After switching off the small DJT effect in the electron-phonon Hamiltonian only a single $E$ mode appears with a smaller optical transition dipole moment. This clearly demonstrates that the small DJT effect does play an important role in understanding the optical features of the singlet states~\cite{Thiering2018}. The simulated PL spectrum from \emph{ab initio} wavefunctions is shown in Fig.~\ref{fig:PL_singlet} that can be directly compared to the low temperature experimental PL spectrum~\cite{Rogers2008}. Clearly, the broad feature with the maximum intensity at $\approx$43~meV can be reproduced (red curve). It was found that the broad feature consists of two vibronic excited levels  [see red ink text in Fig.~\ref{fig:el_ph_spectra}]. The experimental intensity and the shape of this broad feature can be well reproduced by invoking our electron-phonon Hamiltonian (red curve). This theory is further supported by an uniaxial stress experiment on the PL spectrum which showed up the existence of a forbidden state at $\approx$14~meV~\cite{Rogers2015}. This can be naturally explained by the calculated $A_1$ vibronic excited state [see blue ink text in Fig.~\ref{fig:el_ph_spectra}]. This $A_1$ state will play an important role in the temperature dependence of the ISC rate where $\approx$16~meV phonon mode was deduced from the temperature dependent ISC rate measurements in non-stressed diamond samples~\cite{Robledo2011} that should be identical with the optically forbidden vibronic mode.  

The absorption spectrum of the singlets should be calculated in a different manner. The PJT effect produces a barrier energy for the damped DJT APES energy surface of ${}^{1}\widetilde{E}$ state, thus at any event of absorption of the photon, the electron stays one of the distorted global energy minima. The photoexcited electron arrives to a highly symmetric state with equidistant ${}^{1}\widetilde{A_1}$ vibronic levels that look like a perfect quasiharmonic solution. As a consequence, the absorption spectrum can be simulated from a frozen distorted structure in the ground state towards a highly symmetric structure in the excited state~\cite{Thiering2018} that can be calculated within Huang-Rhys approximation. Finally, the simulated shape of the absorption spectrum is very different from that of the PL spectrum~\cite{Thiering2018}, in agreement with the experiments~\cite{Rogers2008, Kehayias2013}. We note that as the APES of the singlets was not directly calculated in this procedure~\cite{Thiering2018}, therefore the sharp feature with energy above the phonon bands of diamond in the absorption spectrum is not reproduced by this method, which feature was associated with the nitrogen-carbon local vibrations~\cite{Kehayias2013}.

\subsection{Magnetic properties}
\label{ssec:magn}

We previously discussed the orbitals of NV center and their interaction with phonons. Group theory analysis was very powerful to identify the electronic structure of the ground state and excited states orbitals. The fine structure of NV center can be further analyzed by invoking the analysis of $C_{3v}$ double group which treats spinor functions. Since NV center has even number of electrons (holes) the single group rows in the $C_{3v}$ double group appears as irreducible representation of the spin states~\cite{Lenef1996, Manson2006, Maze2011, Doherty2011}. In particular, the ${}^3A_2$ ground state splits to $A_1$ $ms=0$ and $E$ $ms=\pm1$ states, whereas the ${}^3E$ state splits to $E_{x,y}$ $ms=0$, $E_{1,2}$ $ms=\pm1$, $A_2$ and $A_1$ $ms=\pm1$. These splittings are intrinsic to the $C_{3v}$-symmetry defect, and not the external fields are responsible for their presence. Basically, spin-orbit and dipolar electron spin -- electron spin interactions may introduce energy spacing between spin levels. In the ${}^3A_2$ ground state, the effective angular momentum ($L$) of the orbital is zero, therefore, the spin-orbit interaction ($\lambda_z \hat{L}_z\hat{S}_z$) is zero, where $\lambda_z$ is the strength of interaction along the symmetry axis $z$. The zero-field-splitting $D$ between $A_1$ and $E$ is therefore caused by dipolar electron spin-spin interaction as already stated in Eq.~\ref{eq:Hspin}. On the other hand, $L=1$ can be envisioned for the ${}^3E$ state, thus both spin-orbit and dipolar electron spin -- electron spin interactions are present. The spin-orbit coupling and the $zz$ component of the dipolar electron spin -- electron spin tensor split the $ms=0$ and $ms=\pm1$ levels in ${}^3E$ state. Further splitting between $A_1$ and $A_2$ $ms=\pm1$ levels is caused by the $x^2-y^2$ component of the dipolar electron spin -- electron spin tensor~\cite{Maze2011, Doherty2011}, whereas its $xz$ component can couple the $E_{x,y}$ $ms=0$ with $E_{1,2}$ $ms=\pm1$ states~\cite{Maze2011}. The latter is responsible for the very weak radiative decay from $ms=0$ towards the ground state via $E_{1,2}$ state (see Sec.~\ref{ssec:rates}), that is the base for realizing the so-called $\Lambda$-scheme for single shot readout~\cite{Robledo2011}. These are the basic zero-field-splitting (ZFS) parameters in the ground and excited state spin manifolds.

The spin states and levels behave in a strikingly different way as a function of temperature~\cite{Batalov2009, Acosta2010}. In the ground state manifold, the ZFS $D$ constant decreases slightly ($-74.2(7)$~kHz/K) but well resolved~\cite{Acosta2010}. On the other hand, the fine structure of ${}^3E$ state can be well observed at cryogenic temperatures with $\lambda_z = 5.33\pm0.03$~GHz~\cite{Batalov2009, Bassett2014} and $D=1.42$~GHz,  $D_{A_1, A_2}=D_{xx}-D_{yy}=3.1$~GHz~\cite{Batalov2009} with $D_{E_{1,2}, E_{x,y}}=D_{xz}=0.2$~GHz coupling parameter~\cite{Tamarat2008} but only ZFS caused by $D_{zz}$ remains observable at $T=20$~K and temperatures above whereas the gap between $A_1$ and $A_2$ as well as the spin-orbit related splitting ($\lambda_z $) completely disappears~\cite{Batalov2009}. Understanding these features requires to calculate the ZFS parameters, i.e., the spin-orbit energy and the dipolar electron spin -- electron spin tensor as well as understanding dynamical effects in ${}^3E$ state. The latter will be discussed in Sec.~\ref{sssec:temperature}.    

The spin Hamiltonian of the spin-spin dipolar interaction may be written as
\begin{equation}
\label{eq:Hss}
\begin{aligned}&
H_{ss} = -\frac{\mu_0 g_e^2 \mu_B^2}{4\pi} 
\sum_{i>j}\frac{3\left(\hat{\mathbf{S}}_i \cdot \mathbf{r}_{ij}\right) \left(\mathbf{r}_{ij} \cdot \hat{\mathbf{S}}_j\right) - \left(\hat{\mathbf{S}}_i \cdot \hat{\mathbf{S}}_j\right) \mathbf{r}_{ij}^2}{\left|\mathbf{r}_{ij}\right|^5} \\&
\equiv \sum_{i>j} \hat{\mathbf{S}}_i D_{ij} \hat{\mathbf{S}}_j  \text{,}
\end{aligned}
\end{equation}
where $\mathbf{r}_{ij}=\mathbf{r}_i-\mathbf{r}_j$. The $3\times3$ $D$-tensor can be diagonalized to find the spectrum and spin eigenstates. In the $C_{3v}$ symmetry, the $D_{zz}$ component lies in the symmetry axis which also sets the quantization axis of the electron spin along the symmetry axis of NV center. In the ground state and excited state manifold, the ZFS is $D = (3/2) D_{zz}$ and the corresponding spin levels can be computed as $D\left( S_z^2 - \frac{S(S+1)}{3}\right)$ for $S_z = \{ 0, \pm1\}$ with $S=1$, thus the components, e.g., $zz$ component, of the tensor should be calculated. The $D$ tensor is associated with the two-particle spin density matrix, $n_2 (r_1, r_2)$, which can be approximated by using the Slater-determinant of the KS wave functions $\phi$ of the considered system, so that $n_2(r_1, r_2) \approx \left|\Phi_{ij}(r_1, r_2)\right|^2$, where $\Phi_{ij}(r_1, r_2) = \frac{1}{\sqrt{2}} \left(\phi_i(r_1)\phi_j(r_2) - \phi_j(r_1)\phi_i(r_2)\right)$ and then
\begin{equation}
\label{eq:Dtensor}
\begin{aligned}&
D_{ab} = \frac{1}{2}\frac{\mu_0}{4\pi}\frac{g_e^2 \mu_B^2}{S(2S-1)}\\&
\sum_{i>j} ^{\text{occupied}} \chi_{ij} \int \left|\Phi_{ij} (r_1, r_2)\right|^2 \frac{r_{}^2\delta_{ab}-3 r_a r_b}{r_{}^5} d^3r_1 d^3r_2 \text{,}
\end{aligned}
\end{equation}     
where $r_{a,b} = \left(\mathbf{r}_1 - \mathbf{r}_2\right)_{a,b}$ and $\chi_{ij}$ is either $1$ or $-1$ for KS $i, j$ states of the same or different spin channels, respectively. Note that in DFT the KS states are not spin restricted, as mentioned previously. Consequently, not only the unpaired KS states but also the rest of the occupied states can contribute to the spin density and the ZFS~\cite{Rayson2008}. The first implementation of Eq.~\refeq{eq:Dtensor} in DFT supercell version with LCAO basis set was reported in Ref.~\onlinecite{Rayson2008}. The first implementation into plane wave supercell code was done by Iv\'ady and co-workers~\cite{Ivady2014} that only used pseudo wave functions in the \textsc{VASP} code. The theory of all-electron PAW version of $D$-tensor was reported by Bodrog and Gali~\cite{Bodrog2014}. The all-electron version was then implemented by Martijn Marsman into \textsc{VASP} code in 2014. Other implementations also followed later: both the pseudo density version~\cite{Seo2017} and all-electron versions, have been recently implemented into \textsc{Quantum Espresso} code~\cite{pwscf, Biktagirov2018} or an adaptive finite elements three-dimensional grid code~\cite{Ghosh2019}. 

The calculated $D$ constant in the ground state was within 3\% with respect to the low-temperature value at $\approx 2.87$~GHz~\cite{Ivady2014} in $\Gamma$-point 512-atom supercell calculation. By increasing the supercell size, this value did not change but larger supercell might be needed for similar defect quantum bits~\cite{Davidsson2018}. $\Delta$SCF method can be used to generate the KS states for calculating the components of the $D$-tensor in the ${}^3E$ excited state. By following previous group theory analysis~\cite{Maze2011, Doherty2011}, the calculated $D=1.61$~GHz, $D_{A_1, A_2}=D_{xx}-D_{yy}=1.95$~GHz, and $D_{E_{1,2}, E_{x,y}}=D_{xz}=0.15$~GHz do not show such an excellent agreement with the derived parameters from PLE measurements~\cite{Fuchs2008, Batalov2009}, which resulted in 1.42~GHz, 3.1~GHz, and 0.2~GHz, respectively. It is not yet clear why there is a larger discrepancy with respect to the experimental data in the excited state than that in the ground state. The $\Delta$SCF method should work well for ${}^3E$ state, and tests on the size of the supercell do not significantly change the results (unpublished results from the author). 

The spin-orbit interaction Hamiltonian in zero-order approximation can be written as
\begin{equation}
\label{eq:HSO_gen}
\hat{H}_{\text{SO}} = \frac{1}{2} \frac{1}{c^2 m_e^2} \sum_i \left( \nabla_i V \times \hat{\mathbf{p}}_i \right) \hat{\mathbf{S}}_i  \text{,}
\end{equation}
where $V$ is the nuclear potential energy, $m_e$ is the electron mass, and $\hat{\mathbf{p}}_i$ and  $\hat{\mathbf{S}}_i$ are the momentum and spin of electron $i$. The elements of the orbital operator vector $\hat{\mathbf{O}} =  \nabla_i V \times \hat{\mathbf{p}}_i $ can be calculated from the KS orbitals which defines the effective angular momentum of NV center. Note that the crystal field of a solid breaks the spherical symmetry of the spin-orbit interaction. In $C_{3v}$ symmetry, the interaction Hamiltonian can be rewritten as~\cite{Maze2011}
\begin{equation}
\label{eq:HSO_C3v}
\hat{H}_{\text{SO}} = \sum_i \lambda_{\perp} \left( \hat{L}_{i,x} \hat{S}_{i,x} +  \hat{L}_{i,y} \hat{S}_{i,y} \right) + \lambda_z \hat{L}_{i,z} \hat{S}_{i,z} \text{,}
\end{equation}
where $\lambda_{\perp}$ and $\lambda_{z}$ are the basal or non-axial and axial parameters of the interaction, respectively. The former flips the orbital and spin state, thus, it does not play a role in the solution of double group of $C_{3v}$ symmetry but can play a role in the ISC by coupling $ms=\pm1$ of the triplet to a singlet state~\cite{Maze2011, Doherty2011}. The latter will only shift spin levels in the ${}^3E$ triplet or may couple the $ms=0$ of the triplet to a singlet state~\cite{Maze2011, Doherty2011}.

\begin{figure}
\includegraphics[width=0.85\columnwidth]{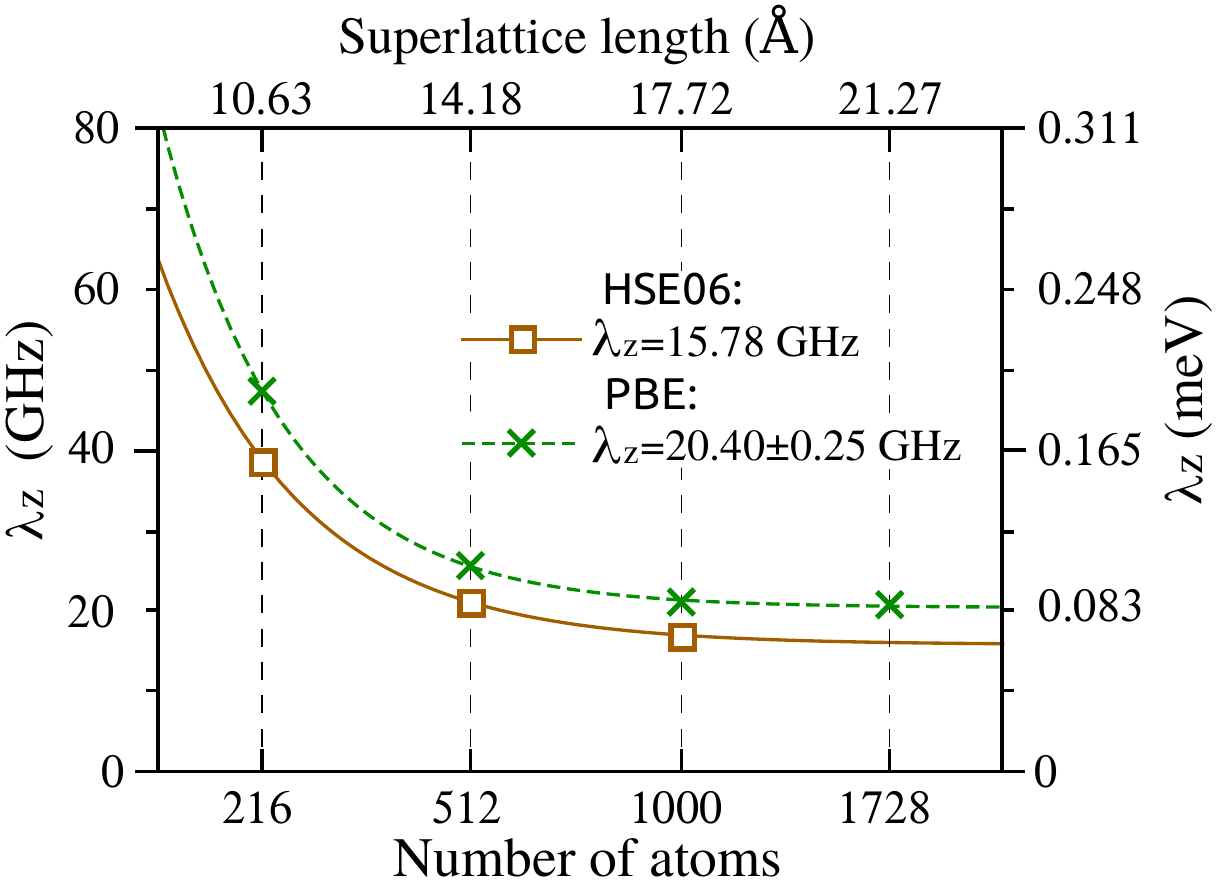}
\caption{\label{fig:SOscal} Convergence of the axial spin-orbit coupling parameter $\lambda_z$ in the excited state of NV center~\cite{Thiering2017}. Horizontal axes shows the supercell size and the number of carbon atoms in the defect free supercells within $\Gamma$-point sampling. DFT PBE calculations prove the exponential decay of the value that can be used to fit the converged value for DFT HSE06 calculations.}  
\end{figure}

The axial spin-orbit interaction strength in the ${}^3E$ excited state can be obtained both from total energy difference calculations and from the splitting of KS states in $\Delta$SCF non-collinear spin-orbit DFT calculations~\cite{Thiering2017}. It has been shown that the two approaches gave the same result. The latter is more practical because it only calls a single calculation but should be done in the $\Gamma$-point of the BZ, as the dispersion and splitting of the defect states in low symmetry $k$-points may be larger than the axial spin-orbit splitting. To accurately determine small values of the axial spin-orbit interaction, the calculations require high numerical convergence and accuracy as the spin-orbit energy falls in the $\mu$eV regime. Finite size effect turned to be crucial for spin-orbit interaction calculations~\cite{Thiering2017} (see Fig.~\ref{fig:SOscal}). Thiering and Gali in Ref.~\onlinecite{Thiering2017} attributed the observed finite size effect to the overlap of the defect states and used an exponential fit to eliminate supercell size dependence of the  $\lambda_z$.

The calculated HSE06 $\lambda_z=15.78$~GHz is about $3\times$ larger than the data derived from PLE measurements at cryogenic temperature~\cite{Batalov2009}. This difference is not a discrepancy but rather the consequence of the DJT nature of the ${}^3E$ state as it is stated in Eq.~\refeq{eq:tilde_Psi}. Ham already suggested that the effective spin-orbit energy will be reduced in such systems with a Ham reduction factor $p$~\cite{Ham1965}. In the particular ${}^3E$ state of NV center,  $p=\sum_{nm}\left[c_{nm}^{2}-d_{nm}^{2}\right]$ which represents the mixture of the $E_+$ component with the $E_-$ component of the ${}^3E$ state that results in the quenching of the effective angular momentum (see the derivation in Ref.~\onlinecite{Thiering2018prx}). The $c_{nm}$ and $d_{nm}$ coefficients were taken from the solution of the $E \otimes e$ DJT electron-phonon Hamiltonian, that finally results in $p=0.304$ with an effective spin-orbit splitting of 4.8~GHz~\cite{Thiering2017}. 

The ISC between ${}^1A_1$ and ${}^3E$ state is associated with the perpendicular component of the spin-orbit operator. Unlike the case of $\lambda_z$, one has to apply an approximation to calculate this, namely, that the KS wave functions building up the ${}^3E$ state and the ${}^1A_1$ multiplet do not change. This enables to compute $\lambda_\perp$ \emph{ab initio} by using the $a_1$ and $e_{+,-}$ KS wave functions of the NV($-$) in the ${}^3E$ excited state as $\langle a_1 | \hat{H}_\text{SO}|e_{+}\rangle/\sqrt{2}$. The converged intrinsic value is $\lambda_\perp = (2 \pi) 56.3$~GHz in rad/s unit which is relatively large, and it is not damped by DJT because ${}^1A_1$ state is not the part of DJT effect. We note that the nitrogen contribution is minor in $\lambda_z$ whereas it is significant in $\lambda_\perp$ that explains the large anisotropy between $\lambda_z$ and $\lambda_\perp$.  

In NV center (see Eq.~\refeq{eq:Hspin}), the ${}^{14}$N nuclear spin has a quadrupole splitting, $C_Q$, that can be written as
\begin{equation}
\label{eq:quadrupole}
C_Q = 3 e Q_{\text{N}} V_{zz}/4 h \text{,}
\end{equation}
where $h$ and $e$ are the Planck-constant and the charge of the electron, and $V_{zz}$ is the gradient of the electric field changes around the nitrogen nucleus in $z$ direction (chosen symmetry axis of the defect). $Q_\text{N}$ is
the nuclear electric quadrupole moment of ${}^{14}$N, which scatters between 0.0193 and 0.0208~barn in the literature~\cite{Stone2005}. $V_{zz}$ can be calculated \emph{ab initio} within the PAW framework. In this calculation the criterion for the wave functions (plane wave cutoff) and the forces should be set accordingly, in order to achieve convergent $V_{zz}$ (see the Supplementary Information in Ref.~\onlinecite{Pfender2017}). The calculated $C_Q$ value for NV center is $-5.019\pm0.19$~MHz, where the uncertainty comes from the uncertainty in the experimental data of $Q_\text{N}$. Nevertheless, the \emph{ab initio} result is very close to the experimental data, and could be used to follow the change in the quadrupole moment in NV defect as a function of the charge state~\cite{Pfender2017}.

The last remaining terms in Eq.~\refeq{eq:Hspin} at zero magnetic field are related to the hyperfine interactions. Hyperfine interaction describes the interaction between the electron spin and the nucleus spins that can be generally written as
\begin{equation}
\label{eq:Hhyperfine}
H_{\text{hyp}} = \hat{\mathbf{S}} A \hat{ \mathbf{I} } \text{,}
\end{equation}
where $A$ is the hyperfine tensor and $\hat{\mathbf{I}}$ is the nuclear spin vector operator. Unlike the case of electron spin -- electron spin interaction, where the Pauli exclusion principle does not allow two electrons with the same spin at the same place, the electron spin may be localized at the place of the nucleus spin. This is called the Fermi-contact term whereas the dipole-dipole term is familiar to what we learnt for the electron spin -- electron spin interaction. These can be respectively written as
\begin{equation} 
\label{eq:Hyp}
\begin{aligned}&
A^{(n)}_{ab} = \frac{ 2 \mu_0 }{3} g_e \mu_{\text{B}} g_{n} \mu_{n} \frac{n_s \! \left( \mathbf{R} \right)}{S} \\& + \frac{ \mu_0 }{4 \pi} g_e \mu_{\text{B}} g_{n} \mu_{n} \frac{1}{S} \int  \frac{ 3 r_a r_b - r^2 \delta_{ab} }{r^5} n_s \! \left( \mathbf{r} \right)  d^3r \text{,}
\end{aligned}
\end{equation}
where $n_s \! \left( \mathbf{r} \right)$ is the electron spin density, $\mathbf{r}$ is the vector between the electron spin and nuclear spin at $\mathbf{R}$, $g_{\text{n}}$ is the nuclear $g$-factor, and $\mu_{\text{n}}$ is the nuclear magneton for a given nucleus $n$.  We note that the nitrogen atom and distant carbon atoms sit in the symmetry axis by creating a hyperfine field that preserves the $C_{3v}$ symmetry. For those nuclear spins, the hyperfine tensor is diagonal with diagonal element $A_{xx} = A_{yy} = A_{\perp}$ and $A_{zz}= A_{\parallel} $. These parameters can be expressed by the Fermi-contact term $a$ and a simplified dipolar coupling term $b$ as  $A_{\perp} = a - b $ and $A_{\parallel} = a+2b$. The other nuclear spins reduce the symmetry, thus all the principal values of the hyperfine tensor differ for those nuclear spins. One can recognize from the definition of the Fermi-contact term that the spin density at the nucleus site should be calculated very accurately. Thus, all-electron methods should be applied: either treating the core electrons explicitly (typical for GTO codes~\cite{Nizovtsev2018}), PAW method for plane wave codes~\cite{Blochl2000, Gali2008, Smeltzer2011, Szasz2013} or finite elements numerical method on a grid~\cite{Ghosh2019}. 

\begin{figure}
\includegraphics[width=0.95\columnwidth, keepaspectratio]{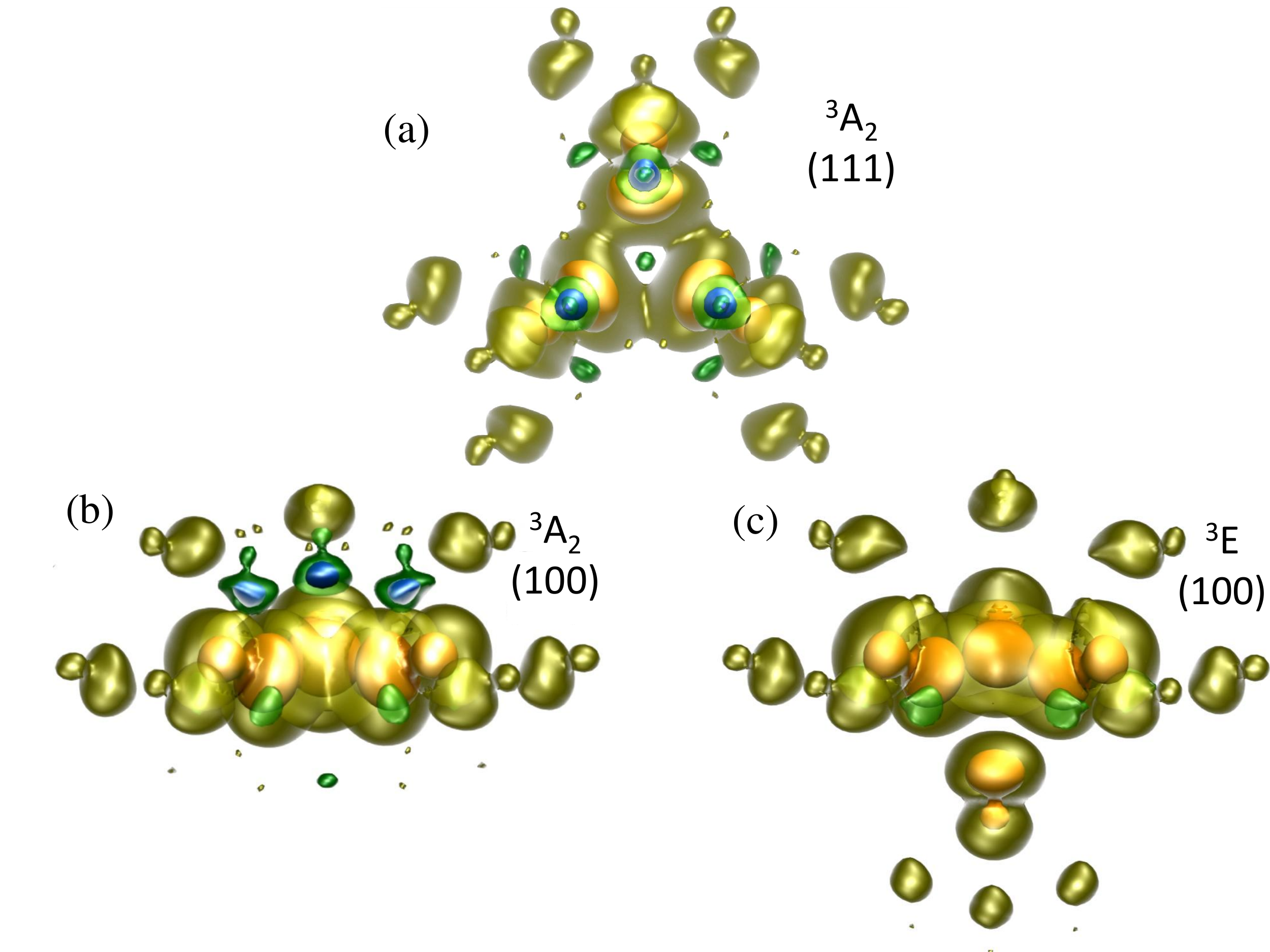}
\caption{\label{fig:spindens} Calculated spin density in a 512-atom supercell in Ref.~\onlinecite{Gali2008}. The isosurfaces of blue, green, yellow, orange and red lobes are $-0.014$, $-0.007$, $+0.014$, $+0.126$, and $+0.478$, respectively. (a) ${}^3A_2$ state top view of (111) plane. The threefold rotation symmetry of $C_{3v}$ point group can be well identified with symmetrically equivalent ions around the vacancy. The red lobes show the position of the three carbon atoms near the vacancy with the largest localization of the spin density. (b) ${}^3A_2$ state side view of (110) plane. The spin density shows an oscillation in the plane of the three nearest carbon atoms around the vacancy. (c)  ${}^3E$ side view of (110) plane. The spin density is significantly localized on the nitrogen atom below the nearest three carbon atoms around the vacancy.}  
\end{figure}

The hyperfine constants of NV center were first determined by plane wave PAW DFT LDA method~\cite{Gali2008}. It was found that the spin density is not evenly distributed around the core of the defect but rather elongated in the plane of the three carbon dangling bonds where the amplitude of the spin density decays like $\sin x / x$ function, where $x$ is the distance from the vacancy [Fig.~\ref{fig:spindens}(a) and (b)]. This spin density distribution has an analog of a Friedel-oscillation with NV center as an impurity. This spin density distribution is responsible for the fact that direct quantum mechanical interaction of NV center with acceptor defects depends is much stronger in the basal plane of the three carbon dangling bonds of the NV center, i.e., in the (111) plane than that along the $\langle111\rangle$ direction~\cite{Chou2018}. Another important consequence of this finding is that the Fermi-contact term cannot be neglected for ${}^{13}$C spins in the (111) plane near the NV center but some proximate ${}^{13}$C spins have small hyperfine constants that lie in the node of the spin density oscillation. In this study~\cite{Gali2008}, some ${}^{13}$C spin quantum bits were identified that were employed in a previous entanglement measurement between the electron spin and nuclear spins of NV center~\cite{Childress2006}. 

The spin density significantly redistributes in the ${}^3E$ excited state and is significantly localized around nitrogen atom [Fig.~\ref{fig:spindens}(c)]. As a consequence, the hyperfine constants on the nitrogen spins increases up to $\approx50$~MHz which broadens the resonance condition of external magnetic fields for transferring the polarization of the electron spin towards the nitrogen nuclear spin in the excited state (see Refs.~\onlinecite{Jacques2009, Gali2009prb13C, Ivady2015}).

This type of calculations was repeated by PAW DFT PBE method~\cite{Gali2009prb13C, Smeltzer2011}, where the agreement between the reported hyperfine constants of ${}^{14}$N and ${}^{13}$C nuclear spins was very good~\cite{Loubser1977, Felton2009}. Later it was found that this was partially fortuitous agreement~\cite{Szasz2013} because of the PAW implementation with frozen core orbitals~\cite{Blochl2000}. The core polarization of carbon $1s$ orbital significantly contributes to the Fermi-contact term, in particular, where the spin density is much localized~\cite{Szasz2013}. Generally, core polarization correction is required in the PAW implementation of accurate hyperfine tensor calculation~\cite{Yazyev2005, Szasz2013}. DFT HSE06 calculations localized the wave functions more around the vacancy than DFT PBE does but core polarization correction will reduce the Fermi-contact term of ${}^{13}$C spins with resulting in an excellent agreement with the experimental data~\cite{Szasz2013}. It was shown~\cite{Szasz2013} that cancellation of errors (DFT PBE underestimation of spin density and neglect of core polarization) does not generally occur for other defect quantum bits, thus implementation of core polarization in the PAW formalism is crucial.
In a recent study~\cite{Nizovtsev2018}, non-flipping proximate ${}^{13}$C spins in NV center going from the ground state to the excited state were identified in an 510-atom molecular cluster model by means of hybrid functional calculations of hyperfine tensors as implemented in an all-electron GTO code that can be used as long-living quantum memories.    

We note that orbital hyperfine interaction may contribute to the hyperfine dipole-dipole interaction in defects with an effective angular momentum of the associated orbitals, e.g., Ref.~\onlinecite{Blugel1987}, which was implemented into a Green-function based code and applied to nickel defects in diamond~\cite{Larico2009} but not to known defect quantum bits.

\subsection{Radiative and non-radiative rates: a complete theory on the optical spinpolarization loop}
\label{ssec:rates}

By having the magneto-optical parameters in hand, one can discuss the \emph{ab initio} description of ODMR and PDMR readout of the NV center in diamond. In the ODMR readout, the interplay between the radiative and ISC decay rates from the ${}^3E$ excited state plays a crucial role that will be first discussed below. In the PDMR readout, the direct photoionization and Auger-recombination processes also enter that we describe next. 

Radiative rates can be readily calculated from KS DFT by taking the corresponding $a_1$ and $e_{x,y}$ KS orbitals of NV center and calculating the $\langle e_{x,y}\left|e\mathbf{\hat{r}}\right|a_1\rangle$ transition dipole matrix element (e.g, Refs.~\onlinecite{Siyushev2013, Thiering2017}). By using the Wigner-Weisskopf theory of fluorescence, the radiative lifetime $\tau$ can be calculated as~\cite{Weisskopf1930}
\begin{equation}
\label{eq:PL_lifetime}
\frac{1}{\tau} = \frac{n_r\omega^3 \left| e\mathbf{r} \right|^2}{3 \pi \epsilon_0 \hbar c^3}
\text{,}
\end{equation}
where $n_r$ is the refractive index of diamond, $\hbar \omega$ is the excitation energy, $c$ is the speed of light, and $e\mathbf{r}$ is the optical transition dipole moment. One can realize that the $e \mathbf{z}$ component will be zero, and only the perpendicular components will give non-zero results as was also found in the experiments~\cite{Alegre2007}. The polarization property of the absorbed an emitted photons is naturally obtained by KS DFT calculations. 
Finally, Eq.~\refeq{eq:PL_lifetime} results in about 8~ns radiative lifetime, which is relatively close to the experimental one (about 12~ns that corresponds to about $(2\pi) 13$~MHz rate, see Refs.~\onlinecite{Batalov2008, Goldman2015prl}). For the triplet optical transition, this is a very good approximation because of the relatively simple character of the electronic ground and excited state. Calculation the optical transition between the singlet states has just become feasible by the DFT+CI-cRPA method~\cite{Bockstedte2018} where the multiplet wave functions are considered in the optical transition dipole moment calculation. The calculated radiative lifetime of ${}^1A_1$ state is 1878~ns that is two orders of magnitude longer than that for the triplet~\cite{Bockstedte2018}. This \emph{ab initio} result sheds light on the experimental fact that not just the competition with non-radiative decay and the low fractional population of the singlets but the inherently tiny optical transition dipole moment is responsible for the very weak infrared PL signal of NV center~\cite{Rogers2008}. This agrees well with the conclusion of absorption measurements between the singlets~\cite{Kehayias2013}. The very recent pump-probe PL measurement has observed 100~ps lifetime of the ${}^1A_1$ state~\cite{Ulbricht2018_IR}. The combination of this recent experimental data with the \emph{ab initio} radiative rate~\cite{Bockstedte2018} and observed absorption~\cite{Kehayias2013} indicates that the non-radiative decay from the ${}^1A_1$ state to the  ${}^{1}E$ is much faster than the radiative one. This should be verified by future \emph{ab initio} calculations of the non-radiative rate between the singlets. 

We note that these calculations inherently assume that the Franck-Condon principles hold for the optical transitions, i.e., the optical transition dipole moment is independent from the ionic movements. This is a valid approximation for the optical transition between the triplets of NV center in diamond but it does not necessarily holds for all types of vibrations and solid state defect quantum bits, in which the Herzberg-Teller interaction is considerable~\cite{Norambuena2016}. We showed above that the fluorescence of the singlets of NV center in diamond does not follow the Franck-Condon principles either, and one should use polaronic wave functions in the calculation of the optical spectrum.

The non-radiative rates are associated with the spin-orbit interaction between the triplet and singlet states mediated by phonons, i.e., ISC processes. Thus, the ISC rates from the ${}^3E$ spin states to the ${}^1A_1$  state and from the ${}^1E$ state to the ${}^3A_2$ spin states should be calculated. In the Franck-Condon type of approximation, the ISC rate from $A_1$ state of ${}^3E$ state may be calculated~\cite{Goldman2015prl} as
\begin{equation}
\label{eq:gamma_A1}
\Gamma_{A_1} = 4 \pi \hbar \lambda_\perp^2 \sum_n \left|\langle \chi_0 | \chi'_{\nu_n}\rangle \right|^2 \delta\left(\nu_n - \Delta\right) \text{,}
\end{equation}
where $\lambda_\perp$ is given in rad/s unit, $\Delta$ is the energy spacing between $|A_1\rangle$ and $|^1A_1\rangle$, and $\delta$ is the Dirac delta function. $|\chi_0\rangle$ is the ground vibrational level of $|A_1\rangle$, and $|\chi'_{\nu_n}\rangle$ are the vibrational levels of $|^1A_1\rangle$ with energies $\nu_n$ above that of $|^1A_1\rangle$. Group theory implies that ISC can only occur between $|A_1\rangle$ and $|^1A_1\rangle$ [see Fig.~\ref{fig:struct}(b)]. However, three different ISC rates were observed in the experiments at cryogenic temperature in which the ratio of two rates is about $0.52\pm0.07$ and the third rate was much smaller than the other two~\cite{Goldman2015prl}.   
We already showed above that the ${}^3E$ state is not a pure electronic state but the $e_{x,y}$ phonons couple the components of ${}^3E$ state. By solving the $E\otimes e$ electron-phonon Hamiltonian and analyzing the solution by group theory, one finds $\widetilde{A}_1$, $\widetilde{A}_2$, and $\widetilde{E}_{1,2}$ vibronic states as eigenstates, where the contribution of the electronic $|A_1\rangle$ in $|\widetilde{A}_1\rangle$, $|\widetilde{E}_{1,2}\rangle$, and $|\widetilde{A}_2\rangle$ is $c_i$, $d_i$ and $f_i$, respectively, where running variable $i$ groups the phonon wave functions for given representations of phonon wave functions, and $n_i$ is the sum of quantum numbers of $e_x$ and $e_y$ phonons~\cite{Thiering2017}. Thiering and Gali demonstrated that DJT nature of $|^3E\rangle$ naturally explains the three observed rates that can be written with modifying Eq.~\refeq{eq:gamma_A1} as
\begin{equation}
\label{eq:G_A1}
\Gamma_{A_{1}}=4\pi\hbar\lambda_{\perp}^{2}\sum_{n=0}^{\infty}\sum_{i=1}^{\infty}\left[c_{i}^{2}\left|\langle \chi_0 | \chi'_{\nu_n}\rangle \right|^2 \delta\left(\nu_n -\Delta+n_{i}\hbar\omega_{e}\right)\right] \text{,}
\end{equation}
\begin{equation}
\label{eq:G_E12}
\Gamma_{E_{12}}=4\pi\hbar\lambda_{\perp}^{2}\sum_{n=0}^{\infty}\sum_{i=1}^{\infty}\left[\frac{d_{i}^{2}}{2}\left|\langle \chi_0 | \chi'_{\nu_n}\rangle \right|^2 \delta\left(\nu_n -\Delta+n_{i}\hbar\omega_{e}\right)\right] \text{,}\\
\end{equation}
\begin{equation}
\label{eq:G_A2}
\Gamma_{A_{2}}=4\pi\hbar\lambda_{\perp}^{2}\sum_{n=0}^{\infty}\sum_{i=1}^{\infty}\left[f_{i}^{2}\left|\langle \chi_0 | \chi'_{\nu_n}\rangle \right|^2 \delta\left(\nu_n -\Delta+n_{i}\hbar\omega_{e}\right)\right] \text{,}
\end{equation}
where $|\chi'_{\nu_n}\rangle$ are now restricted to the $a_1$ vibrational levels of $|^1A_1\rangle$~\cite{Thiering2017}. The \emph{ab initio} solution yielded  $f_1<0.001$ which explains the very low $\Gamma_{A_{2}}$ at cryogenic temperature. On the other hand, $c_1=0.578$ and $d_1=0.331$ are comparable. As a consequence, $\Gamma_{A_{1}}/\Gamma_{E_{12}}\approx0.5$ for $\Delta$=0.4~eV, which agrees well with the experimental findings~\cite{Goldman2015prl, Thiering2017}. On the other hand, the absolute values of the $\Gamma_{A_{1}}$ and other rates were significantly larger than the experimental data by taking $\lambda_{\perp}$=56~GHz from first principles spin-orbit matrix element from $a_1$ and $e_{x,y}$ KS wave functions. It is suspected that KS wave functions building up ${}^3E$ state and ${}^1A_1$ state may change, so it could be not a good approximation to calculate $\lambda_{\perp}$ from KS wave functions fixed at ground state electronic configuration.  
 
\begin{figure}
\includegraphics[width=0.7\columnwidth]{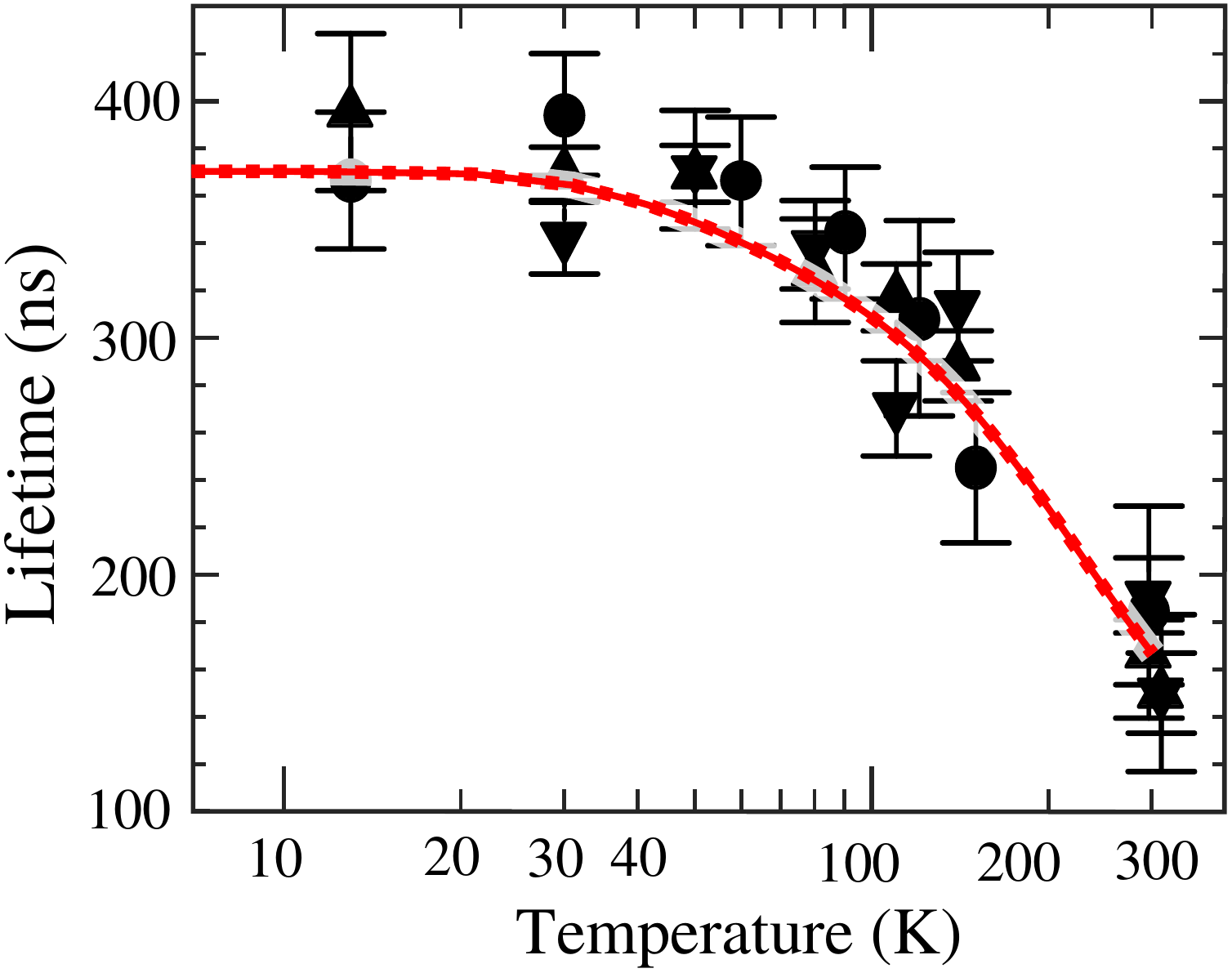} 
\caption{\label{fig:lifetime_T}The calculated lifetime (Ref.~\onlinecite{Thiering2018}) of the singlet shelving state is plotted as a function of the temperature with the observed lifetimes for two single NV centers (dot and triangle data points with uncertainties) taken from Ref.~\onlinecite{Robledo2011}.} 
\end{figure}

The next step is to calculate the ISC rate from ${}^1E$ state to the ${}^3A_2$ spin states in the lower branch that will also determine the lifetime of ${}^1E$ state. It is known from experiments that the lifetime of $|^1E\rangle$ is temperature dependent, and it varies between about 371~ns and 165~ns going from cryogenic temperature to room temperature~\cite{Robledo2011njp}. The temperature dependence could be well understood by a stimulated phonon emission process with an energy of $16.6\pm0.9$~meV~\cite{Robledo2011njp}. On the other hand, the ODMR contrast does not seem to vary significantly as a function of temperature. From PL decay measurements it was derived that the decay rates towards $ms=0$ and $ms=\pm1$ favors $ms=0$ only by about 20\% which was a surprising result but does not contradict with the measured ODMR contrast because of the high spin selectivity in the upper branch~\cite{Robledo2011njp}. Indeed, the nature of the ${}^1E$ state is complex: it is a polaronic  ${}^1\widetilde{E}$ state in which the character of ${}^1A_1$ and ${}^1E^\prime$ states appear. The former and latter links  ${}^1\widetilde{E}$ to $ms=0$ and $ms\pm1$ spin states of ${}^3A_2$, respectively, with the corresponding ISC rates of $\Gamma_z$ and $\Gamma_\perp = \Gamma_\pm + \Gamma_\mp$, respectively.  These rates may be described as~\cite{Thiering2018}
\begin{equation}
\label{eq:G_z}
\Gamma_{z}= \frac{2\pi C^2}{\hbar}\sum_{i}^{\infty}4\lambda_{z}^{2} d_{i}^{2}\left|\langle...\left|\chi_{i}\left(e_\pm\right)\right\rangle \right|^{2}\delta\bigl(\Sigma-n_{i}\hbar\omega_{e}\bigr) \text{,}
\end{equation} 
where the summation over all vibration wave functions of ${}^3A_2$ collapses to the number of $|\chi_{i}(e_\pm)\rangle$ vibration modes in the phonon overlap integral. Here, $d_i$ coefficient is responsible to the contribution of electronic ${}^1A_1$ state in $|{}^{1}\widetilde{E}\rangle$ that is linked to $ms=0$ of ${}^3A_2$ by $\lambda_z$. $\Sigma$ is the energy spacing between ${}^3A_2$ and ${}^1\widetilde{E}$ states of about 0.4~eV, and $\left(1-C^2\right)\approx0.1$ is the contribution of ${}^1E^\prime$ state in ${}^1\widetilde{E}$ state. The other rates can be expressed as~\cite{Thiering2018}
\begin{equation}
\label{eq:G_pm}
\Gamma_{\pm}=\frac{2\pi (1-C^{2})}{\hbar}\sum_{i}^{\infty}\lambda_{\perp}^{2}c_{i}^{2}\left|\langle...\left|\chi_{i}\left(a_{1}\right)\right\rangle \right|^{2}\delta\left(\Sigma-n_{i}\hbar\omega_{e}\right)
\end{equation}
and
\begin{equation}
\label{eq:G_mp}
\Gamma_{\mp}=\frac{2\pi (1-C^{2})}{\hbar}\sum_{i}^{\infty}\lambda_{\perp}^{2}f_{i}^{2}\left|\langle...\left|\chi_{i}\left(e_{\mp}\right)\right\rangle \right|^{2}\delta\left(\Sigma-n_{i}\hbar\omega_{e}\right) \text{.}
\end{equation}
It was again found that $\lambda_\perp/\lambda_z$=1.2 provides consistent absolute total ISC rate ($T_E^{-1} = \Gamma_z + \Gamma_\perp$) of about 2.7~MHz with $\Sigma\approx0.4$ energy gap, and the $\lambda_\perp$ derived from \emph{ab initio} spin-orbit matrix element between KS $a_1$ and $e_{x,y}$ states is largely overestimated. The calculated $\Gamma_z/\Gamma_\perp$ is about 5.5 which means that \emph{ab initio} theory predicts~\cite{Thiering2018} about 84\% selectivity towards $ms=0$ state over $ms=\pm1$ state in the ISC process in the lower branch of the optical spinpolarization loop. In a recent experimental study~\cite{Kalb2018}, the refined modeling of the signals resulted in $\Gamma_z/\Gamma_\perp = 4 \pm 0.5$ which agrees well with the \emph{ab initio} theory. This means that relatively high spin selection appears in the ISC process between ${}^1\widetilde{E}$ state and ${}^3A_2$ ground spin states. Finally, the temperature dependence of the lifetime $T_E$ of the ${}^1\widetilde{E}$ state can be calculated with using the calculated vibronic levels of ${}^1\widetilde{E}$ and the ISC theory in Eqs.~\eqref{eq:G_z}-\eqref{eq:G_mp} (see Fig.~\ref{fig:lifetime_T}). 
The Boltzmann occupation of vibronic levels was used at the given temperature, in order to compute ISC rates in Ref.~\onlinecite{Thiering2018}. A very good agreement was found with the experimental data~\cite{Robledo2011} as the calculated lifetime is reduced from 370~ns at cryogenic temperatures down to 171~ns at room temperature to be compared to 371$\pm$6~ns and 165$\pm$10~ns, respectively. \emph{Ab initio} theory calculations showed quantitatively that the vibronic state associated with the optically forbidden phonon feature at $\approx$14~meV in the PL spectrum plays a key role in the temperature dependence of the ISC rates~\cite{Thiering2018} where this connection was previously hinted in Ref.~\onlinecite{Doherty2013}. The calculated $\Gamma_z/\Gamma_\perp$ is only reduced by $\sim$5\% going from cryogenic temperature to room temperature in the simulations~\cite{Thiering2018} which means that the spinpolarization efficiency per single optical cycle does not degrade significantly as a function of temperature. These results demonstrate that \emph{ab initio} theory developed by Thiering and Gali can account for the intricate details of the ISC processes in NV center in diamond and reproduce the basic experimental data.

\emph{We emphasize that Eqs.~\eqref{eq:G_A1}-\eqref{eq:G_mp} complete the theory of optical spinpolarization loop of NV center in diamond where the corresponding matrix elements and coupling parameters could be mostly determined from first principles calculations~\cite{Thiering2017, Thiering2018, Bockstedte2018}.}  The remaining issues are discussed in Sec.~\ref{sec:outlook}. 

Next we discuss the photoionization of NV center and related defects that are important in the PDMR readout of NV center's spin state. The idea of PDMR readout is based on the standard ODMR measurements of single NV center. The phonon sideband photoexcitation of NV center (typically 532-nm wavelength) with high laser power can lead to unintentional ionization of the defect via absorption of a second photon before decay from the ${}^3E$ state towards the ground state. \emph{Ab initio} theory predicts that $\approx2.7$~eV requires to ionize NV center from the ground state to neutral NV defect at cryogenic temperature. The absorption of the second photon at the ${}^3E$ state likely occurs at the zero-point-energy since the phonon relaxation in the ${}^3E$ state is extremely fast~\cite{Ulbricht2018}, which is 1.945~eV above the ground state (ZPL energy of NV center). Thus, absorption of the second photon would promote the electron to  
$\approx4.3$~eV above the ground state level of NV center which is much larger than the ionization threshold. The two-photon absorption and the corresponding ionization of NV center will convert it to neutral NV~\cite{Gaebel2006}. \emph{Ab initio} theory proved that this two-photon ionization is an Auger-recombination process~\cite{Siyushev2013}. It is well-known in semiconductor physics that two types of ionization of defects may occur: (i) direct ionization or (ii) Auger-recombination. In the Auger-recombination multiple defect levels and states may play a role. In NV center, the direct ionization from the $e_{x,y}$ orbital in the ${}^3E$  state competes with the Auger-process where the high-energy electron promoted to the conduction band will fall back to the hole left in the $a_1$ state, and the energy gain is used to kick-out an electron from the $e_{x,y}$ level~\cite{Siyushev2013}. The Auger-recombination rate, i.e., inverse process of impact ionization, can be written as a Coulomb-interaction between the initial exciton state and all the possible final exciton states. The initial state $i$ is an excitation from orbital $j$ to $a$ with
spin $\downarrow$, here $j$ is the hole on the $a_1$ level and $a$ is a state
close to the conduction band edge, which can be simplified as $\phi_{j\downarrow}\phi_{a\downarrow}$ electron-hole KS wave functions.
The energy of this exciton is approximated as
$\epsilon_{a\downarrow}-\epsilon_{j\downarrow}$, where
$\epsilon_{r\sigma}$ is the KS level of state
$r\sigma$. The final exciton state is approximated similarly as $\phi_{k\uparrow}\phi_{b\uparrow}$ electron-hole KS wave functions,
where $k$ denotes the hole on the double degenerate $e_{x,y}$ orbital, $b$
is the electron above the conduction band minimum, and the energy of the exciton is $\epsilon_{b\uparrow}-\epsilon_{k\uparrow}$. In order to obtain the rate of transition, we need to sum over all possible final states: the degenerate $e_{x,y}$ level (index $k$) and the electron in the conduction band minimum (index $b$).
The final formula may be given as~(see Supplementary Materials in Ref.~\onlinecite{Siyushev2013} and also Ref.~\onlinecite{Voros2013})
\begin{equation}
\Gamma_{a}=\frac{2\pi}{\hbar}\sum_{bk}\left|V_{k\uparrow b\uparrow a\downarrow j\downarrow}\right|^2
\times
\delta[(\epsilon_{a\downarrow}-\epsilon_{j\downarrow})-(\epsilon_{b\uparrow}-\epsilon_{k\uparrow})] \text{.}
\end{equation}
We obtain a single $\Gamma$ by averaging over initial states:
\begin{equation}
\Gamma=\frac{\sum_a\Gamma_{a}\delta[\epsilon_{a\downarrow}-\epsilon_{j\downarrow}-\hbar \omega]}{\sum_a\delta[\epsilon_{a\downarrow}-\epsilon_{j\downarrow}-\hbar \omega]}\text{,}
\end{equation}
where $\hbar \omega$ is the sum of two energy pulses (two-photon absorption) and $V$ can be generally expressed as
\begin{equation}
\begin{aligned} &
V_{r\sigma''s\sigma''u\sigma'''t\sigma'''}=\\&\int\int d^3rd^3r'\phi_{r\sigma''}^*(\vec{r})\phi_{s\sigma''}(\vec{r})\frac{4\pi}{\epsilon_r\epsilon_0|\vec{r}-\vec{r}'|}\phi_{u\sigma'''}^*(\vec{r}')\phi_{t\sigma'''}(\vec{r}')\text{,}
\end{aligned}
\end{equation}
where $\phi_{s,r,u,t}$ are the corresponding KS wave functions. V\"or\"os and co-workers implemented~\cite{Voros2013} the computation of this rate into a plane-wave supercell code~\cite{pwscf} which was applied for NV center in diamond and $\epsilon_r$ was simply estimated from the refraction index of diamond in this particular calculation~\cite{Siyushev2013}. This could be the first \emph{ab initio} calculation of the Auger-recombination rate for point defects in solids. The direct ionization rate can be calculated like the usual absorption rate with optical transition dipole operator just the final state is at the conduction band minimum. The calculated Auger-recombination rate was $800$~ps whereas the direct ionization rate was $0.5$~$\mu$s. A very important consequence of this finding is that the two-photon absorption of NV center results in the \emph{ground state} of neutral NV defect, thus no PL signal of neutral NV defect is expected. The typical 532-nm excitation wavelength can excite the neutral NV defect for which the same type of two-photon absorption occurs, and finally it ends at the ground state of NV center~\cite{Siyushev2013}. It was found that the two-photon absorption of neutral NV defect is very effective at the excitation wavelength of its ZPL wavelength at 575~nm (yellow light). One can conclude that the full cycle of photoexcitation and ionization of NV center will lead to emission of an electron and a hole, and the final state is the ground state of NV center. 

Since the ejection of the electron goes through the ${}^3E$ excited state in which the spin states have different lifetimes, the rate of ionization (absorption probability of the second photon) should be also spin-dependent. This idea was tested on a diamond device where electrodes were built on the diamond structure for observation of the photocurrent upon photoionization of the NV centers~\cite{Bourgeois2015}. They observed a contrast in the photocurrent as they hit the microwave resonance of the electron spin in the ground state of ensemble NV centers~\cite{Bourgeois2015, Brandt2017}, i.e., PDMR readout. The full \emph{ab initio} description of a PDMR cycle would require to calculate the Auger-recombination rate for the neutral NV defect, which has not yet been reported. On the other hand, the emission of the electron from NV center and its spin-selectivity is now fully understood by using \emph{ab initio} theory. CI-cRPA calculations showed little or no photoionization from the ${}^1\widetilde{E}$ state towards the conduction band edge by 532-nm green excitation~\cite{Bockstedte2018}. Thus, all the important rates are in hand, in order to calculate the PDMR contrast, as the rates for ODMR readout can be used in combination with the Auger-recombination rate. 

In the first PDMR measurements~\cite{Bourgeois2015}, the PDMR contrast was relatively low compared to ODMR contrast because of the background current coming from mostly the P1 center in diamond. Therefore, it is highly important to understand the photoionization of NV center and other parasitic defects, in order to optimize the PDMR readout technique. Recent \emph{ab initio} calculations have found that green light excitation much favors the ionization of P1 center over that of NV center, however, the relative ionization rates can be pushed towards increasing that of NV center if blue light excitation is applied~\cite{Bourgeois2016}. In these calculations, the valence and conduction bands were included in the computation of the absorption rates (not just the band edges), therefore, the BZ sampling should go up to $6\times6\times6$ MP k-point set for convergent calculations in a 512-atom supercell~\cite{Bourgeois2016}. Indeed, dual-beam photoionization of the diamond sample with ensemble of NV centers resulted in an $3\times$ enhanced PDMR contrast over the single-beam photoionization scheme, where blue-light was employed for direct photoionization whereas green-light was applied to induce spin-selective non-radiative decay from the $^3E$ excited state~\cite{Bourgeois2016}. This again proves the predictive power of \emph{ab initio} calculations and its role for improvement the control and readout of solid state defect quantum bits. Later on it has been found that lock-in techniques with usual green-beam excitation can result in also an enhanced PDMR readout contrast~\cite{Gulka2017}. Other parasitic defects producing photoionization bands in the diamond samples could be identified by \emph{ab initio} calculations~\cite{Londero2018}, which can be rather used as a resource, i.e., PDMR quantum bits, that might not be visible in the ODMR readout but could be effective in the PDMR readout~\cite{Londero2018}. 

We note that the carrier capture cross section rates of NV defect and/or parasitic defects in diamond can be important, in understanding the atomistic processes and improving further the PDMR readout. \emph{Ab initio} framework already exists to calculate these rates~\cite{Alkauskas2014carrier} but not yet has been applied to NV center in diamond. The carrier capture cross section together with the Auger-recombination process may play an important role in the electroluminescence of NV defect, that only could successfully employed for the neutral NV defect~\cite{Mizuochi2012}.

\subsection{Effect of external perturbation}
\label{ssec:extpert}

\subsubsection{Magnetic field and hyperpolarization}
Having the computed magneto-optical parameters in hand, one can calculate the coupling parameters to various external fields and temperature. The most obvious type of external perturbation is the constant magnetic field that appears in Eq.~\refeq{eq:Hspin} known as Zeeman-effect. Generally, the interaction between the electron spin and magnetic field cannot be simply described by the $g$-factor of a the free electron, $g_e$, but rather by a $g$-tensor, where the Cartesian elements of the $g$-tensor can be obtained from the second derivative of the relativistic many-electron energy (e.g., Ref.~\onlinecite{Ziegler1997}). Implementation of $g$-tensor in supercell plane wave codes exist  (e.g., gauge including projector augmented wave approach~\cite{Pickard2001, Pickard2002, Ziegler2009, pwscf}), however, it has not been applied to NV center in diamond but other defects (e.g., Refs.~\onlinecite{Bardeleben2015, Bardeleben2016}). For NV center in diamond, the $g$-tensor does not significantly deviate from that of the free electron. Under external magnetic field, the Zeeman-effect appears both in the ${}^3A_2$ ground state and ${}^3E$ excited state but the $g$-tensor in the excited state should differ at low temperatures because spin-orbit interaction can contribute to the $g$-tensor in the excited state but it is almost negligible in the ground state. As a consequence, PLE measurements at cryogenic temperatures should be able to identify the difference in the $g$-tensor in the ground state and excited state in future Zeeman experiments.   

When constant (positive) magnetic field is applied along the symmetry axis of the NV center with the magnitude near the $D$-constant then the $ms=0$ and $ms=-1$ levels of the triplet state approach each other. The perpendicular component of the hyperfine constant of nitrogen spin opens a gap between these levels, so called level anticrossing, and the electron and spin states are coupled and rotate with a rate of  $\left| A_\perp \right| / (\sqrt{2} \hbar)$~\cite{Jacques2009, Gali2009prb}. As a consequence, optical pumping will result in highly spinpolarized nuclear spin state of nitrogen. For off-axis ${}^{13}$C spins, the process is similar but more complicated because of lowering the symmetry of the system. As a consequence, the spinpolarization of those nuclear spins is generally not so effective as discussed in details in Ref.~\onlinecite{Ivady2015}. Nevertheless, the nitrogen and carbon nuclear spins can be much well polarized at room temperature that could be available by traditional techniques (Boltzmann occupation of the lowest spin level split by Zeeman effect at large constant magnetic field and low temperatures). This is called optical dynamic nuclear spin polarization, and if majority of the carbon nuclear spins was polarized then hyperpolarization of diamond is achieved. In these studies, the full hyperfine tensor of all the carbon spins is needed that were taken from first principles HSE06 calculations~\cite{Ivady2015}. This shows the importance of \emph{ab initio} calculations in understanding spin related phenomena.

\subsubsection{Electric field and strain}
In the presence of electric field and strain, the ground state spin Hamiltonian in Eq.~\refeq{eq:Hspin} is modified. Again, there are symmetry conserving interactions that only shifts the levels whereas symmetry breaking interaction may mix the spin states. Accordingly, the $z$-component of the electric field vector $E$ only shifts the spin levels (transforms as fully symmetry $A_1$ representation; $E_z$) whereas the $x$ and $y$ components (transform as $E$ representation) can mix the $ms=\pm1$ levels and $ms=0, \pm1$ levels that can be written as 
\begin{equation}
\label{eq:Efield}
\begin{aligned}
\hat{H}_\text{el} &=  H_{\text{el} 0} + H_{\text{el} 1}+H_{\text{el} 2} 
\\&
= h d_z E_z S_z^2 \\&
+ h d'_\perp \left[  E_x \left( S_xS_z + S_zS_x \right) + E_y \left( S_yS_z + S_zS_y\right)  \right] \\& 
- h d_\perp E_x \left(S^2_x - S^2_y\right)  + h d_\perp E_y \left(S_xS_y + S_yS_x\right) \text{,}
\end{aligned}
\end{equation}
where $h$ is the Planck- constant. Eq.~\refeq{eq:Efield} had been known as a linear Stark-shift for $C_{3v}$ $S=1$ EPR centers, and was derived specifically for NV center by Doherty and co-workers in Ref.~\onlinecite{Doherty2012}. The coefficients $d_\perp = 17 \, \text{Hz}\, \text{cm}/\text{V}$ and 
$d_z = 0.35\, \text{Hz}\, \text{cm}/\text{V}$ have been inferred
in the experiment of Ref.~\onlinecite{vanOort1990}. However, to our knowledge, the coefficient $d'_\perp$ has not been quantified experimentally or theoretically;
nevertheless it is expected~\cite{Doherty2012} to have the same order of magnitude as $d_\perp$. We note that the effect of $d'_\perp$ is often neglected as in the zero magnetic field it is suppressed by the relatively large $D=2.87$~GHz gap between the $ms=0$ and $ms=\pm1$ levels. On the other hand, under constant magnetic fields close to the ground state level anticrossing point this effect can be dominating whereas the mixing between $ms=\pm1$ ($d_\perp$ term) will be suppressed. 

First principles calculations of electric field is not straightforward in the three-dimensional supercell model as the electric field breaks the periodicity of the system in the direction of the electric field. One possible solution is to apply a slab model for modeling the NV center with a surface that is perpendicular to the direction of the electric field. In Ref.~\onlinecite{Falk2014}, the NV-center in diamond was modeled by a 990-atom 2-nm thick diamond (111) slab, where the the applied electric field was set to 0-0.1~V/\AA\ along the symmetry axis of the defect in diamond slab with -OH termination. In this interval, the response was found to this perturbation to be linear on the calculated $D$-constant and resulted in $d_z = 0.76\, \text{Hz}\, \text{cm}/\text{V}$ (see Ref.~\onlinecite{Falk2014}) which falls to the order of magnitude deduced from experiments. The disadvantages of this method are that (i) the calculation of the perpendicular component of the coupling parameters requires another slab model with an appropriate surface orientation and termination, and (ii) the surface termination may induce "artificial" polarization around the NV center even when is buried deepest in the slab model. 

Recently, an alternative approach was used to study the effect of electric fields on the properties of NV center~\cite{Udvarhelyi2019}. In this particular case, the coupling of external electric field to the optical transition was examined which is also known as Stark-shift~\cite{Udvarhelyi2018}. Stark-shift is generally an undesirable effect for quantum emitters because it can cause spectral diffusion (e.g., Ref.~\onlinecite{Siyushev2013}): during optical excitation of NV center nearby defects may be ionized that creates different electric fields around NV center that results in a shift of ZPL of NV center in subsequent optical excitation cycles. Another consequence is that zero (magnetic) field ODMR signals of \emph{ensemble} NV centers will show a splitting due to the various charged defects around the NV centers~\cite{Mittiga2018}. For the determining the Stark-shift of NV center, the permanent dipole moments should be calculated in the ${}^3E$ excited state and the ${}^3A_2$ ground state.

The alternative approach of calculating this quantity relies on the modern theory of polarization~\cite{KingSmith1993, Vanderbilt1993, Resta1994, Gajdos2006} that can be applied to periodic systems. The polarization $\mathbf{p}$ can be written as
\begin{equation}
\mathbf{p}=\frac{ife}{8\pi^{3}}\sum_{n=1}^{M}\int_{\text{BZ}}d\mathbf{k}\left<u_{\mathbf{k}n}\right|\nabla_{\mathbf{k}}\left|u_{\mathbf{k}n}\right>\text{,}
\end{equation}
where $M$ is the number of occupied bands, $f$ is the occupation number, $e$ is the elementary charge of the electron, and the wave functions ($u_{\mathbf{k}}$) are cell periodic and periodic in the reciprocal space. Using density functional perturbation theory (DFPT), $\nabla_{\mathbf{k}}\left|u_{\mathbf{k}n}\right>$ can be calculated from the Sternheimer equations with similar self-consistent iterations as in the self-consistent field DFT
\begin{equation}
\left(H_{\mathbf{k}}-\epsilon_{\mathbf{k}n}\right)\nabla_{\mathbf{k}}\left|u_{\mathbf{k}n}\right>=-\frac{\partial \left(H_{\mathbf{k}}-\epsilon_{\mathbf{k}n}\right)}{\partial \mathbf{k}}\left|u_{\mathbf{k}n}\right>\text{.}
\end{equation}
By applying this theory to NV center, the coupling parameter of the electric field to the optical transition is $\Delta\mathbf{p} = \left|\langle ^3E|\mathbf{p}| ^3E\rangle - \langle ^3A_2|\mathbf{p}| ^3A_2\rangle\right| = 0.903 e\AA \approx 4.33$~Debye, and it directs towards the vacancy-nitrogen line along $\langle111\rangle$ direction. This value corresponds to $2.18$~MHz~cm/V which means that a point-like charge (e.g., positively charged N$_\text{s}$) at 50~nm distance from NV center shifts the ZPL energy by $2.4$~GHz. This result agrees well with previous experimental results~\cite{Tamarat2006}, where $\left|\Delta\mathbf{p}\right| \approx 1.5$~Debye was deduced for NV center in diamond.

Basically, the advanced theory of polarization enables to calculate the $d_z$, $d_\perp$ and $d'_\perp$ components of Eq.~\refeq{eq:Efield} in a three-dimensional supercell model as the electric field can be added to the Hamiltonian at an arbitrary direction. Nevertheless, this has not yet been applied to NV center in diamond.       

The ${}^3E$ excited state is much more sensitive to the presence of electric field or strain as the degenerate orbitals may split in the presence of symmetry breaking fields, and they are intertwined~\cite{Maze2011, Doherty2011}. Therefore, the effect of strain is first discussed before discussing the effect of electric field in the ${}^3E$ excited state.

We note that the effect of hydrostatic pressure on the spin levels in the ground state was first studied in experiments~\cite{Doherty2014} and \emph{ab initio} theory~\cite{Ivady2014} before determining the spin-stress coupling parameters. In the theoretical study, the lattice constant of the 512-atom supercell was varied, in order to induce the appropriate pressure on NV center and the $D$-constant was calculated within pseudo wave function approximation~\cite{Ivady2014}. The experimentally deduced linear coupling coefficient is 14.58~MHz/GPa that was measured at room temperature~\cite{Doherty2014}. The results of \emph{ab initio} simulations are valid at $T=0$~K, and yielded 10.3~MHz/GPa (see Ref.~\onlinecite{Ivady2014}). It was found in the \emph{ab initio} simulations, that the local relaxation of ions and adaption of the wave functions to the new ionic positions upon external pressure plays a crucial role in the final response, thus \emph{ab initio} simulations are essential in the calculation of the spin-stress parameters. We further note that the pressure shifts the energy of the ZPL optical transition too by 5.5~meV/GPa~\cite{Kobayashi1993}. DFT PBE $\Delta$SCF calculations yielded 5.75~meV/GPa shift~\cite{Deng2014}, in good agreement with the experimental data. Furthermore, they predicted that the emission in the phonon sideband will be significantly suppressed and redistributed towards the ZPL emission under giant (100~GPa) pressure~\cite{Deng2014}. This has not yet been confirmed in experiments to our knowledge. Next, we turn to the spin-strain coupling parameters.

The spin-strain coupling in the ground state of NV center is more complicated then spin-electric field coupling as strain is a tensor and can result in more coupling coefficients than the electric field does. Indeed, the correct spin-strain coupling coefficients have been only recently derived which contains \emph{six} independent real coupling-strength parameters
$h_{41}$, $h_{43}$, $h_{25}$, $h_{26}$, $h_{15}$, $h_{16}$,
and has the following form~\cite{Udvarhelyi2018}:
\begin{subequations}
\begin{eqnarray}
\label{eq:spinstrain}
H_{\varepsilon} &=& H_{\varepsilon 0} + H_{\varepsilon 1}+H_{\varepsilon 2},
\\
H_{\varepsilon 0} /h&=&
[h_{41} (\varepsilon_{xx} + \varepsilon_{yy}) 
+
h_{43} \varepsilon_{zz}]S_z^2,
\label{eq:h0}
\\
\label{eq:h1}
H_{ \varepsilon 1}/h &=& \nonumber
\frac 1 2
\left[h_{26} \varepsilon_{zx}
- \frac 1 2 h_{25} (\varepsilon_{xx} - \varepsilon_{yy})
\right]
 \{ S_x,S_z\}
\\
&+&
\frac 1 2 
\left(
h_{26}  \varepsilon_{yz} 
+  h_{25} \varepsilon_{xy}
\right)
\{ S_y,S_z\},
\\
H_{\varepsilon 2} /h&=&  \frac 1 2 \left[
	 h_{16}  \varepsilon_{zx}
	- \frac 1 2 h_{15} (\varepsilon_{xx} - \varepsilon_{yy})
\right](S_y^2-S_x^2) \nonumber
\\
&+& \frac 1 2 (
 h_{16} \varepsilon_{yz} + h_{15} \varepsilon_{xy}
) \{S_x, S_y\}\text{,}
\end{eqnarray}
\end{subequations}
where $\varepsilon_{ij} = (\partial u_i/\partial x_j +\partial
u_j/\partial x_i)/2$ denotes the strain tensor and ${\bf u}({\bf r})$
is the displacement field. Similarly to Eq.~\eqref{eq:Efield}, the subscripts $0$, $1$, and $2$ here refer to the difference in the electron spin quantum numbers $ms$ connected by the corresponding Hamiltonian. The spin-stress
Hamiltonian $H_\sigma$ is analogous to Eq.~\eqref{eq:spinstrain}, with the 
substitutions $\varepsilon \mapsto \sigma$ and $h \mapsto g$. The conversion from $h$ to $g$ can be calculated by using the stiffness tensor ($C$) of diamond which connects the external stress to the internal strain in diamond (e.g., Ref.~\onlinecite{Udvarhelyi2018}) with the parameters $C_{11}=1076~\mathrm{GPa}$, $C_{12}=125~\mathrm{GPa}$, $C_{44}=576~\mathrm{GPa}$ in the cubic reference frame. 

The six spin-strain coupling-strength parameters of Eq.~\refeq{eq:spinstrain} were determined using numerical DFT calculations~\cite{Udvarhelyi2018}. 
To model the structure subject to mechanical strain, described by the strain tensor $\varepsilon$, the cubic supercell was deformed to a parallelepiped, 
whose edge vectors are obtained by transforming the
undeformed edge vectors with the matrix 
$1+\varepsilon$ in the cubic reference frame,
and allow the atomic positions to relax.
For each strain configuration, 
the elements of the 
$3 \times 3$
ZFS matrix $D$,
were calculated using the VASP
implementation by Martijn Marsman with
the PAW formalism~\cite{Bodrog2014}.

We illustrate this methodology to obtain the six spin-strain
coupling-strength coefficients with the example of $h_{16}$. 
To determine $h_{16}$,
the supercell using a strain tensor is deformed whose only 
non-vanishing element is $\varepsilon_{yz}$,
and obtain the $D$ matrix from the calculation.
Due to Eq.~\eqref{eq:spinstrain}, 
the chosen strain configuration implies that
the Hamiltonian has the form
\begin{eqnarray}
H = 
\frac 1 2 \varepsilon_{yz}
\mathbf{S}^T \cdot \left( \begin{array}{ccc}
0 &  h_{16}  & 0 \\
h_{16} & 0 &  h_{26} \\
0 &  h_{26} & 0
\end{array} \right)
\mathbf{S}.
\end{eqnarray}
This, together with the above definition of the $D$ matrix, 
yields 
\begin{eqnarray}
\label{eq:dtoh}
h_{16} = 2 
\left.
	\frac{\partial D_{xy}}{\partial \varepsilon_{yz}}
\right|_{\varepsilon = 0}\text{.}
\end{eqnarray}
The numerical error of our DFT calculations was estimated by using a linear fit to the data points that were taken from 11 equidistant values of 
$\varepsilon_{yz}$ between $-0.01$ and $+0.01$ (see Ref.~\onlinecite{Udvarhelyi2018}). The final results for all the coupling parameters are summarized in Table~\ref{tab:DFTresults}.
\begin{table}
\caption{
 Spin-strain ($h$) and spin-stress ($g$) coupling-strength parameters of the ground state as obtained from density functional theory (Ref.~\onlinecite{Udvarhelyi2018}). Results are rounded to significant digits. The negative sign means compression in this convention.}
\begin{ruledtabular}
\begin{tabular}{cl|cl}
\multicolumn{2}{l|}{parameter  (MHz/strain)} & \multicolumn{2}{l}{parameter (MHz/GPa)} \\\hline
$h_{43}$ & $2300\pm200$ & $g_{43}$ & $2.4\pm0.2$\\
$h_{41}$ & $-6420\pm90$ & $g_{41}$ & $-5.17\pm0.07$\\
$h_{25}$ & $-2600\pm80$ & $g_{25}$ & $-2.17\pm0.07$\\
$h_{26}$ & $-2830\pm70$ & $g_{26}$ & $-2.58\pm0.06$\\
$h_{15}$ & $5700\pm200$ & $g_{15}$ & $3.6\pm0.1$\\
$h_{16}$ & $19660\pm90$ & $g_{16}$ & $18.98\pm0.09$\\
\end{tabular}
\end{ruledtabular}
\label{tab:DFTresults}
\end{table}

In Table~\ref{tab:comparison}, we compare the numerical DFT 
results of Table~\ref{tab:DFTresults} to the experimental results of Ref.~\onlinecite{Barson2017}. In Ref.~\onlinecite{Barson2017},
four out of the six independent spin-stress coupling-strength
parameters of the spin-stress interaction Hamiltonian
were measured. Ref.~\onlinecite{Barson2017} 
defines these 4 spin-stress coupling-strength parameters,
denoted as $a_1$, $a_2$, $b$, $c$, 
in a `hybrid' representation, where the spin-stress
Hamiltonian is expressed in terms of the NV-frame components
of the spin vector ($S_x$, $S_y$, $S_z$) and 
the cubic-frame components
of the stress tensor ($\sigma_{XX}$, $\sigma_{XY}$, etc).
To be able to make a comparison between the DFT results
and the experimental ones, 
the notations of Ref.~\onlinecite{Barson2017} were taken,
and $d$, $e$, $\mathcal{N}_x$,
$\mathcal{N}_y$ were introduced, to express
the spin-stress Hamiltonian 
$H_{\sigma}$
in this hybrid representation:
\begin{subequations}
\begin{eqnarray}
H_{\sigma 0} /h &=& \mathcal M_z S_z^2, \\
H_{\sigma 1} /h &=& 
  \mathcal{N}_x \{S_x,S_z\} 
  + \mathcal{N}_y \{S_y,S_z\},
\\
H_{\sigma 2} /h &=& 
  - \mathcal{M}_x (S_x^2-S_y^2)
  + \mathcal{M}_y \{S_x,S_y\}, 
\end{eqnarray}
\end{subequations}
where
\begin{subequations}
\begin{eqnarray}
\mathcal{M}_z &=&
a_1 (\sigma_{XX} + \sigma_{YY} + \sigma_{ZZ}) \nonumber \\
&+& 
2a_2 (\sigma_{YZ} + \sigma_{ZX} + \sigma_{XY}),
\\
\mathcal{N}_x &=&
  d(2 \sigma_{ZZ} - \sigma_{XX} - \sigma_{YY}) \nonumber \\
  &+&e(2\sigma_{XY} - \sigma_{YZ} - \sigma_{ZX}),
\\
\mathcal{N}_y &=& 
\sqrt{3} \left[
  d(\sigma_{XX} - \sigma_{YY})
  + e(\sigma_{YZ} - \sigma_{ZX})
\right],
\\
\mathcal{M}_x &=&
  b(2 \sigma_{ZZ} - \sigma_{XX} - \sigma_{YY}) \nonumber \\
  &+&c(2\sigma_{XY} - \sigma_{YZ} - \sigma_{ZX}),
\\
\mathcal{M}_y &=& 
\sqrt{3} \left[
  b(\sigma_{XX} - \sigma_{YY})
  + c(\sigma_{YZ} - \sigma_{ZX})
\right].
\end{eqnarray}
\end{subequations}
The relations between the hybrid-representation parameters ($a_1$, $a_2$, $b$, $c$, $d$, $e$) and the NV-frame parameters ($g_{41}$, etc)
are given in the first two columns of Table~\ref{tab:comparison}.
Importantly, $H_{\sigma0}$ and $H_{\sigma2}$ is identical
to the spin-stress Hamiltonian in Eqs. (1) and (2) of 
Ref.~\onlinecite{Barson2017}.
\begin{table}
\caption{
Spin-stress coupling-strength parameters in the ground state: Comparison of density functional theory (Ref.~\onlinecite{Udvarhelyi2018}) and experimental (Ref.~\onlinecite{Barson2017}) results.
Parameters in the hybrid representation ($a_1$, $a_2$, etc.)
are expressed in terms of the parameters in the NV-frame
representation ($g_{41}$, etc) in the second column. 
Par. and exp. are abbreviations for `parameters' and `experimental results'.  The negative sign means compression in this convention.}
\begin{ruledtabular}
\begin{tabular}{c|c|ll}
par. & relation & DFT (MHz/GPa) & exp. (MHz/GPa) \\\hline
$a_{1}$ & $\frac{2g_{41}+g_{43}}{3}$ & $-2.66\pm0.07$ & $-4.4\pm0.2$\\
$a_{2}$ & $\frac{-g_{41}+g_{43}}{3}$ & $2.51\pm0.06$ & $3.7\pm0.2$\\
$b$ & $\frac{-g_{15}+\sqrt{2}g_{16}}{12}$ & $1.94\pm0.02$ & $2.3\pm0.3$\\
$c$ & $\frac{-2g_{15}-\sqrt{2}g_{16}}{12}$ & $-2.83\pm0.03$ & $-3.5\pm0.3$\\
$d$ & $\frac{-g_{25}+\sqrt{2}g_{26}}{12}$ & $-0.12\pm0.01$ & -\\
$e$ & $\frac{-2g_{25}-\sqrt{2}g_{26}}{12}$ & $0.66\pm0.01$ & -
\end{tabular}
\end{ruledtabular}
\label{tab:comparison}
\end{table}
It was found (Table~\ref{tab:comparison}) that $d$ and $e$ parameters are in the same order of magnitude as the other four parameters, thus this full six-parameter description is required to deduce accurate parameters from experimental data. From the calculated spin-stress coupling coefficients in the ground state, the ODMR contrast and basic sample dependent data (such as coherence time of the NV center's spin) the sensitivity of strain quantum sensors can be deduced based on Hahn-echo measurements at a given direction of the applied stress~\cite{Udvarhelyi2018pra}. Next, we turn to the spin-strain interaction in the optically allowed excited state.

The ${}^3E$ excited state can be written in the $\{ A_1, A_2, E_x, E_y, E_1, E_2 \}$ manifold, and the strain will affect these as~\cite{Maze2011, Doherty2011}
\begin{equation}
\label{eq:Estrain}
\left( \begin{tabular}{cc|cc|cc}
  &  &  &  & $\delta_{E1}^a$ & $-i\delta_{E2}^b$ \\ 
 &   & &  & $-i\delta_{E2}^b$ & $\delta_{E1}^a$ \\ \hline
 &  & $\delta_{E1}^a$ & $\delta_{E2}^b$  & & $$ \\
 & & $\delta_{E2}^b$ & $-\delta_{E1}^a$  & $$ & \\ \hline
$\delta_{E1}^a$ & $i\delta_{E2}^b$ & & $$ &  & \\ 
$i\delta_{E2}^b$ & $\delta_{E1}^a$ & $$ & & &  
\end{tabular}
\right) \text{,} 
\end{equation}
where $\delta_{E1}^a = (\varepsilon_{xx}-\varepsilon_{yy})/2$,
$\delta_{E2}^a=(\varepsilon_{xy}+\varepsilon_{yx})/2$, $\delta_{E1}^b =
(\varepsilon_{xz}+\varepsilon_{zx})/2$, and $\delta_{E2}^b=(\varepsilon_{yz}+\varepsilon_{zy})/2$. The degenerate levels will split due to symmetry-breaking strain components and may also mix the corresponding states. The symmetry conserving strain ($\varepsilon_{zz}$) only shifts all the levels with the same energy. The effect of the electric field is very similar to that of strain in Eq.~\refeq{eq:Estrain} and may be given as~\cite{Maze2011, Doherty2011}
\begin{eqnarray} 
\label{eq:Eefield}
h d_z E_z + h d_\perp
\left( \begin{tabular}{cc|cc|cc}
$$  &  &  &  & $E_x$ & $-iE_y$ \\ 
 &  $$ & &  & $-iE_y$ & $E_x$ \\ \hline
 &  &  $E_x$ & $E_y$  & & $$ \\
 & & $E_y$ & $ -E_x$  & $$ & \\ \hline
$E_x$ & $iE_y$ & &  & $$ & \\ 
$iE_y$ & $E_x$ & $$ & & &   $$
\end{tabular}
\right) \text{.}
\end{eqnarray}
In Ref.~\onlinecite{Maze2011}, DFT PBE simulation on molecular cluster model was applied together with electric fields, in order to estimate the $d_z$ and $d_\perp$ coefficients. The effect of the electric field was separated to \emph{ionic} effect, i.e., piezo effect, when strain appears due to the presence of electric field, and \emph{electronic} effect, when the cloud of the electrons and the corresponding potential changes due to the applied electric fields (see Ref.~\onlinecite{Maze2011} and Supplementary Materials in Ref.~\onlinecite{Falk2014}). In these simulations, the change in the geometry and splitting of the degenerate orbitals as a function of the applied electric field can be used to derive the order of magnitude for $d_z$ and $d_\perp$. It was found that $h d_z\approx6.6\,\text{GHz (MV/m)}^{-1}$ and $h d_\perp\approx0.6\,\text{GHz (MV/m)}^{-1}$ (see Appendices C and D in Ref.~\onlinecite{Maze2011}). 

As mentioned above the electric field is intertwined with strain via piezo effect, thus electric field may be applied to suppress the effect of strain (c.f., Eqs.~\refeq{eq:Estrain} and \refeq{eq:Eefield}), and tune two individual NV centers' ZPL energies into the same position~\cite{Bassett2011, Pfaff2014}. It was also found that if symmetry breaking strain is present and a symmetry breaking electric field is applied to NV center at the same time then the optical transition will show quadratic effect as a function of the strength of the electric field with anticrossing feature approaching the zero electric field, however, linear behavior shows up at very low strain fields~\cite{Tamarat2008, Maze2011}. This explains the early measurements on the electric field dependence of the ZPL lines where single NV centers showed often linear characteristics upon the applied electric fields but also quadratic features were observed for other single NV centers~\cite{Tamarat2006}. It was also shown that the optical transition from the $A_2$ state towards the $ms=\pm1$ ground state will preserve the circular polarization at low strain fields which was successfully used in entanglement schemes~\cite{Togan2010}. Here we note that we already discussed above that phonons can couple $A_1$ state into $A_2$. This coupling is very small at cryogenic temperature~\cite{Thiering2017} but it becomes substantial at $T=10$~K and above as can be inferred from the deduced ISC rates from the $ms=\pm1$ ${}^3E$ states towards the ${}^1A_1$ state~\cite{Goldman2015prl}. Thus, the afore-mentioned conclusion about the strain dependence of the circular polarization in the optical transition between $A_2$ state and the $ms=\pm1$ ${}^3A_2$ state is valid at very low temperatures.   

\subsubsection{Temperature}
\label{sssec:temperature}

The temperature dependence of the lifetime of the ${}^1\widetilde{E}$ state and the corresponding ISC rate was already discussed in Sec.~\ref{ssec:rates} which could be described by the temperature occupation of a single effective phonon mode. On the other hand, the temperature dependence of the properties ${}^3E$ state is associated with such dynamical effects~\cite{Rogers2009, Plakhotnik2014, Plakhotnik2015}, where the single effective phonon approach is not valid. 

First, we discuss the basic spin Hamiltonian of the ${}^3E$ excited state.
The orthorombic component of the $D$ constant appears in the presence of the perpendicular component of the strain ($\epsilon_\perp = \sqrt{\epsilon_{xx}^2 + \epsilon_{yy}^2}$) which is damped by coupling of the $E_+$ and $E_-$ states by acoustic $E$ phonons. The final spin Hamiltonian of dipolar spin-spin interaction in the $^3E$ state can be expressed as follows~\cite{Plakhotnik2014},
\begin{equation}
H = D \left( S_z^2 - \frac{2}{3} \right) + D_\perp R(T) \left( S_x^2 - S_y^2\right) \text{,} 
\end{equation} 
where $D_\perp = (D_{xx}-D_{yy}) / 2$ and $R(T)= (1 - e ^{-h\epsilon_\perp/k_\text{B}T}) / (1+ e^{-h\epsilon_\perp/k_\text{B}T})$ is the temperature reduction factor. In the limit of zero strain $D_\perp$ disappears. At finite non-zero strain, $D_\perp$ has the largest value in the limit of $T=0$~K and it approaches zero in the limit of $k_\text{B}T \gg h\epsilon_\perp$. The temperature dependence of $D$ component in the ${}^3E$ state is similar to that in the ${}^3A_2$ ground state~(e.g., Ref.~\onlinecite{Plakhotnik2014}), which will be discussed below. Similarly to $D_\perp$, the spin-orbit splitting in the ${}^3E$ state is reduced by increasing the temperature because Raman-scattering of acoustic $E$ phonons that couples $E_+$ and $E_-$ states~\cite{Plakhotnik2014}. This model was applied to understand the temperature dependence of ODMR and ZPL linewidths~\cite{Plakhotnik2015}, and they found consistent results for the observed temperature dependence of the ZPL linewidth~\cite{Fu2009} with assuming quadratic interaction with acoustic $A_1$ phonons in the optical transition~\cite{Plakhotnik2015}, in contrast to the pure DJT model within single effective phonon solution~\cite{Abtew2011}. The afore-mentioned mechanisms are responsible for the temperature dependence of ISC rates between the $ms=\pm1$ states and ${}^1A_1$ state~\cite{Goldman2015prl}. Here we note that the Debye-Waller factor of NV center was found also temperature dependent~\cite{Plakhotnik2014}, that may be utilized for all-optical thermometry with NV center in nanodiamonds.

All the afore-mentioned results are based on the electronic solution of the ${}^3E$ state~\cite{Plakhotnik2014}. On the other hand, the strong electron-phonon coupling due to DJT results in already strongly coupled vibronic solution even at $T=0$~K in the ${}^3E$ state~\cite{Thiering2017}. This implies that the phonon scattering between the vibronic levels and states should be considered which can result in additional scattering channels compared to those from pure electronic solution, e.g., participation of acoustic $A_1$ phonons in the scattering between $\widetilde{A}_1$ and $\widetilde{E}_{1,2}$ states. In addition, the solution of DJT system beyond single effective phonon mode is also temperature dependent which may further increase the complexity of the problem. In the derivation of temperature dependent rates, a cutoff frequency was applied for the Raman-process of $E$ phonons and quadratic interaction with $A_1$ phonons at about 14~meV and 38~meV~(see Ref.~\onlinecite{Plakhotnik2015}), respectively, which closely agrees with the calculated energy gap between the first excited state ($A_1$) and ground state ($E$) vibronic levels in the singlet ${}^1{E}$ and triplet ${}^3{E}$ states~\cite{Thiering2018, Thiering2017}, respectively, which defines the tunneling rate of the electron in those states~\cite{Bersuker2006}.     
 
The temperature dependence of the $D$-constant in the ground state can neither be described by a simple model in which the change in the $D$-constant is attributed to the lattice constants increase at elevated temperatures because this effect is minor in diamond as was confirmed in \emph{ab initio} simulations~\cite{Ivady2014}. Rather, dynamical effects with electron-phonon coupling should be responsible for this effect as explained by a phenomenological approach in Ref.~\onlinecite{Doherty2014}. All these findings imply that the \emph{ab initio} calculation of the temperature dependence of the magneto-optical properties in quasi-degenerate states requires the calculation of either single photon emission and absorption, Raman-process or even more complicated scattering processes between the corresponding orbitals and spin states.

A recent example for \emph{ab initio} simulation of the phonon scattering due to electron-phonon coupling between two-level system is the calculation of electron spin flipping time ($T_1$) or rate ($1/T_1$) in Ref.~\onlinecite{Gugler2018}. It was found previously in experiments~\cite{Jarmola2012} that the rate goes as $\sim T$ or as a constant at $4<T<50$~K depending on the spin density in the diamond sample and the applied magnetic fields, as $\sim \exp (1/T)$ at $T<50<200$~K and as $\sim T^5$ at $T>200$~K (see Refs.~\onlinecite{Takahashi2008, Jarmola2012, Norambuena2018, Astner2018}). The different temperature regimes can be identified as single phonon absorption and emission, Orbach-type process, and second-order Raman-scattering process, respectively~\cite{Jarmola2012, Norambuena2018}. In high purity diamond samples with no external magnetic fields, the $T_1$ exceeded 8 hours for NV electron spin at 15~mK temperature~\cite{Astner2018} which is determined by phononic vacuum fluctuations.

\emph{Ab initio} simulations computed $T_1$ time of the ${}^3A_2$ ground state spin at the very low temperature regime where the single phonon absorption and emission is relevant~\cite{Gugler2018}. At zero perturbation with perfect $C_{3v}$ symmetry, the system can be described as a two-level system where the upper $ms=\pm1$ level is double degenerate with an energy gap of the $D$-constant which separates it from the lower $ms=0$ level. The $D$-tensor depends on the distance of the electron spins localized around the vacancy. During the vibration of atoms the distance between the atoms dynamically change so the $D$-tensor dynamically changes. This effect may result in a spin-flip because of the spin raising and lowering operators that appear in the dipolar electron spin -- electron spin operator (e.g., $S_i S_j$ in Eq.~\refeq{eq:Hss} and Ref.~\onlinecite{Waller1932}). In the \emph{ab initio} simulation, the derivative of the $D$-tensor elements is computed as a function of the ion coordinates along the normal coordinates of the phonon modes for each phonon mode which effectively changes the distance between the electron spins localized on the carbon dangling bonds in the NV center~\cite{Gugler2018}. In order to accelerate the calculation of $D$-tensor components, the total spin density was maximally localized on the $e_{x,y}$ orbitals by Wannier-localization~\cite{Marzari2012}. The final rate at $T\rightarrow 0$~K limit depends on the phonon density of states that may depend on the type of defects and defects' concentrations which shift the phonon density of states towards lower energies~\cite{Gugler2018}. The final calculated $\Gamma_0$ rates fall between $2\times 10^{-5} - 3\times 10^{-5}$~s$^{-1}$ depending on the choice of the phonon density states which are close to the experimental one at $3.47(16) \times 10^{-5}$~s$^{-1}$ in Ref.~\onlinecite{Astner2018}. The temperature dependence can be written as $\Gamma_0 \times \left\{1 + 3 / \left[ \exp ( \hbar D / k_\text{B}T) -1 \right] \right\}$ which goes linearly with temperature because $\hbar D$ is only $138$~mK.      

\section{Outlook}
\label{sec:outlook}

The \emph{ab initio} theory on NV quantum bit in diamond is close to complete but some issues have to be solved. For full description of the ISC rates, the adiabatic potential energy surface should be calculated also for the highly correlated singlet states. This requires to extend the CI-cRPA or similar methods for calculating the quantum mechanical forces acting on the ions. This may reveal the issue of the critical spin-orbit matrix element associated with the perpendicular component of the spin-orbit coupling between the triplet and singlet states. An accurate approach in the calculation of spin density matrix might provide consistent dipolar electron spin - electron spin coupling parameters, which can be important for improving the calculation of the $D$-tensor components in the electronic excited state. 

In general, the calculation of temperature dependence of the magneto-optical properties is still should be further elaborated. Although, the Jahn-Teller theory developed by Bersuker and Ham can be successfully applied to calculate the effective electron-phonon coupling for degenerate states but one has to go beyond the single effective phonon approach for studying the accurate temperature dependence together with the calculation of Raman-scattering process between the resultant states, an analog to the calculations in Ref.~\onlinecite{Gugler2018}. One possible route along this line is the application of the many-body perturbation theory of the electron-phonon coupling~\cite{Marini2008, Cannuccia2011} that was successfully applied to the simulation of the ionization potential of diamondoids~\cite{Gali2016}.       

We further note that the spin-orbit energy is much smaller than the electron-phonon energy for NV center in diamond, thus it can be treated as a perturbation. In other defect quantum bits the spin-orbit and electron-phonon energies are in the same order of magnitude, thus the spin-orbit and electron-phonon coupling should be solved simultaneously~\cite{Thiering2018prx}. This is highly non-trivial with going beyond an effective single phonon approach in the electron-phonon calculations. Spin-orbit interaction may contribute to the zero-field splitting either in the first order (e.g., Ref.~\onlinecite{Thiering2018prx}) or the second order which should be considered for defects with relatively large spin-orbit energies.

Despite the persisting issues raised above, recent progress in the description of solid state defect quantum bits has significantly contributed to the optimization of quantum bit operation and prediction of key properties that could be harnessed in experiments. We note that the full description of quantum bit cannot be separated from the description of the environment which means the interaction with other defects' charge and spin that are present either in bulk or close to the surface, beside other external perturbations such as strain, electric and magnetic fields and temperature. Thus, description of charge and spin dynamics caused by these species should be connected to the full characterization of the target defect quantum bit, in order to compute the readout contrasts, the longitudonal spin relaxation time and the coherence time. By computing these key quantities, the sensitivity of quantum sensing protocols and optimization of quantum control can be designed and may guide future experimental studies. Only this comprehensive approach makes the \emph{ab initio} search for alternative solid state defect quantum bits reliable and powerful that might be superior for a given quantum technology application.

\section{Summary}
\label{sec:summary}

In this paper, we reviewed the \emph{ab initio} theory on the nitrogen-vacancy center in diamond which is an exemplary solid state point defect quantum bit. We summarized the methods to calculate the key magneto-optical properties of this quantum bit, and derived the decay rates and coupling parameters to different types of external perturbation and temperature. We briefly mentioned the missing pieces for complete \emph{ab initio} description of the operation of this quantum bit. The complete description of solid state quantum bits would make possible to convert the phenomenological description of the spin control and quantum sensing to fully \emph{ab initio} solution or, at least, a solution based on \emph{ab initio} wave functions, densities and density matrices.    

\section{Acknowledgment}
A.G.\ acknowledges the Hungarian NKFIH grants No.~KKP129866 of the National Excellence Program of Quantum-coherent materials project, No. 127889 of the EU QuantERA Q-Magine project, EU H2020 Quantum Technology Flagship project ASTERIQS (Grant No.~820394) as well as the National Quantum Technology Program (Grant No. 2017-1.2.1-NKP-2017-00001).


%

\end{document}